\documentclass[usegraphicx,useAMS,usenatbib]{mn2e}
\usepackage{rotating,graphicx,graphics,subfigure,general_cite,epsf,longtable,amsmath,amssymb,sidecap}

\newcommand{\hi}{H\,{\sc i}}
\newcommand{\degree}{$^{\circ}$}
\newcommand{\kms}{km s$^{-1}$}
\newcommand{\etal}{et al.}
\newcommand{\rah}{$^h$}
\newcommand{\ram}{$^m$}
\newcommand{\ras}{$^s$}
\title[AGES: observations of the NGC 628 group] {The Arecibo Galaxy Environment Survey:
    Precursor Observations of the NGC 628 group}
\author[Auld et al.]
       {R. Auld,$^1$, R. F. Minchin$^2$, J. I. Davies$^1$,
    B. Catinella$^2$,W. van Driel$^{3}$,
    P. A. Henning$^{4}$,\newauthor S. Linder$^{5}$,
    E. Momjian$^2$,E. Muller$^6$, K. O'Neil$^{7}$,
    S. Sabatini$^{8}$, S. Schneider$^9$,\newauthor
    G. Bothun$^{10}$, L. Cortese$^1$, M. Disney$^1$, G. L. Hoffman$^{11}$,
    M. Putman$^{12}$,\newauthor J. L. Rosenberg$^{13}$, M. Baes$^{14}$, W. J. G. de Blok$^{15}$,A. Boselli$^{16}$, 
     E. Brinks$^{17}$, N. Brosch$^{18}$,\newauthor 
    J. Irwin$^{19}$,I. D. Karachentsev$^{20}$, V. A. Kilborn$^{6,21}$,
    B. Koribalski$^6$, K. Spekkens$^{22}$ \\
    $^1$Department of Physics \& Astronomy, Cardiff University, UK\\
    $^2$Arecibo Observatory, HC3 Box 53995, Arecibo, PR 00612,
    USA\\
    $^{3}$Observatoire de Meudon, 5 Place Jules Janssen, 92195 Meudon,
    France\\
    $^{4}$Institute for Astrophysics, University of New Mexico, 800
    Yale Blvd, NE, Albuquerque, NM 87131, USA\\
    $^{5}$Hamburger Sternwarte, Universita\"t, Gojenbergsweg 112, 21029
    Hamburg, Germany\\
    $^6$Australia National Telescope Facility, CSIRO, P.O. Box 76,
    Epping, NSW 1710, Australia\\
    $^{7}$National Radio Astronomy Observatory, Green Bank, WV
    24944, USA\\
    $^{8}$INAF-OAR, via di Frascati 33, 00040 Monteporzio Catone,
    Roma, Italy\\
    $^9$Department of Astronomy, University of Massachusetts,
    Amherst, MA 01003, USA\\
    $^10$Department of Physics, University of Oregon, Eugene, OR 97403,
    USA\\
    $^{11}$Hugel Science Center, Lafayette College, Easton, PA 18042,
    USA\\
    $^{12}$Department of Astronomy, University of Michigan, Ann Arbor,
    MI 48109, USA\\
    $^{13}$Harvard-Smithsonian Center for Astrophysics, 60 Garden
    Street, MS 65, Cambridge, MA 02138-1516, USA\\
    $^{14}$European Southern Observatory, Casilla 19001, Santiago 19,
    Chile\\
    $^{15}$RSAA, Australian
    National University, Mount Stromlo Observatory, Cotter Road,
    Weston Creek, ACT 2611, Australia\\
    $^{16}$Laboratoire d'Astrophysique, Traverse du Siphon, BP8, 13376
    Marseille, France\\
    $^{17}$Centre for Astrophysics Research, Science \& Technology
    Research Institute, University of Hertfordshire, Hatfield, AL10
    9AB, UK\\
    $^{18}$The Wise Observatory and the School of Physics and Astronomy,
    Raymond and Beverly Sackler Faculty of Exact Sciences,\\ Tel Aviv
    University, Tel Aviv 69978, Israel\\
    $^{19}$Department of Physics, Queen's University, Kingston,
    Ontario K7L 3N6, Canada\\
    $^{20}$Special Astrophysical Observatory, Russian Academy of
    Sciences, Niszhnij Arkhyz 369167, Zelencukskaya,
    Karachai-Cherkessia, Russia\\
    $^{21}$Centre for Astrophysics and Supercomputing, Swinburne
    University of Technology, P.O. Box 218, Hawthorn, VIC 3122,
    Australia\\
    $^{22}$Jansky Fellow, NRAO, Department of Physics and Astronomy, Rutgers, The State University of New Jersey, Piscataway, NJ 08854, USA\\}
\date{Accepted 0000 December 00.
      Received 0000 December 00;
      in original form 2005 June 07}

\begin{document}
\date{Accepted 0000 December 00.
      Received 0000 December 00;
      in original form 2004 July 26}

\pagerange{\pageref{firstpage}--\pageref{lastpage}} \pubyear{2002}

\maketitle

\label{firstpage}
\clearpage

\begin{abstract}

The Arecibo Galaxy Environment Survey (AGES) is one of several \hi\/ surveys
utilising the new Arecibo L-band Feed Array (ALFA) fitted to the 305m
radio telescope at Arecibo\footnote{The Arecibo Observatory is
  part of the National Astronomy and Ionosphere Center, which is
  operated by Cornell University under a cooperative agreement with
  the National Science Foundation}. The survey is specifically designed to
investigate various galactic environments to higher sensitivity,
higher velocity resolution and higher spatial resolution than previous
fully sampled, 21 cm multibeam surveys. The emphasis is on making
detailed observations of nearby objects although the large
system bandwidth (100 MHz) will allow us to quantify the \hi\/
properties over a large instantaneous velocity range.

In this paper we describe the survey and its goals and present the
results from the precursor observations of a 5\degree$\times$1\degree\/ region containing the
nearby ($\sim 10$ Mpc) NGC 628 group. We have detected all the group
galaxies in the region including the low mass
(M$_{HI}\sim10^7$M$_{\odot}$) dwarf, dw0137+1541 (Briggs, 1986). The
fluxes and velocities for these galaxies compare well with previously
published data. There is no intra-group neutral gas
detected down to a limiting column density of 2$\times10^{18}$cm$^{-2}$. 

In addition to the group galaxies we have detected 22 galaxies beyond
the NGC 628 group, 9 of which are previously uncatalogued. We present 
the \hi\/ data for these objects and also SuperCOSMOS images
for possible optical galaxies that might be associated with the \hi\/ signal. We have used V/V$_{max}$
analysis to model how many galaxies beyond 1000 \kms\/ should be detected and
compare this with our results. The predicted number of detectable
galaxies varies depending on the \hi\/ mass function (HIMF) used in
the analysis. Unfortunately the precursor survey area is
too small to determine whether this is saying anything fundamental
about the HIMF or simply highlighting the effect of low number statistics. This is just one of many questions that will be addressed by
the complete AGES survey.
\end{abstract}


\section{Introduction}
\label{sec:intro}
With the advent of 21cm multibeam receivers on single dish
telescopes, it has become possible to carry out fully sampled surveys
over large areas of sky. The \hi\/ Parkes All Sky Survey (HIPASS:
Barnes \etal\/ 2001) and
its sister survey the \hi\/ Jodrell All Sky Survey (HIJASS: Lang \etal\/
2001) have
recently completed mapping the whole sky south of $\delta$ =+25\degree\/
as well as selected fields north of $\delta$=+25\degree. These surveys
have been used to construct \hi\/ selected 
catalogues of galaxies (Koribalski \etal\/ 2004, Meyer \etal\/ 2004), identify extended \hi\/ structures and High Velocity
Clouds (HVCs) in the Local Group (Putman \etal\/ 2002; Putman
\etal\/ 2003), place limits on the faint end of the HIMF (Kilborn,
Webster \& Staveley-Smith 1999; Zwaan \etal\/ 2005), to identify a population of gas-rich
galaxies (Minchin \etal\/ 2003), to place limits on the number
of previously undetected \hi\/ clouds and extended \hi\/ features
(Ryder \etal\/ 2001; Davies \etal\/ 2001; Ryan-Weber \etal\/ 2004), to measure the \hi\/ properties of cluster galaxies
(Waugh \etal\/ 2002; Davies \etal\/ 2004) and to measure
the cosmic mass density of neutral gas (Zwaan \etal\/
2003). Although these surveys have been very successful in many ways
they have suffered from rather poor sensitivity ($\sim$13 mJy beam$^{-1}$),
velocity resolution ($\sim 18$ \kms) and spatial resolution (HIPASS: $\sim
15$\arcmin, HIJASS: $\sim 12$\arcmin). They have however set a
benchmark against which other surveys will be measured. 

The installation of the Arecibo L-Band Feed Array (ALFA) at Arecibo presents the opportunity
to survey the sky with higher sensitivity and higher spatial
resolution than has been achievable previously. In addition to the
receiver array, new back-ends have been designed specifically for Galactic,
extragalactic and pulsar observers. Three consortia have been formed 
to gain the full potential from ALFA: PALFA for Pulsar research, GALFA
for Galactic science and EALFA is the
extragalactic consortium. The new back-end for extragalactic astronomers 
provides not only increased spectral resolution but also more bandwidth than
was available for previous \hi\/ surveys. 

The Arecibo Galaxy Environment Survey (AGES) forms one of four
working groups within EALFA and will study environmental effects on
\hi\/ characteristics. The Arecibo Legacy
Fast ALFA, ALFALFA, survey (Giovanelli \etal\/ 2005) covers a larger
area ($\sim$7000 sq. deg.) but is much shallower than AGES
(T$_{int}\sim$24s). AGES will use 2000 hours of telescope time over
the next 5 years and will go deeper than ALFALFA to produce a much
more sensitive survey (T$_{int}\sim$300s) over a smaller area
($\sim$200 sq. deg.).

The Zone of Avoidance survey, ZOA (Henning \etal\/ in prep.) will
concentrate on the low-Galactic latitude ($b<5$\degree) portion of the
sky visible to Arecibo. By performing commensal
observations with the Galactic ALFA working group (GALFA) and the
Pulsar ALFA working group (PALFA), they will reach a T$_{int}\sim 270$s. 

The ALFA Ultra Deep Survey, AUDS (Freudling \etal\/ in prep.) will
reach an integration time of $\sim$75 hours, over a small area
($\sim$0.4 sq. deg.) and will search for objects more distant than the
previous surveys will be able to detect.

The paper is organised as follows: first we describe the aims of the AGES survey
and the survey fields. Then in section \ref{sec:survey} we discuss the observing strategy and the
data reduction. In section \ref{sec:obs} we describe the precursor
observations of the field around NGC 628 and present preliminary analysis. In \ref{sec:results} we present
the results from the NGC 628 group and discuss objects detected beyond
the NGC 628 group. Conclusions of the AGES precursor observations are
given in section \ref{sec:conc}.

\section {The Arecibo Galaxy Environment Survey (AGES)}
\label{sec:survey}
Exploiting the improvements in sensitivity, velocity resolution and spatial
resolution offered by the ALFA system, the
Arecibo Galaxy Environment Survey (AGES) aims to study
the atomic hydrogen properties of different galactic environments to
sensitive limits; low \hi\/ masses ($5\times10^6$M$_{\odot}$)\footnote{at the distance
of the Virgo Cluster (16 Mpc)} and low column 
densities ($3\times10^{18}$cm$^{-2}$). These environments
range from apparent voids in the large scale structure of galaxies,
 via isolated spiral galaxies and their halos, to galaxy-rich
regions associated with galaxy clusters and filamentary
structures. Our intentions are: to explicitly investigate the HIMF in
each environment, to measure the spatial distribution of \hi\/ selected
galaxies, to identify individual low mass and low column density
objects, to determine the low column density extent of large
galaxies and to compare our results with expectations derived from QSO
absorption line studies. In addition we aim to explore the nature of
High Velocity Clouds (HVCs) and their possible link to dwarf galaxies,
measure the contribution of neutral gas to the global baryonic mass
density, identify gaseous tidal features as signatures of mergers and
interactions and compare our results with numerical models of galaxy
formation.

While ALFALFA and AGES share similar goals, by going deeper
AGES will be able to reach lower mass limits and column density limits
than ALFALFA at any given distance. An added benefit of choosing smaller
fields is that deep multi-wavelength comparison surveys become much more
feasible. This is extremely important as it will allow us to study not
only objects of interest but also the surrounding environment which
may be influencing their evolution. This environment may stretch over
length scales of degrees, particularly for nearby galaxies, requiring
sensitive telescopes with large fields of view. The Hubble Space Telescope,
MEGACAM on the Canada-France-Hawaii Telescope, GALEX and UKIRT amongst
others will be used for deep observations of each of the AGES fields
in optical, UV, and NIR bands.

\subsection{The Arecibo L-band Feed Array (ALFA)}
Here we summarise the properties of ALFA. For a more complete
description the reader is referred to Giovanelli \etal\/ (2005). ALFA operates between 1.225 and 1.525 GHz and consists of a
cluster of seven cooled dual-polarisation feed horns. The outer six feeds are
arranged in a hexagonal pattern around the central beam. The digital
back-end signal processors are Wide-band Arecibo Pulsar Processors
(WAPPs) that have been upgraded to perform spectral line
observing. The WAPPs were configured to cover 100 MHz
observing bandwidth with 4096 channels, giving a channel spacing
of 24.4 kHz $\equiv$ 5.15 \kms\/ at the rest frequency of \hi\/. At 1.4
GHz the mean half power beam width is 3.4\arcmin and the mean system
temperature is 30 K. 

\subsection{The AGES fields}
\label{sec:fields}

\begin{figure*}
\includegraphics[width=1.0\textwidth]{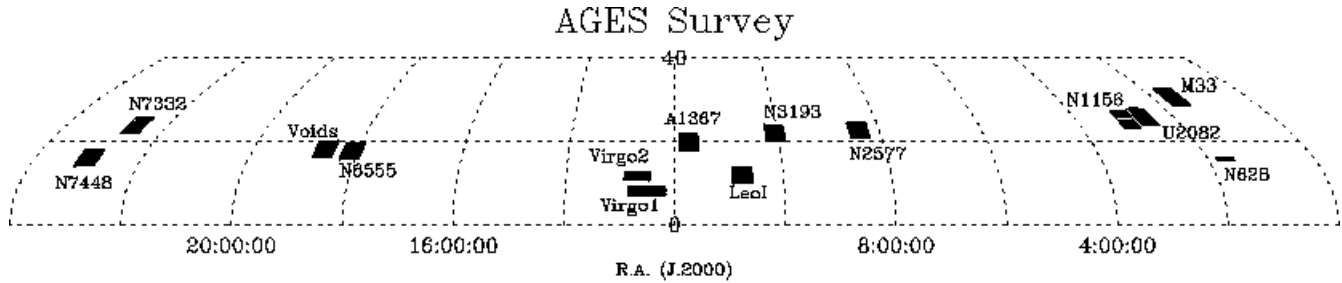}
  \caption{A northern sky map showing the location of the AGES fields. \label{fig0}}
\end{figure*}

Because Arecibo is a fixed-dish telescope the observable sky is
restricted in declination ($-1^{\circ} \la \delta \la +38^{\circ}$)
on-source time is limited to typically a 2 hour window centred on the
meridian on any night. To optimise scheduling, AGES fields were chosen across a range of RAs
(Fig. \ref{fig0}). Most areas comprise a 5\degree$\times$4\degree\/
field with an integration time of 300s per point in two
polarisations. The rms noise in the final data is estimated to be
between 0.5--1 mJy/beam per channel for a channel separation of 5.15
\kms. At the distance of the Virgo cluster ($\sim$16 Mpc) this will
enable us to reach an \hi\/ mass limit of $5\times10^6 $M$_{\odot}$
($W_{50}$ = 30 \kms), and will permit us to detect \hi\/ masses as low
as $5\times10^7$M$_{\odot}$ out to 3 times this distance. This is of
particular significance as the HIMF is poorly constrained below 10$^8$M$_{\odot}$.

\subsubsection{The Virgo cluster}

We have selected two regions of the Virgo cluster to make a comparison
of the \hi\/ properties of cluster galaxies with those in the
field. Both regions avoid the inner 1\degree\/ of the cluster
as the large continuum source associated with M87 limits the dynamic
range in the area of about 1 degree around it. Also previous
observations indicate very little \hi\/ within the cluster core (Davies
\etal\/ 2004). 

The first region is 10\degree$\times$2\degree\/ centred on ($\alpha, \delta$)
= 12\rah\/ 30\ram\/ 00\ras, 8\degree\/ 00\arcmin\/ 00\arcsec. This region
will be imaged as part of the UKIDSS large area survey (Lawrence
\etal\/ in prep.) in $J, H$ and $K$ bands. We also have deep $B$-band
images of the region taken with the Isaac Newton Telescope, on La
Palma. To complement these data, this region will also be observed by
GALEX (Martin \etal\/ 2005) in the near (2300 \AA) and far ultraviolet
(1500 \AA) later in 2006. NIR bands are less susceptible to
extinction and are thus more accurate measures of stellar mass. Optical-NIR
colours can also be used for studying galaxy metallicity and star
formation histories. UV is a tracer of ongoing star formation and will
allow us to compare current star formation rates with integrated star
formation histories. The UV is also useful for searching for signs of
star formation in tidal features.

The second region is a 5\degree$\times$1\degree\/ E--W strip centred on
($\alpha, \delta$) = 12\rah\/ 48\ram\/ 00\ras, 11\degree\/ 36\arcmin\/
00\arcsec. This strip extends radially towards the cluster edge and
will be used to observe changes in galaxy properties with radius from
the cluster centre, expanding on the work of Sabatini \etal\/
(2003). This strip extends through the galaxy grouping known as
sub-cluster A that is associated with M87 (Binggeli \etal\/ 1993). The
results of these observations will be useful for comparing with models
of how the cluster environment affects galaxy evolution (tidal
stripping, ram pressure stripping etc.).

We will also search for low mass galaxy companions or cluster dwarf
galaxies as well as previously unidentified \hi\/ clouds. Another
interesting study will be to see how baryonic material (stars,
X-ray gas, neutral gas) in the cluster is distributed and compare this
with the environment surrounding field galaxies. 

\subsubsection{The local void}
 We have selected a 5\degree$\times$4\degree\/ region centred on
 ($\alpha, \delta$) = 18\rah\/ 38\ram\/ 00\ras, 18\degree\/ 00\arcmin\/
 00\arcsec. We will search for \hi\/ signatures that might be
 associated with very low surface brightness galaxies or with \hi\/
 clouds devoid of stars. The properties of galaxies detected will
 provide an interesting comparison with those in more dense
 environments.

\subsubsection{M33 and the Perseus-Pisces filament}
This will comprise a 5\degree$\times$4\degree\/ region centred on
 ($\alpha, \delta$) = 01\rah\/ 34\ram\/ 00\ras, 30\degree\/ 40\arcmin\/
 00\arcsec . The aim is to map in detail the environment of
 M33 to search for tidal bridges, HVCs etc. that are signatures of
 diffuse hydrogen in the Local Group. Westmeier \etal\/ (2005) and Braun \&
 Thilker (2005) have already observed numerous HVCs around M31 and
 M33, as well as a \hi\/ bridge connecting M31 and M33.
 
 M33 will occupy about 1 sq. deg. at the centre of the field. At the
 distance of M33, the ALFA beam  width (3.4\arcmin) will be about 0.6
 kpc, providing us with superb  spatial resolution. The \hi\/ mass
 sensitivity will be as low 2$\times  10^4 $M$_{\odot}$. The
 Perseus-Pisces filament will become visible behind M33 at
 approximately 4000--6000 \kms.

\subsubsection{The cluster A1367}
A1367 is a spiral-rich, dynamically young cluster at a velocity of
around 6500 \kms, and is currently forming at the intersection of two
large scale filaments (Cortese \etal\/ 2004). Recent optical and X-ray
observations (i.e. Gavazzi \etal 2001; Sun \& Murray, 2002)
suggest that ram pressure stripping and tidal interactions are
strongly affecting the evolution of cluster galaxies, making this
cluster the ideal place to study the environmental effects on the gas
content. The whole cluster will just about fill a
5\degree$\times$4\degree\/ region centred on  ($\alpha, \delta$) =
11\rah\/ 44\ram\/ 00\ras, 19\degree\/ 50\arcmin\/  00\arcsec . Given the
morphology of the cluster we expect many detections including new
objects that are more prominent in \hi\/ than other wavelengths. The
\hi\/ mass detection limit at this distance will be approximately
2$\times 10^8$ M$_{\odot}$.

\subsubsection{The Leo I group}
This group lies at about 1000 \kms\/ and we will survey a
 5\degree$\times$4\degree\/ region centred on ($\alpha, \delta$) = 10\rah\/
 45\ram\/ 00\ras, 11\degree\/ 48\arcmin\/ 00\arcsec. The group is of
 particular interest because of the relatively large number of
 early-type galaxies (e.g. NGC 3377 \& NGC 3379) and will make a good
 comparison with spiral rich groups. We should reach a lower \hi\/
 mass limit of $\sim 2\times 10^6$ M$_{\odot}$ at its distance of 10 Mpc.

\subsubsection{The NGC 7448 group}
NGC 7448 is a Sbc spiral galaxy at a velocity of $\sim$ 2200 \kms, with
a number of late and early type companions. The group will serve as a
good contrast to the Leo I group. We will study a
 5\degree$\times$4\degree\/ area centred on ($\alpha, \delta$) = 23\rah\/
 00\ram\/ 00\ras, 15\degree\/ 59\arcmin\/ 00\arcsec. At the distance
 of this group (30.6 Mpc) we should reach a \hi\/ mass limit of
 1.8$\times 10^6$ M$_{\odot}$.

\subsubsection{The NGC 3193 group}
NGC 3193, in contrast to NGC 7448, is an elliptical galaxy and there are
another 9 known group members all part of a well defined galaxy
filament. NGC 3193 has a velocity of 1362 \kms\/ and we will be
surveying an area of 5\degree$\times$4\degree\/ centred on ($\alpha,
\delta$) = 10\rah\/ 03\ram\/ 00\ras, 21\degree\/ 53\arcmin\/ 00\arcsec. The
lower \hi\/ mass limit at this distance (18.9 Mpc) will be approximately 7.0$\times 10^6$ M$_{\odot}$.

\subsubsection{Individual galaxies}
This will consist of observations of pairs of galaxies (principle
galaxy either early or late type) and very isolated galaxies such as
NGC 1156 and UGC 2082. NGC 1156 has no neighbouring galaxies within
10\degree\/ (Karachentsev, 1996), UGC 2082 has no neighbour within 5
\degree\/ in the Nearby Galaxies Atlas (Tully \& Fisher, 1987). Details 
are shown in Table \ref{tbl2}.

\subsubsection{AGESVOLUME}

In addition to the data collected from the targets mentioned above, the large
system bandwidth (100 MHz) will allow us to simultaneously sample a
much deeper volume of the universe and search for more distant galaxies
along the line of sight out to $cz\sim19000$ \kms. The detection limit
at this distance will be M$_{HI}=1.4\times10^{9}$M$_{\odot}$. This is
the AGESVOLUME and will be a very important part of the project,
providing information on the HIMF, the baryonic mass density of
neutral gas and the distribution of \hi\/.

\begin{table*}
  \centering
%  \begin{minipage}[c]{145mm}
  \caption{\hi\/ Survey fields for individual galaxies.\label{tbl2}}
  \begin{tabular}{lcccccccc}
    Galaxy & RA & dec. & Type & $V_{sys}$ & Distance & \hi\/ mass
      limit & Comments & Survey\\
      &($h~m~s$) &(\degree\/ \arcmin\/ \arcsec\/)&& (\kms) &
      (Mpc)&(M$_{\odot}$)&&area\\
      \hline
      NGC 6555 & 18 06 00 & 17 30 00 & Sc,
      & 2225 &30.9 &1.9$\times 10^{7}$&paired with& 5\degree$\times$4\degree\\
      &&&face-on&&&&NGC 6548\\
      
      NGC 2577&08 24 00&22 30 00&S0&2057&28.6 &1.6$\times 10^{7}$&paired with&5\degree$\times$4\degree\\
      &&&&&&&UGC 4375&\\
      NGC 7332&22 36 00&23 48 00&S0 (pec)&375&5.2 &5$\times 10^{5}$&paired with&2.5\degree$\times$2\degree\/\\
      &&&&&&&NGC 7339&\\
      UGC2082&02 36 00&25 48 00&Sc,
      &696&9.7 &1.8$\times 10^{6}$&very&2.5\degree\/$\times$2\degree\/\\
      &&&edge-on&&&&isolated&\\
      NGC 1156 &03  00 00&25 12 00&Irr&375&5.2 &5$\times 10^{5}$&
      very&2.5\degree$\times$2\degree\/\\
      &&&&&&&isolated&\\
      \\
      \hline\\
  \end{tabular}
  %  \end{minipage}
\end{table*}

\subsection{Observing strategy}
\label{sec:surveyobs}
A number of observing strategies were considered for the survey, which
involved pointed observations, drift scanning, driven scanning
techniques or combinations thereof. The successful observing method
had to strike the right balance between data quality and observing
efficiency.

Giovanelli \etal\/ (2005) describe in detail an observing mode known
as drift scanning. As the name suggests, the array is kept at a fixed
azimuth and elevation while the sky drifts overhead. The benefits of
this technique are that the telescope gain remains constant and the
system temperature only varies slowly over a single scan. Standing
waves set up between the receiver and the dish cause ripples to appear
in the spectrum baseline. These waves are quasi-periodic and depend on
the polarisation, the geometry of the dish and the location of
continuum sources in the main beam and sidelobes. Hence each beam and
polarisation will have a different baseline ripple at any given
time. This baseline ripple can cause confusion with low signal sources
so limiting its effect is highly desirable. The drift scan technique
has the advantage that this ripple is more or less constant over the
course of the scan. It is then possible to accurately estimate the
ripple and remove it from the data during the data reduction. The
benefits made the drift scan highly desirable for our own
observations. 

The Earth's rotation rate governs the on source time in a drift
scan. For Arecibo this means that each point in the sky takes 12s
to cross the beam at 1.4 GHz. 25 separate scans are then required
 per point to reach the 300s integration time. Since the technique
 used in ALFALFA only allows for scans taken along the meridian
 (Giovanelli \etal\/ 2005) it was necessary to adapt the method to
 allow us to conduct scans before and after transit, and hence build
 up sky coverage more quickly. 

\begin{figure}
  \begin{center}
  \subfigure{
    \includegraphics[width=0.2\textwidth]{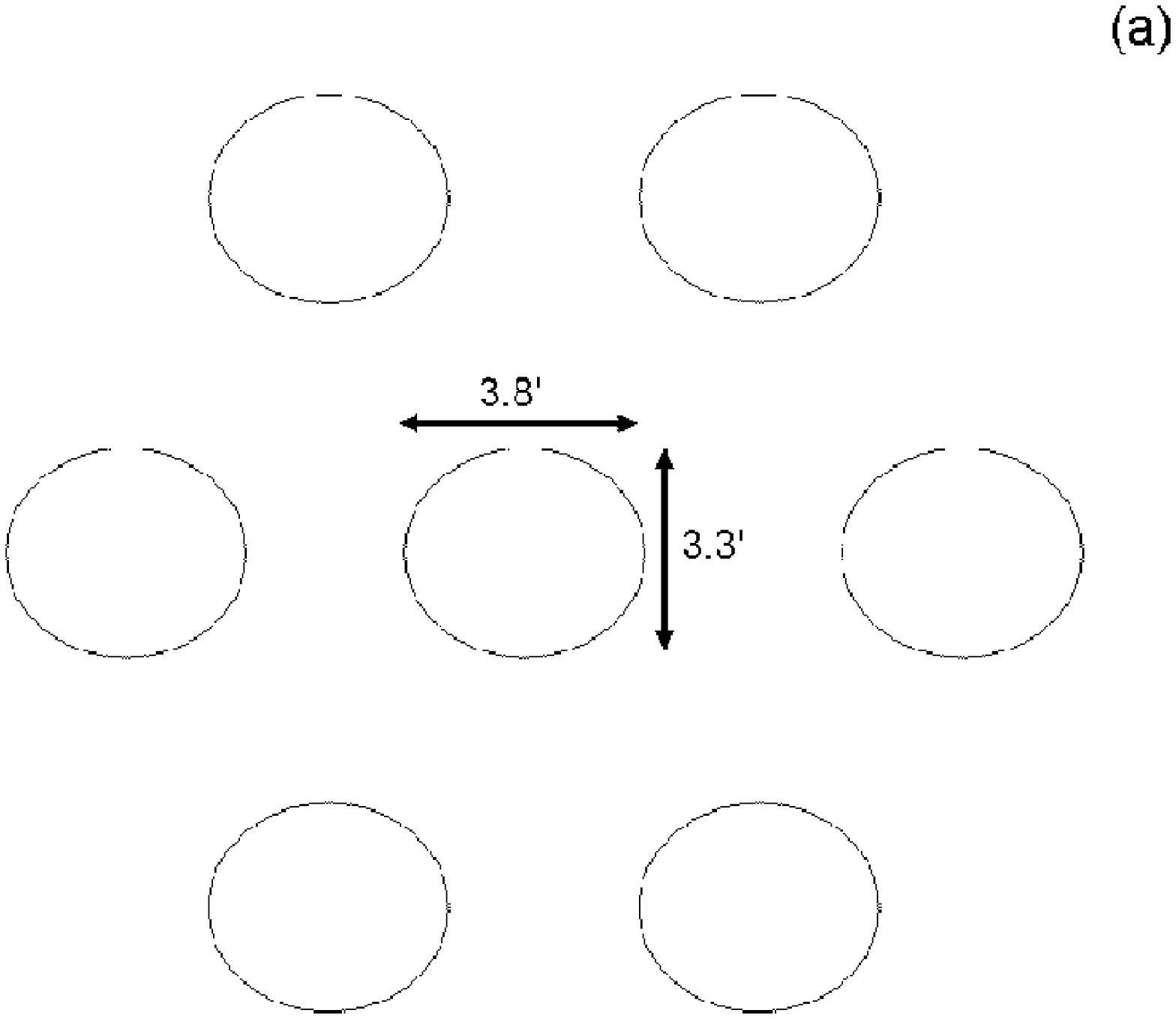}
    \label{fig02a}}
  \subfigure{
    \includegraphics[width=0.2\textwidth]{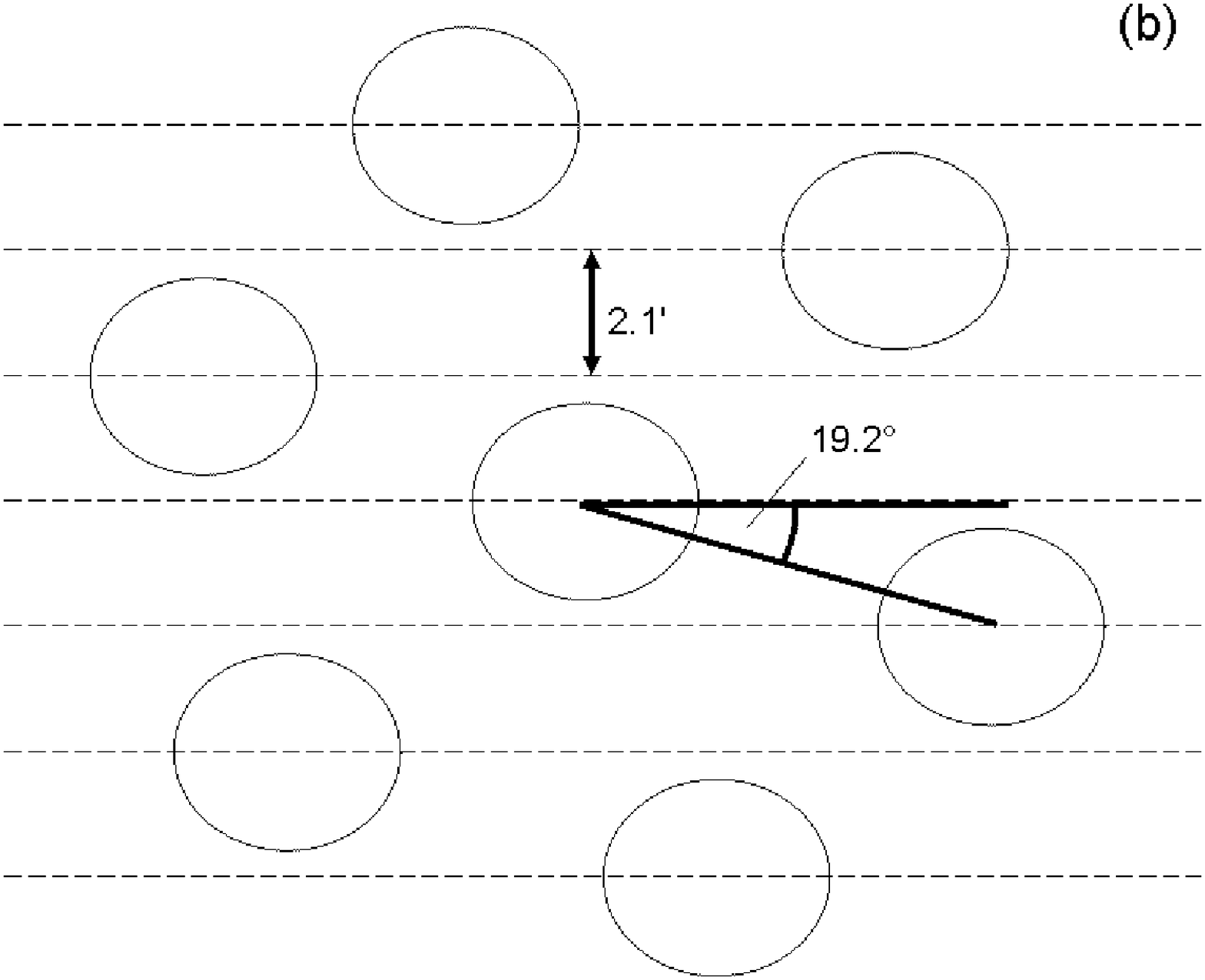}
    \label{fig02b}}
  \caption{An illustration of the projected ALFA footprint, (a)
  unrotated with respect to the direction of the scan, and (b) rotated
  to produce equidistant beams of 2.1\arcmin\/ separation. Dotted
  lines indicate the individual beam tracks across the sky. \label{fig02}}
  \end{center}
\end{figure}

In order to compensate for the change
 in parallactic angle, and thus achieve uniform sky coverage, ALFA
 must be rotated before every scan. The geometry of the Gregorian
 system projects a slightly elliptical beam pattern on the sky. The
 projected beams are also slightly elliptical (see
Fig. \ref{fig02}). As a result of these
 effects, once the array is rotated to produce equidistant beam
 tracks, the beam separation is slightly larger than that required for
 Nyquist sampling. To attain fully sampled sky coverage it is
 necessary to stagger the declination of individual scans by
 $\sim$1\arcmin\/ (half the beam separation). An IDL routine has been
 developed to calculate the rotation angle and the starting
 coordinates and LSTs in advance of every observing run. This
 staggering is arranged such that every beam covers the same patch of
 sky helping to combat variations in gain between the beams. Another
 benefit of this method is to reduce the impact of sidelobe
 contamination. This effect is discussed separately in the next section. 

\subsection{Data reduction}
Data reduction will be performed using the {\sc aips++} packages
{\sc livedata} and {\sc gridzilla} (Barnes \etal\/ 2001), developed by the Australian
Telescope National Facility (ATNF). These packages were originally
designed for the HIPASS and HIJASS surveys but were modified to accept
the Arecibo CIMAFITS file format. {\sc livedata} performs bandpass
estimation and removal, Doppler tracking and calibrates the residual
spectrum. It also has the ability to apply spectral smoothing if
required. Bandpass calibration can be performed using a variety of
algorithms, all based around the median statistic. The median has
the advantage that it is more robust to outliers than the mean
statistic. This property makes the median estimated bandpass more
resistant to radio frequency interference (RFI).

{\sc gridzilla} is a gridding package that co-adds all the spectra
using a suitable algorithm, to produce 3-D datacubes. The user has
full control over which beams and polarisations to select, the
frequency range, the image size and geometry, pixel size and data
validation parameters as well as the gridding parameters
themselves. Various weighting functions are available for gridding the
data. The output datacube has two spatial dimensions and the spectral
dimension can be chosen by the user to be frequency, wavelength or
velocity (numerous conventions are available for each choice). Full
details of the bandpass estimation and gridding technique are given in
Barnes \etal\/ (2001).

\begin{figure}
  \subfigure{
    \includegraphics[width=0.22\textwidth]{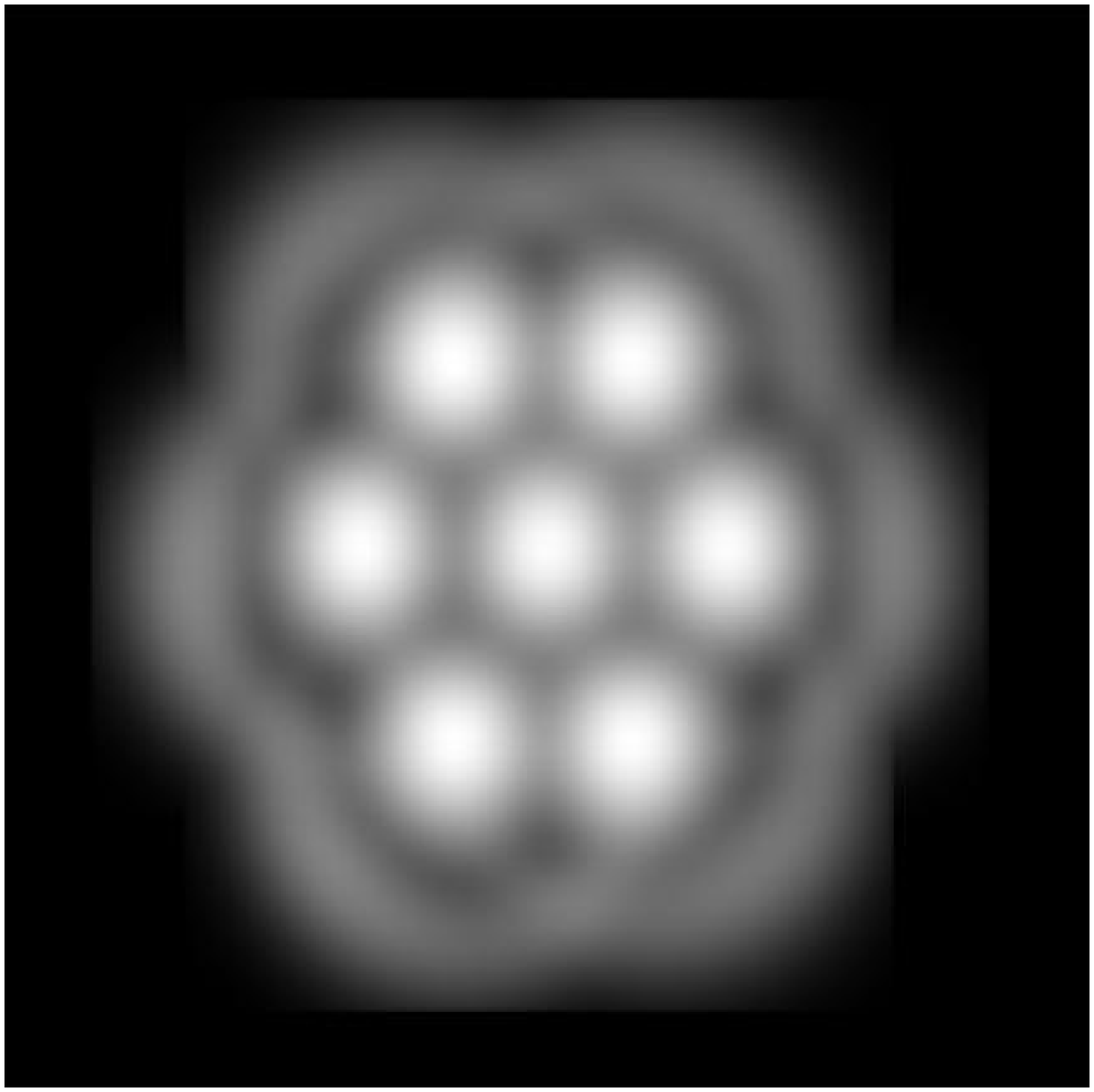}
    \label{fig03a}}
  \subfigure{
    \includegraphics[width=0.22\textwidth]{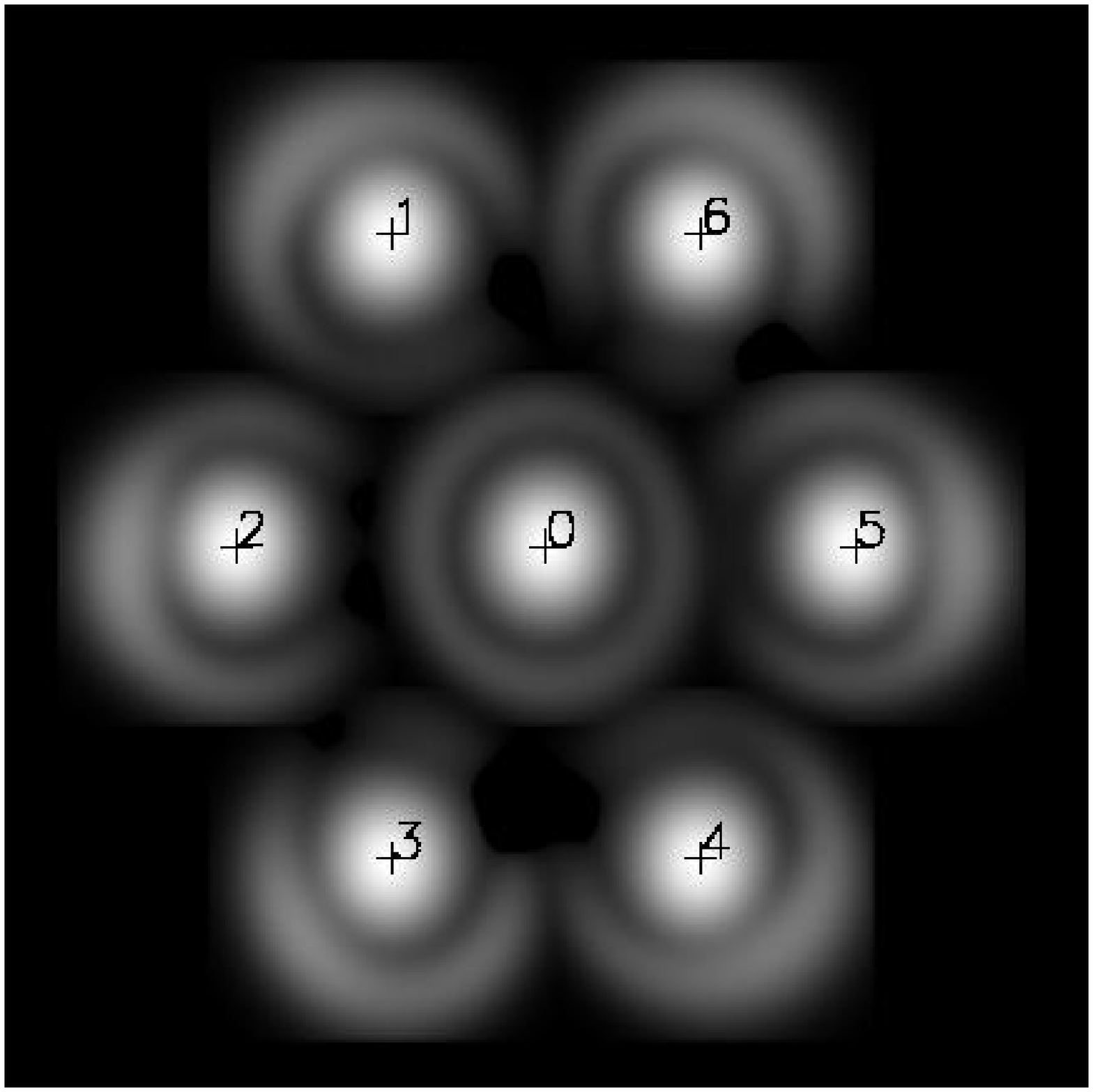}
    \label{fig03b}}
\caption{An illustration of the ALFA beam footprint to emphasise the
  strength and symmetry of the sidelobes. (a) A highly stretched image
  of the combined response of all seven ALFA beams, the image size is
  32\arcmin$\times$32\arcmin. (b) The beam separations have been
  stretched to 2.5 times
  the true separations to highlight the sidelobes of the beams. The
  angular scales of each beam is correct and unchanged. The
  image size in this case is 48\arcmin$\times$48\arcmin. (Figures
  appear courtesy of Carl Heiles.)\label{fig03}}
\end{figure}

The sidelobes for the central beam are symmetric, while sidelobe
levels for each of the outer beams are highly asymmetric, as shown in
Fig. \ref{fig03}. Our ability to regain low column density emission
could be highly compromised if they are not corrected, particularly
near bright objects or very extended objects. In the
final, gridded data, each beam has contributed equally to each pixel
apart from at the extreme edges of the map where sky coverage is not
complete. Fig. \ref{fig03b} highlights the radial alignment of the
peaks of each of the sidelobes. When median gridding is applied
this means that in order for a pixel to be significantly contaminated
by sidelobe emission, the source must appear in more than three of the
beams. Point sources can therefore be considered free of sidelobe
contamination. It should be noted that there is likely to be some low
level contamination in extended objects and we are investigating ways
to reduce this.

\section{Precursor observations}
\label{sec:obs}
The observations were taken using the 305m radio telescope located at
Arecibo, Puerto Rico. We were allocated a total of 86 hours which was
split between different observing strategies, as part of a `shared
risk' strategy. The drift scan observations
received 42.5 hours and we discuss only these observations in this
paper. The observations were taken over 22 nights between November
19th 2004 and December 18th 2004. No radar blanking was used because
it was not functional at the time, leading to several sources of RFI
as discussed later. System malfunctions were inevitable during a
testing run such as this, but only four nights' of observing were
affected. This corresponded to approximately 10 hours lost time. 

\subsection{The NGC 628 group}
The group is centred round a large, face-on spiral, NGC 628 and the peculiar
spiral NGC 660. NGC 628 is associated with several companions: UGC 1104,
UGC 1171, UGC 1176(DDO13), UGC A20 and KDG10. NGC 660 also has a couple of
companions: UGC 1195 and UGC 1200 (Fig. \ref{fig1}). Most of the companions are star
forming dwarf irregulars and together these galaxies form
a gas-rich group analogous to the Local Group (Sharina \etal, 1996). 
The recessional velocity of NGC 628 is $\sim$660 \kms\/ (Huchra, Vogeley
\& Geller, 1999; de Vaucouleurs, 1991; Kamphuis \& Briggs, 1992)
placing it at approximately 10 Mpc away (assuming no peculiar motion and
$H_o$ = 75 \kms\/ Mpc$^{-1}$, as used in Briggs, 1986). Supernovae
measurements (Hendry \etal\/ 2005) give a distance measurement of
9.3$\pm$1.8 Mpc.

NGC 628 has a very narrow velocity width, $W_{50} = 56$
\kms\/ (Kamphuis \& Briggs 1992). It is close enough to allow detection of low-mass \hi\/ objects with
only M$_{HI} = 2\times 10^6 $M$_{\odot}$ of neutral hydrogen, 
but separated enough from local Galactic gas to be able to detect any high
velocity gas associated with the galaxy. Kamphuis \& Briggs (1992)
looked at this galaxy at 21cm with the VLA and discovered high
velocity gas on the outskirts of NGC 628 but nothing that could be
considered HVCs. A few years earlier another survey (Briggs, 1986) revealed a new, low-mass companion, dw0137+1541 (M$_{HI} = 7
\times 10^6 $M$_{\odot}$). This group has also been covered by the HIPASS
northern extension. The interesting aspects of the
group and the availability of comparison observations from other
instruments made it a natural choice for precursor observations using
a new instrument. 

\subsection{Observations and data reduction}
\label{sec:reduc}
\begin{figure}
\includegraphics[width=0.48\textwidth]{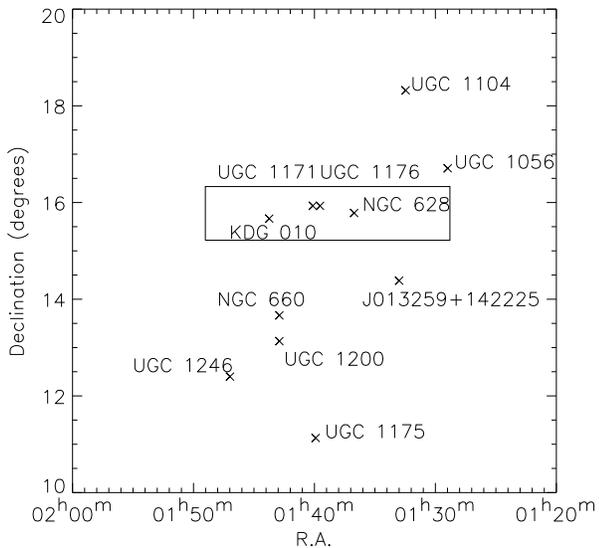}
  \caption{The NGC 628 group with the observed field outlined. Group
  members are denoted by a cross, dw0137+1541 has not been included
  due to its proximity to UGC 1176 and UGC 1171. \label{fig1}}
\end{figure}

A 5\degree$\times$1\degree\/ field was chosen centred on a position
halfway between NGC 628 and the nearest dwarf UGC 1176
(Fig. \ref{fig1}). The region includes NGC 628, UGC 1176, UGC 1171,
KDG010 and dw0137+1541. The observing band was centred on 1381 MHz,
giving a heliocentric velocity range of $-$2270 \kms\/ to $+$20079 \kms\/. The
roll off at the edge of the bandpass due to the filter reduces the
sensitivity for about 1000 \kms\/ at either end. 

\begin{figure}
  \begin{center}
  \subfigure{
    \includegraphics[width=0.48\textwidth]{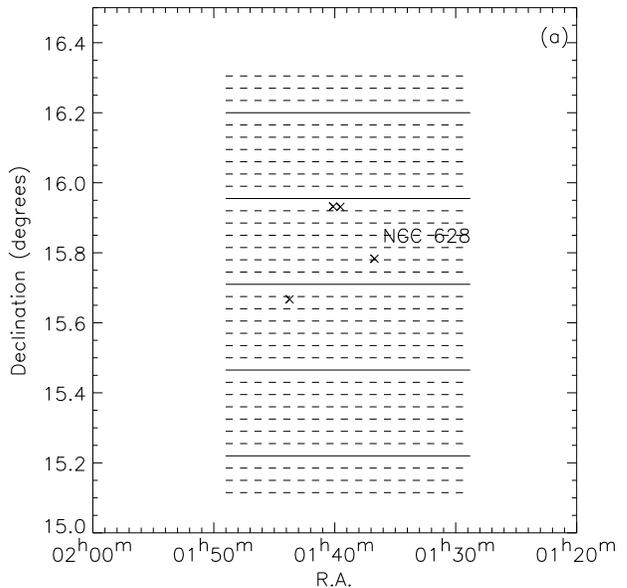}
    \label{fig2a}}
  \subfigure{
    \includegraphics[width=0.48\textwidth]{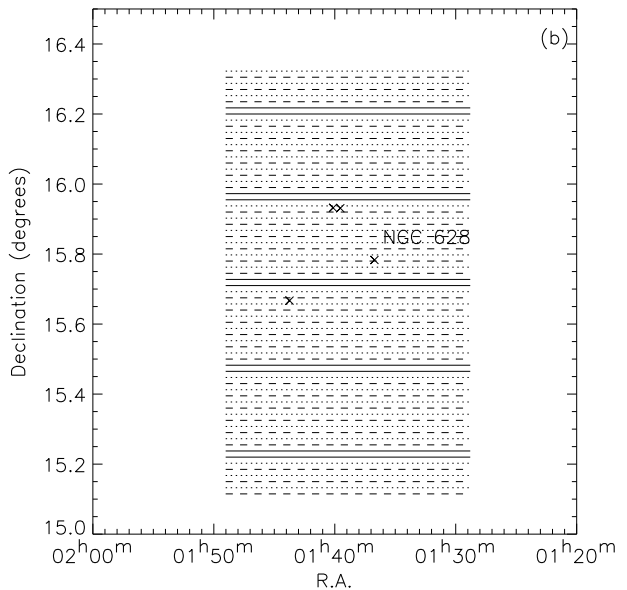}
    \label{fig2b}}
  \caption{An illustration of how the sky coverage is built up over two
    nights of observations. The declination axis has been stretched
    for convenience. The solid lines represent the central beam
    tracks, the dashed lines mark the outer beam
    tracks. On the first night (a), 5 separate drift scans
    are performed to cover the 5\degree$\times$1\degree\/
    area. On the second night (b), exactly the same pattern is reproduced
    offset by half the beam separation (the outer beams from the
    second night at denoted by dotted lines). On subsequent nights the two
    patterns are alternated to build up integration time and also
    maintain uniform sky coverage. Due to observation scheduling it
    was often necessary to drop the northern-most scan. Galaxies are
    marked with a cross, with NGC 628 marked for guidance. Refer to
    Fig. \ref{fig1} to locate other group members.  \label{fig2}}
  \end{center}
\end{figure}

The strategy consisted of two sets of scans. The first set covered the
entire field on one night while the second set were offset by half a
beam width as shown in Fig. \ref{fig2}. This observing strategy differs
slightly from those described in section \ref{sec:surveyobs} since in
this technique individual points in the sky are only covered by one
or two beams. This was a necessary sacrifice since the tools that
assisted in the design of the survey strategy were still in development.

\begin{figure*}
\includegraphics[width=1.0\textwidth]{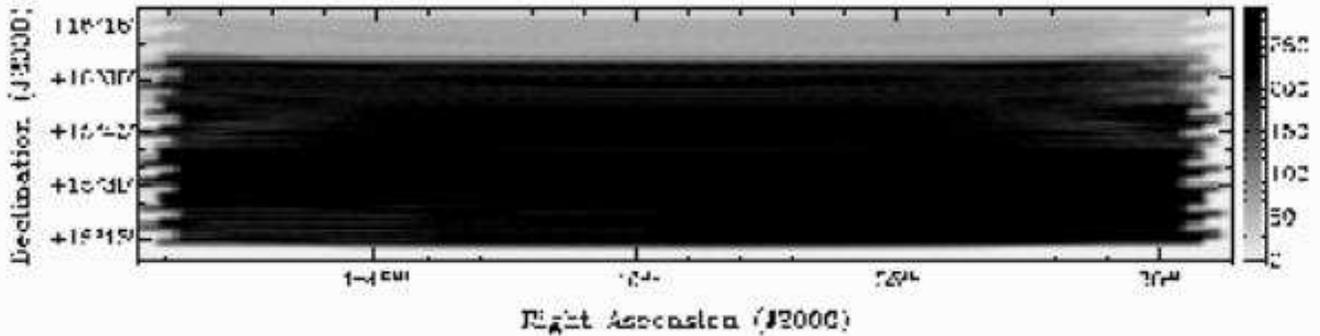}
  \caption{A map showing the sky coverage in the final
  data. The number of spectra that contributed to each pixel is shown as a greyscale from least (light), to
  most (dark). The region above 16\degree\/ 06\arcmin\/ 00\arcsec\/
  was less well covered than the remaining region due to observing
  time constraints.\label{fig01}}
\end{figure*}

Observations were planned such that the central drift scan was
always conducted close to the meridian. Due to the time limits imposed
by the telescope schedule, this often meant only four of the five scans
were completed. With
the available time we were able to cover most of the region to a uniform
depth using 16 nights' data. Due to observing constraints, the region
north of 16\degree\/ 06\arcmin\/ 00\arcsec\/ was often omitted from
the observing run, and hence this region was only observed
successfully on 6 nights. This is illustrated by Fig. \ref{fig01},
which shows the number of spectra that were obtained for each pixel
in the final, gridded data.
This was reflected in the noise quality of the final
data which was significantly poorer for the more sparsely covered
region (above $\sim$16\degree 06\arcmin 00\arcsec). Calibration was
performed at the beginning of every scan using a high-temperature
noise diode that was injected into the beam for a duration of 1s. We
integrated every 1s for each of the 7 beams, both polarisations and
4096 channels as 4-bit floating point numbers. Over a 20 minute drift
scan this equates to a total file size of $\sim$ 395 Mb.  

After data reduction, the resulting {\sc fits} datacube from {\sc gridzilla} has dimensions of RA, dec and
velocity/frequency. The 5\degree$\times$1\degree\/ region was gridded
using a median gridding technique into 1\arcmin$\times$1\arcmin\/ pixels, each of which contains a 4096
channel spectrum. This produced a $\sim$365 Mb file. Initial
investigations of the effects of RFI and the behaviour of noise
in the data were conducted using {\sc idl}. Detailed analysis of the
NGC 628 group was performed using {\sc miriad}, {\sc gipsy} and {\sc
  karma}. 

\subsection{Radio Frequency Interference (RFI)}
Man-made RFI is now, unfortunately, a constant source of contamination
at most radio observatories. At Arecibo in the {\it L}-band there are
a number of RFI sources. The L3 GPS satellite appears at approximately
1381 MHz. There are also several narrow contributions from aircraft radar at
FAA and Punta Salinas in the range 1200--1381 MHz. While RFI from Punta
Salinas was not noticed during the observations, two sources were
noticed at 1350 MHz (FAA radar) and 1381 MHz (L3 GPS). These
interference sources were present throughout the observing runs but
other sources were only noticed occasionally. The median filtering
applied by {\sc livedata} and {\sc gridzilla} is very robust to
intermittent sources of RFI but can do nothing for constant
sources. As a result several channels were contaminated around 1351
MHz and 1381 MHz. These channels were included in the final data, to
allow for the possibility of detecting bright sources that might still
be visible.

\begin{figure}
  \begin{center}
    \subfigure{
      \includegraphics[width=0.5\textwidth]{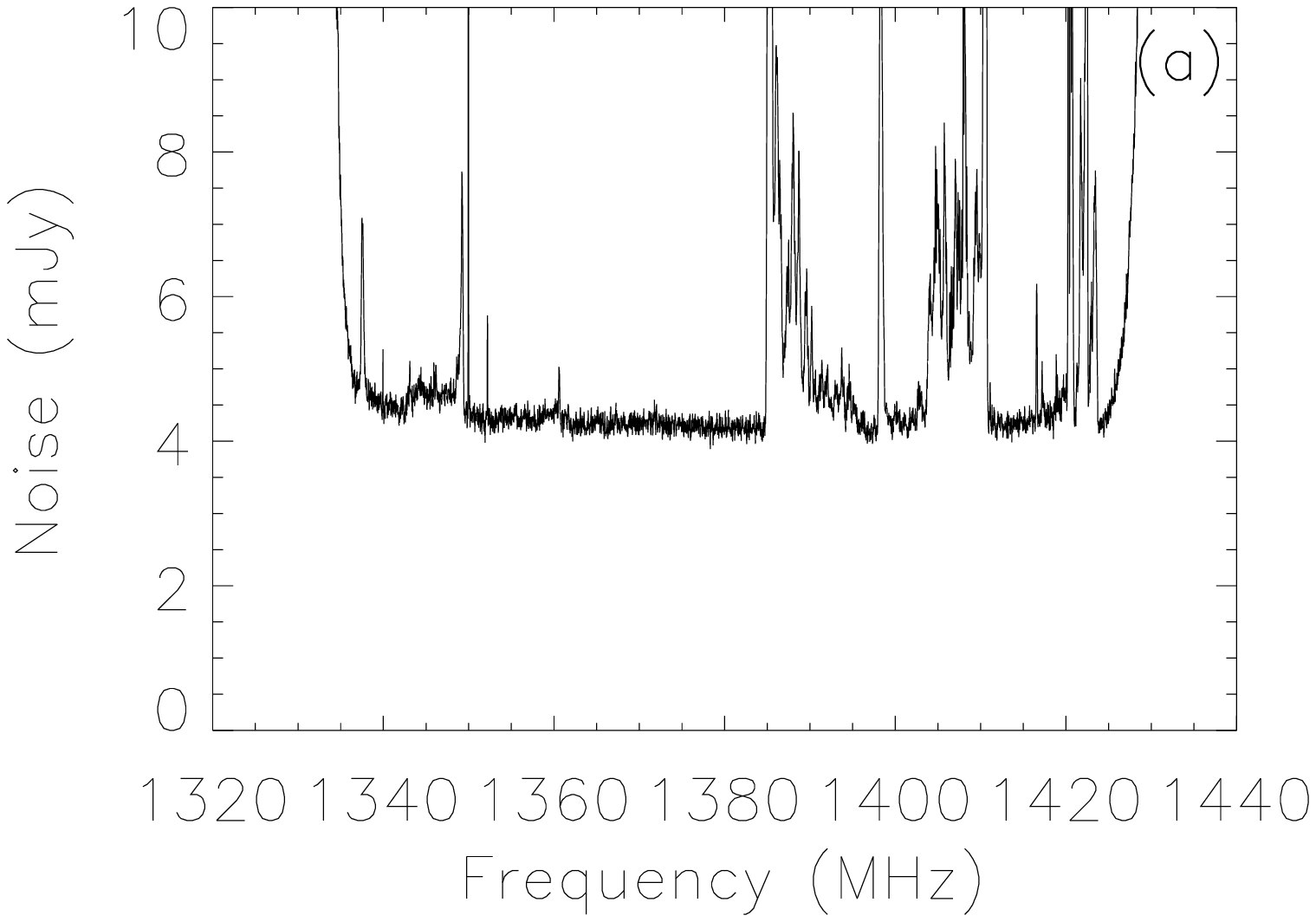}
      \label{fig3a}}
  \subfigure{
    \includegraphics[width=0.5\textwidth]{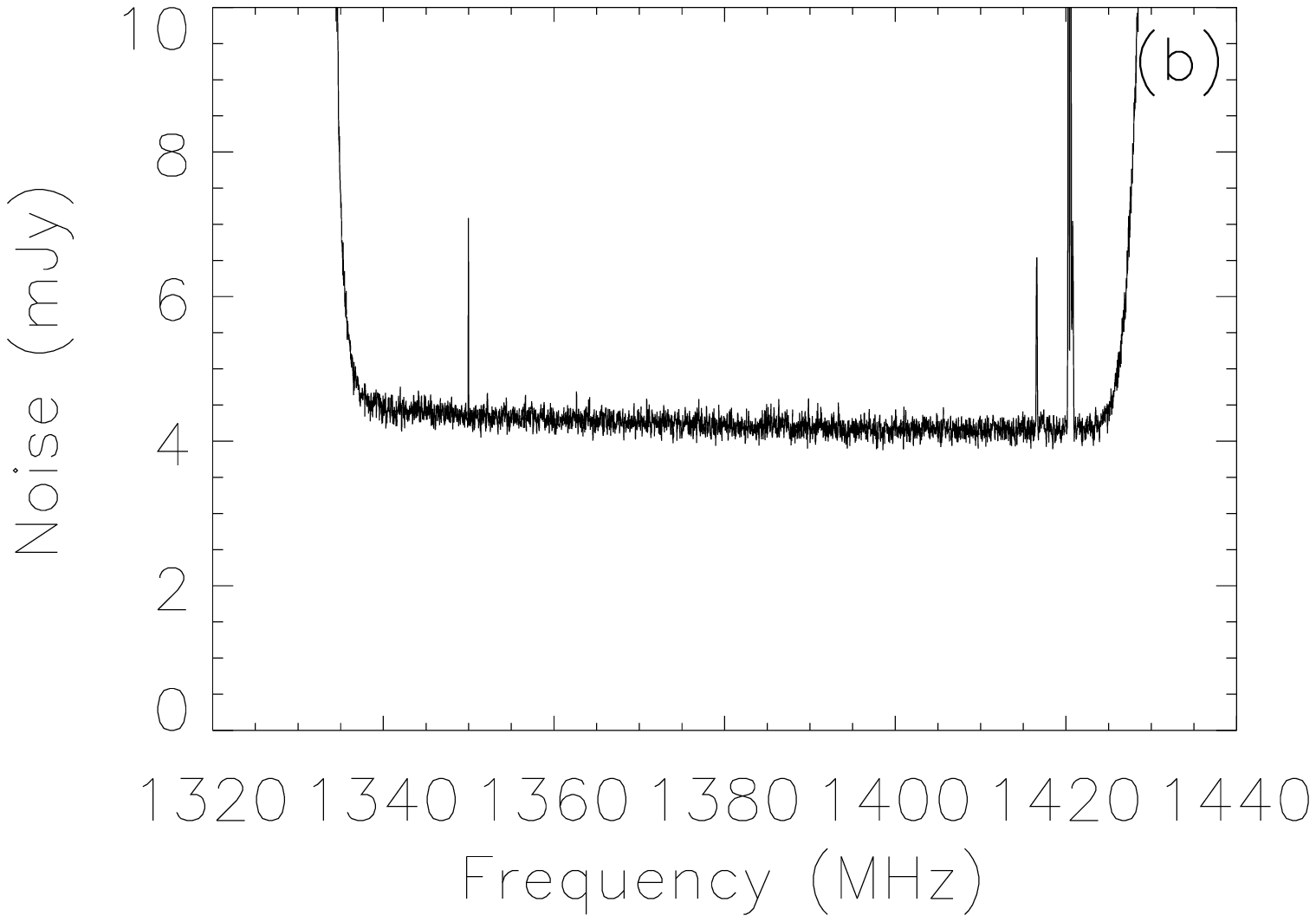}
    \label{fig3b}}
  \caption{The impact of the persistent, frequency-varying RFI on the
    data quality. Each figure shows the rms noise per channel from
    observations on consecutive nights
    (a) with the interference present and (b) without the
    interference. Both datasets represents averages taken over the
    same source-free patch of sky.\label{fig3}}
  \end{center}
\end{figure}

In addition to these sources of RFI there was also a varying source. While the RFI was quite narrow, usually occupying a few
channels, it would shift frequencies in a quasi-periodic fashion
contaminating up to 60 MHz of the band. There was also a harmonic that
would drift in an out of our bandwidth. Figs. \ref{fig3a} \& \ref{fig3b}
are two graphs showing the noise in each channel of the spectrometer,
for two identical regions of sky taken over two consecutive
nights. The first night's data are contaminated by this transient RFI
source, the second night's data are free of it. The effect on the
quality of the final data is significant, causing a rise in the
noise level throughout the band. The RFI source has been attributed to
equipment in the focus cabin and has now been eliminated.

\subsection{Noise behaviour}
\begin{figure}
  \includegraphics[width=0.49\textwidth]{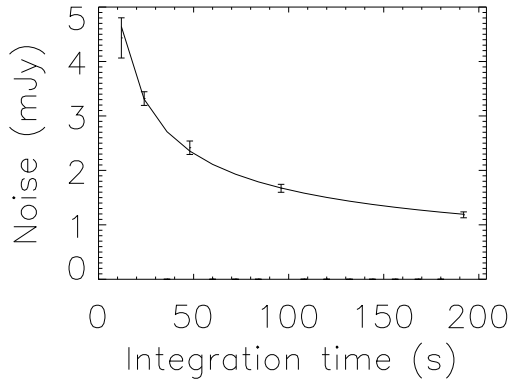}
  \caption{The effect of co-adding successive nights observations
  results in the expected reduction in the rms noise value,
  i.e. $\sigma \propto t^{-0.5}$.\label{fig4}}
\end{figure}

By combining multiple observations of the survey field it is possible
to build up the effective integration time and hence increase the
depth of the survey. One would expect
that as more observations are taken, the noise level in the combined
data should fall. Datacubes were produced for 1, 2, 4, 8, and 16 nights'
observations. Since the integration time each night was the same
(12s beam$^{-1}$), this
was equivalent to 12, 24, 48, 96, and 192s beam$^{-1}$. A robust mean noise value was
then calculated for each datacube, using the {\sc idl} astro library
routine {\sc resistant\_mean} within a large region of sky that
didn't contain sources over the inner 3696 channels. The data were
fitted using a least-squares fit. In theory we would expect the noise
level to be, $\sigma$ $\propto t^{-0.5}$ where $t$ is the
integration time. Fig. \ref{fig4} shows how the rms noise level
varies with the integration time. The line fit indicates the noise decreases as
$t^{-0.49 \pm 0.06}$ which is in very good agreement with the
$t^{-0.5}$ dependence as predicted. This represents 16
nights' data which provided uniform coverage for most of the
region. Following this predicted trend, a final noise level of 0.95
mJy for a velocity resolution of 10.3 \kms\/ could have been achieved
if all 25 nights' observing (300s beam$^{-1}$) had been successful.

\begin{figure}
  \begin{center}
    \subfigure{
      \includegraphics[width=0.49\textwidth]{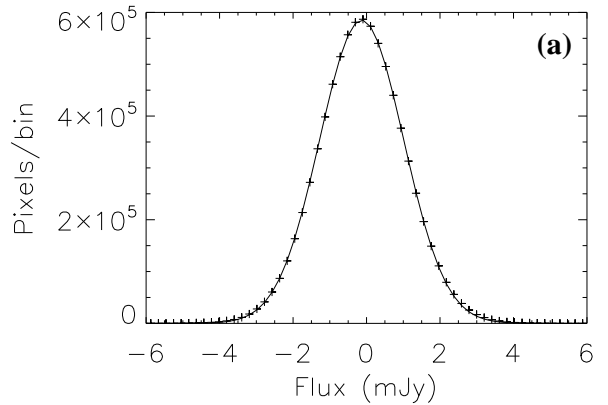}
      \label{fig5a}}
  \subfigure{
    \includegraphics[width=0.49\textwidth]{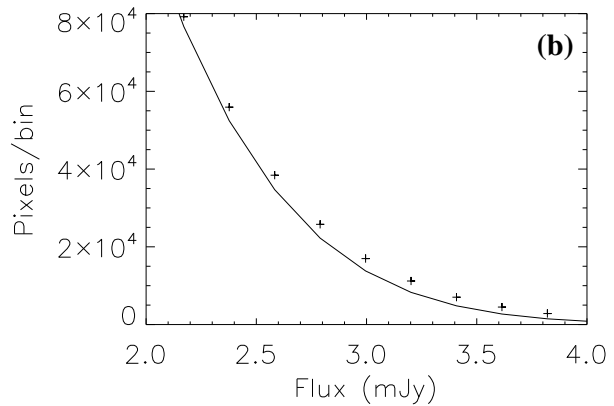}
    \label{fig5b}}
  \caption{(a) The distribution of flux values within an apparently
  blank patch of sky in the reduced datacube is well fit by a Gaussian
  distribution, centred on -0.1 mJy. The data points are represented
  by crosses and the solid line is the Gaussian fit (b) There is a
  departure from Gaussianity between 2.5--4.0 mJy corresponding to
  2.3--3.6$\sigma$. This is possibly due to very low-level sources
  within the region that were not detected by eye.
  \label{fig5}}
  \end{center}
\end{figure}

To investigate the Gaussianity of the noise in the final co-added data
(t$_{int}=192s$),
a source-free region of sky was chosen and the noise in each
channel for each pixel was recorded. The values were binned into 100
equally spaced bins. The noise distribution about the mean value was
then compared to a Gaussian distribution. The values lay
very close to a Gaussian, distributed about zero (Fig. \ref{fig5}). From
Fig. \ref{fig5b} there is a slight excess in noise
at $\sim$3.0 mJy ($\sim3\sigma$). This is probably due to numerous sources
lying just below the detection limit. 

\subsection{Baseline stability}
The quality of the final data is highly dependent on the stability of
our baselines. It is extremely important therefore to examine the
baseline before any fitting is applied to it, and confirm that the
drift-scanning technique is producing the stable baselines expected. Observations were chosen from a night in which there was no
contamination by the persistent, variable RFI. Comparisons were then
made from spectra taken at the beginning of a scan to those at the end
of the scan.

\begin{figure}
  \begin{center}
    \includegraphics[width=0.5\textwidth]{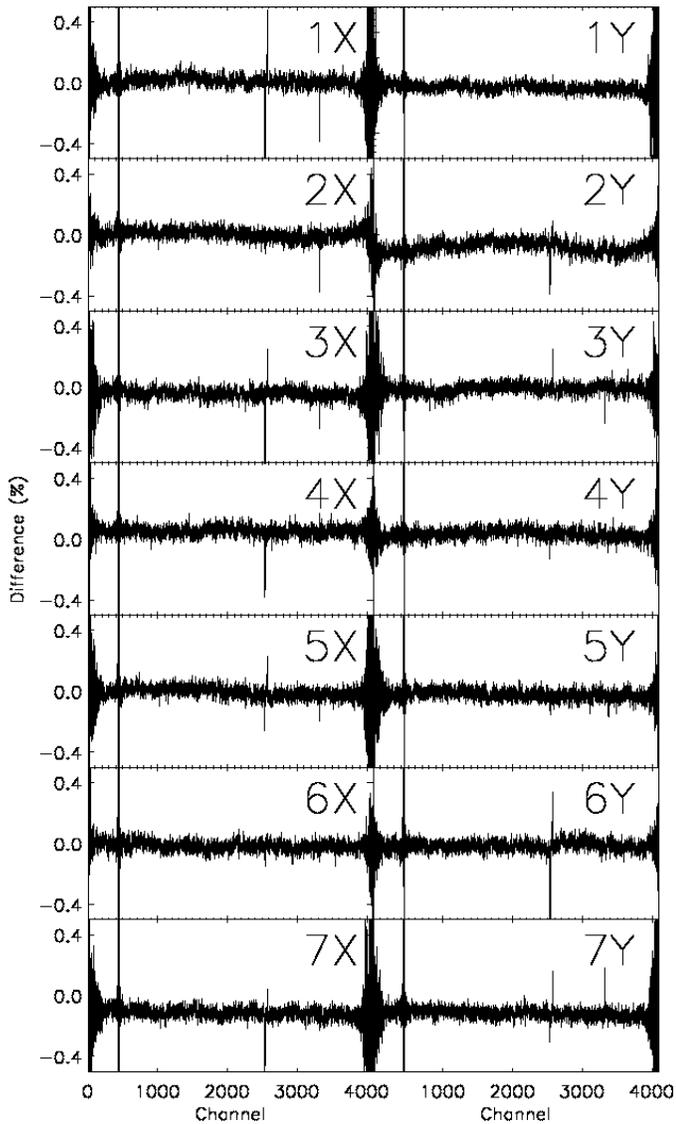}
    \caption{Percentage variation of the raw bandpass over a single scan
    for each individual beam and polarisation. Each raw bandpass
    remains unvarying to within $\sim$0.1\%. \label{fig18}}
  \end{center}
\end{figure}

Fig. \ref{fig18} shows, for each individual beam and polarisation, the
percentage difference between the first 100 spectra and the last 100
spectra within one scan. The baseline remains stable to within 0.1\%
over the duration of a scan. 

\section{Results}
\label{sec:results}
\begin{figure*}
   \subfigure{
     \includegraphics[width=0.45\textwidth]{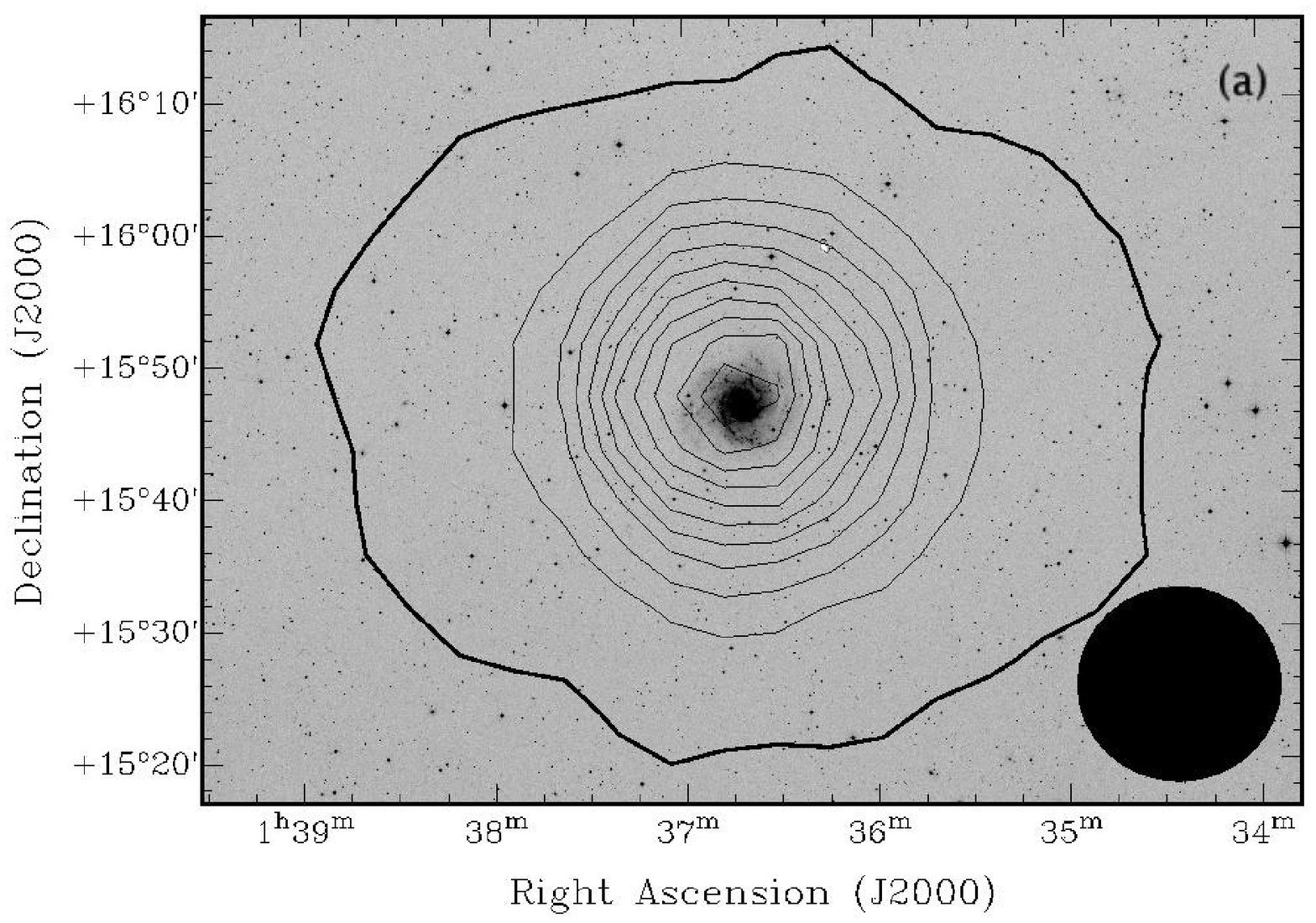}}
   \subfigure{
     \includegraphics[angle=0,width=0.45\textwidth]{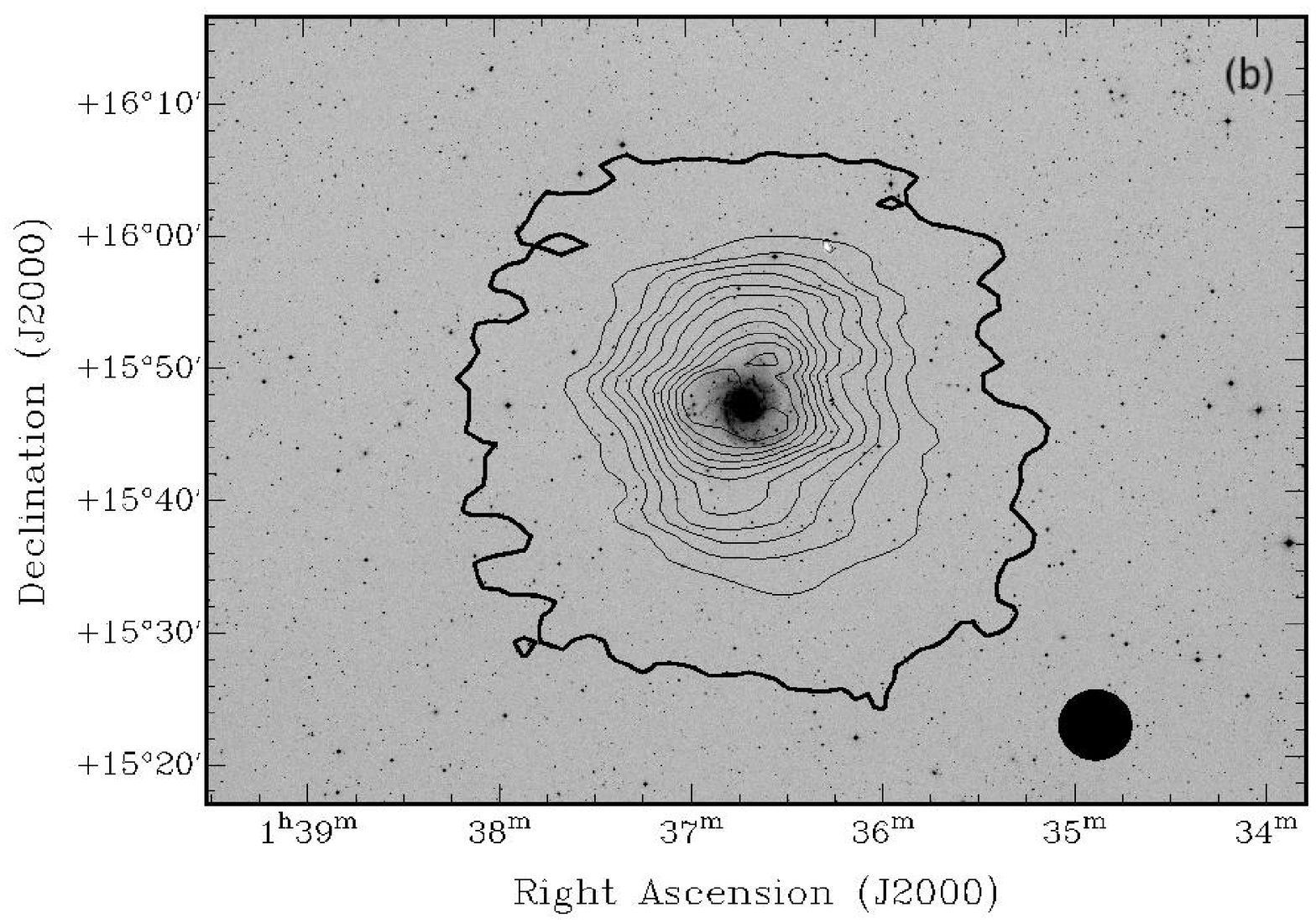}}
   \subfigure{
     \includegraphics[angle=0,width=0.5\textwidth]{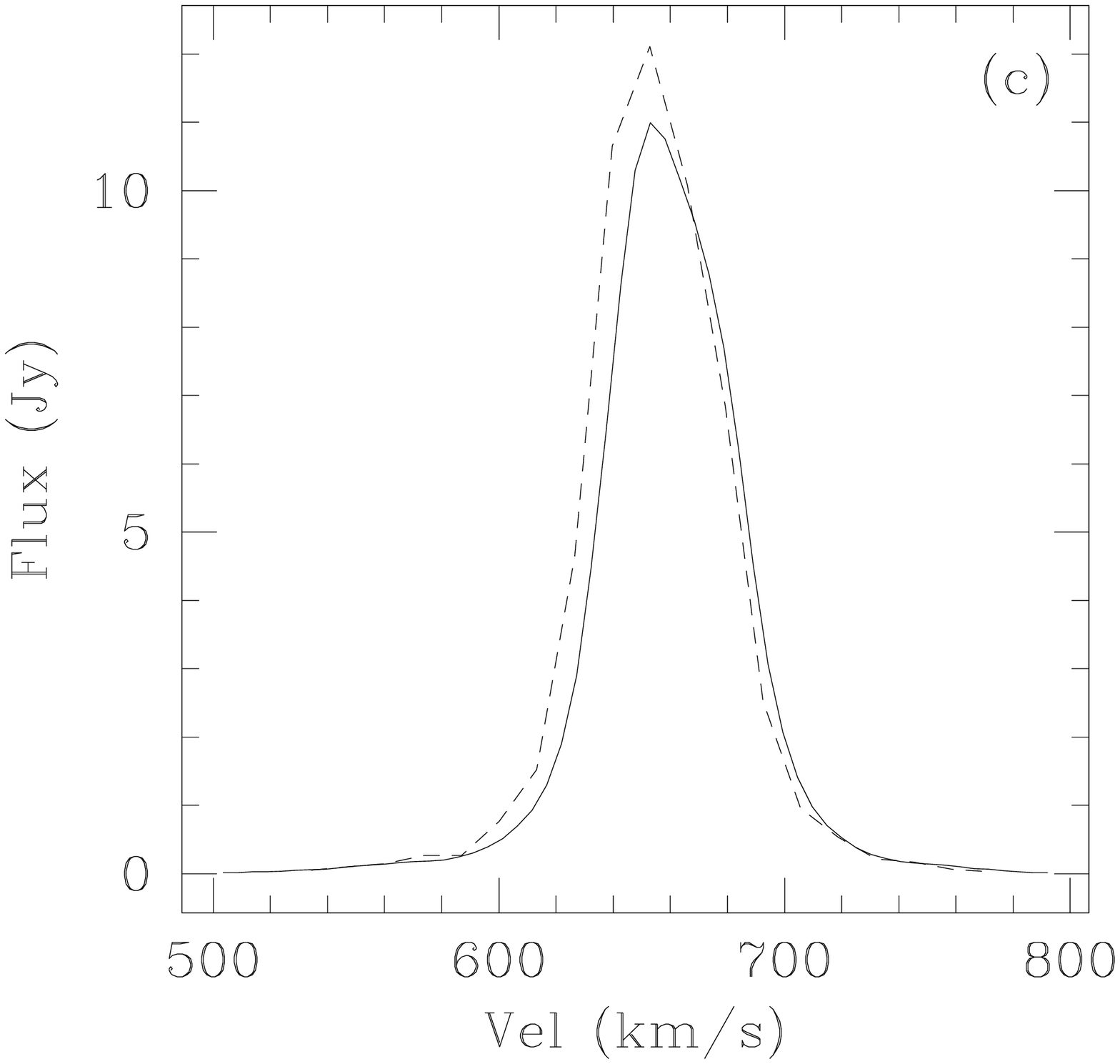}}
   \subfigure{
     \includegraphics[angle=0,width=0.45\textwidth]{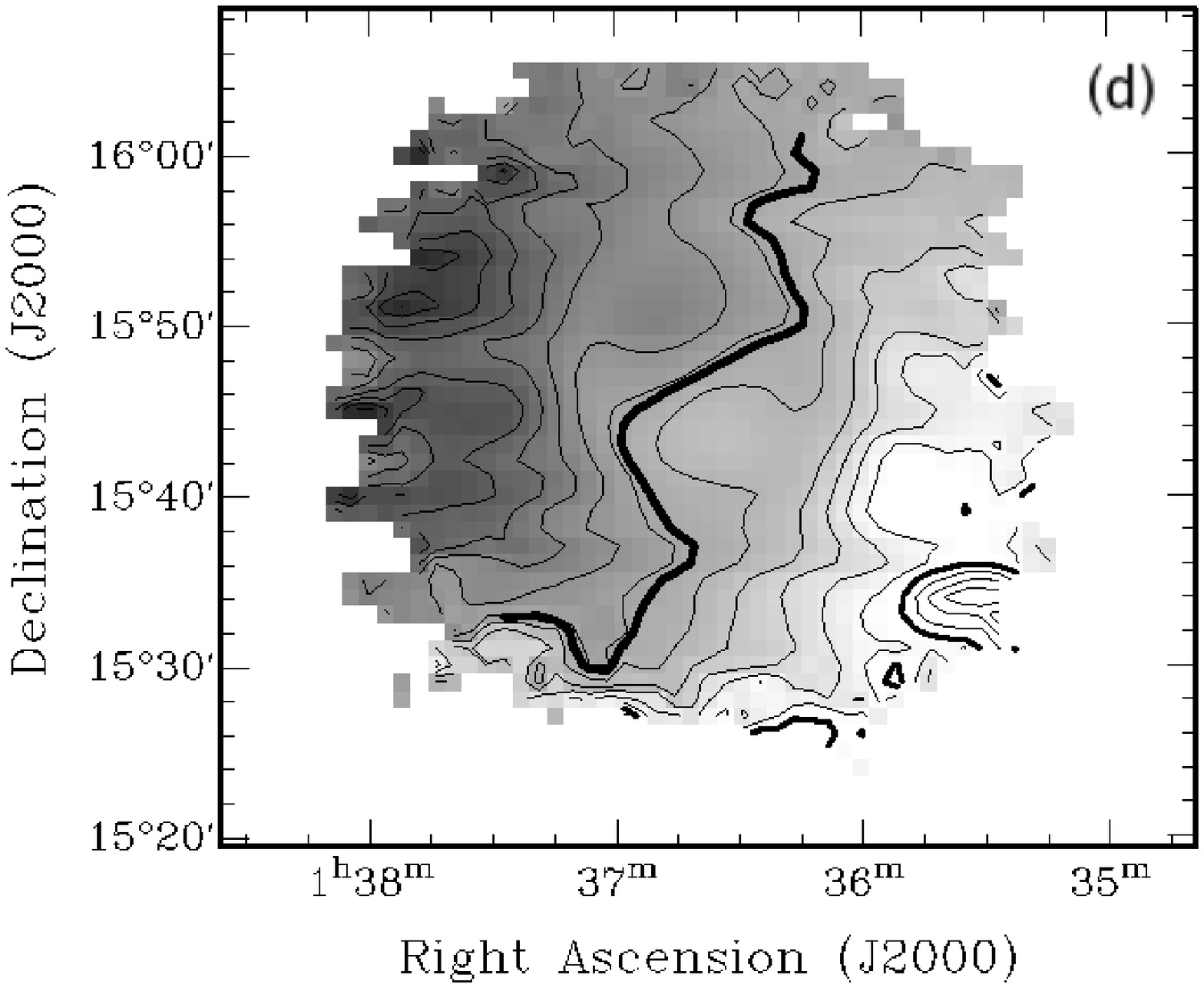}}
   \caption{{\bf NGC 628} (a) HIPASS \hi\/ contours overlaid on a POSS
  II image. Contours ($\times 10^{18}$ cm$^{-2}$): {\bf 2.8}
  (3$\sigma$), 50, 100, 150, 200, 250, 300, 350, 400, 450, 500. The
  solid circle illustrates the size of the HIPASS beam. (b)
  AGES \hi\/ contours overlaid on a POSS II image. Contours ($\times
  10^{18}$ cm$^{-2}$): {\bf 2.2} (3$\sigma$), 100, 200, 300, 400, 500,
  600, 700, 800, 900, 1000, 1100. The solid circle illustrates the size of the
  AGES beam.(c) \hi\/ velocity profiles: HIPASS is shown as the
  dashed line, AGES as the solid line. (d) Velocity field from AGES
  data. Isovelocity contours increase from light to dark: 635, 640,
  645, 650, 655,   {\bf 659}, 660, 665, 670, 675, 680, 685, 690
  \kms. \label{fig6}}
 \end{figure*}

\begin{figure*}
  \begin{center}
  \subfigure{
    \includegraphics[width=0.48\textwidth]{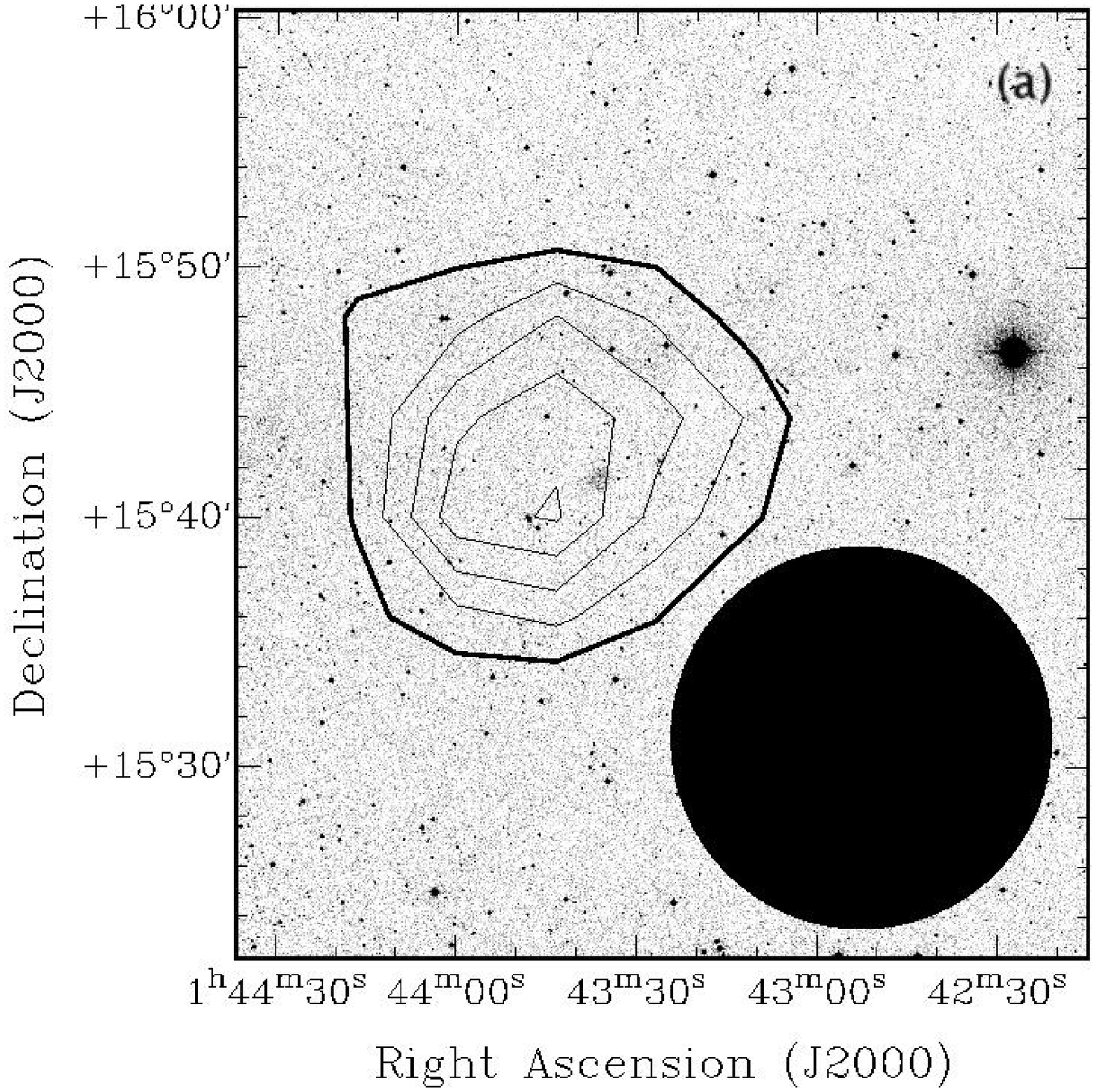}
    \label{fig7a}}
  \subfigure{
    \includegraphics[width=0.48\textwidth]{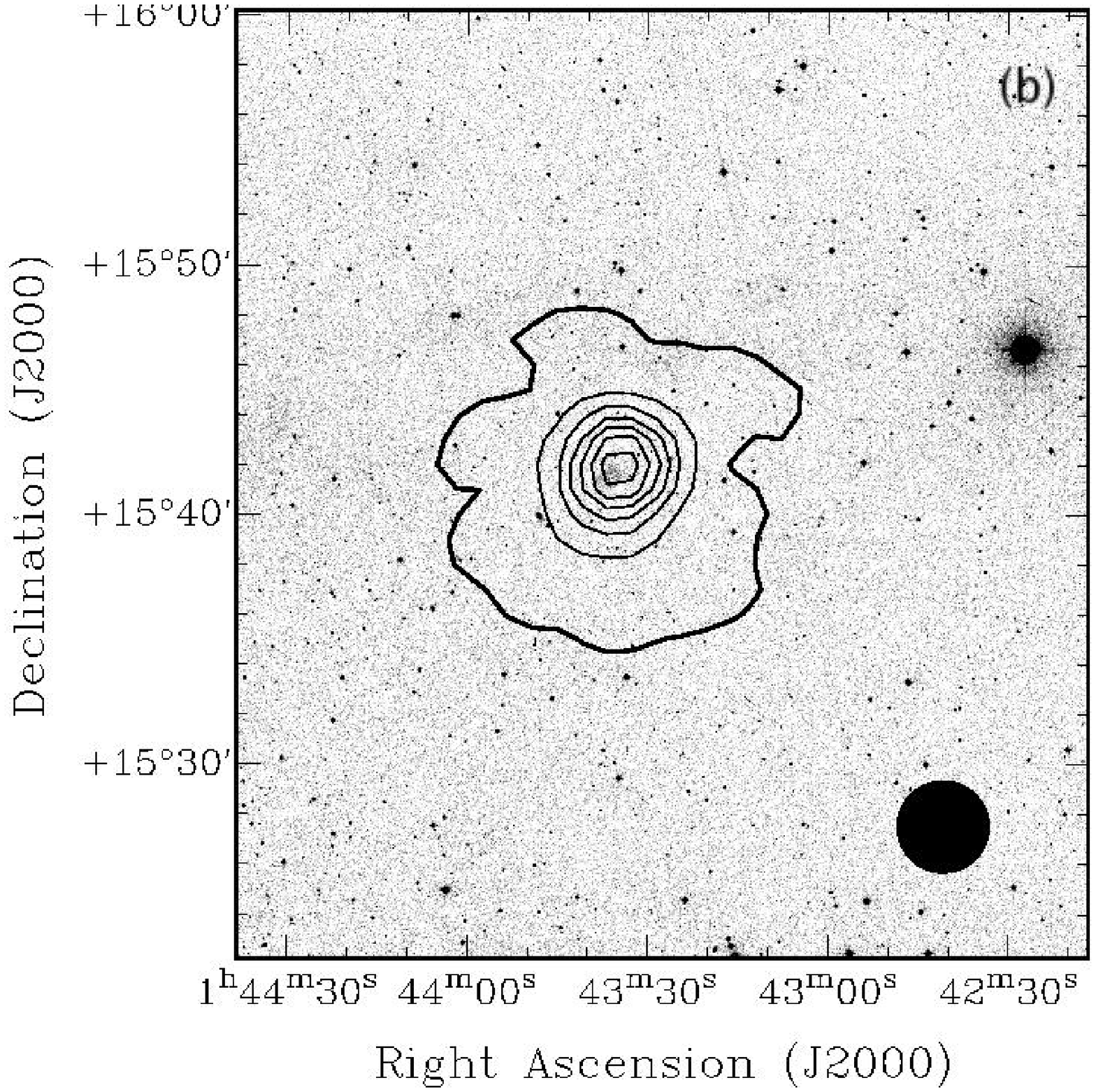}
    \label{fig7b}}
  \subfigure{
    \includegraphics[width=0.48\textwidth]{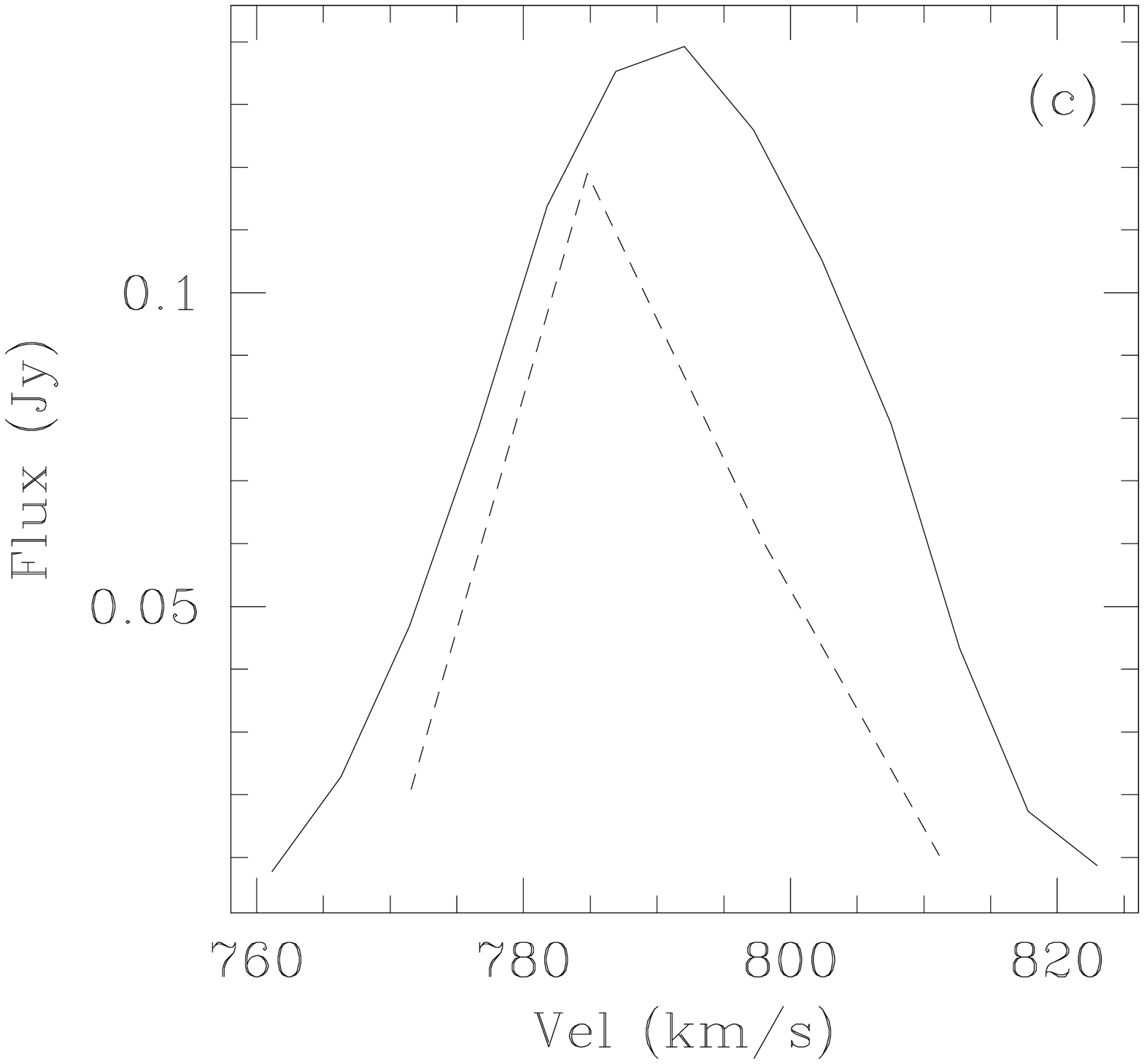}
    \label{fig7c}}
  \caption{{\bf KDG 010} (a) HIPASS \hi\/ contours overlaid on a POSS
  II image. Contour values: 1, {\bf 2} (3$\sigma$), 3, 4, 5
  ($\times10^{18}$cm$^{-2}$). The
  solid circle illustrates the size of the HIPASS beam. (b) AGES \hi\/
  contours overlaid on a POSS II image. Contour values: {\bf 2} (3$\sigma$), 20, 40, 60, 80
  ($\times10^{18}$cm$^{-2}$). The solid circle illustrates the size of the
  AGES beam. (c) \hi\/ velocity profiles: HIPASS is
  shown as the dashed line, AGES as the solid line. \label{fig7}}
  \end{center}
\end{figure*}

\begin{figure*}
   \subfigure{
     \includegraphics[angle=0,width=0.45\textwidth]{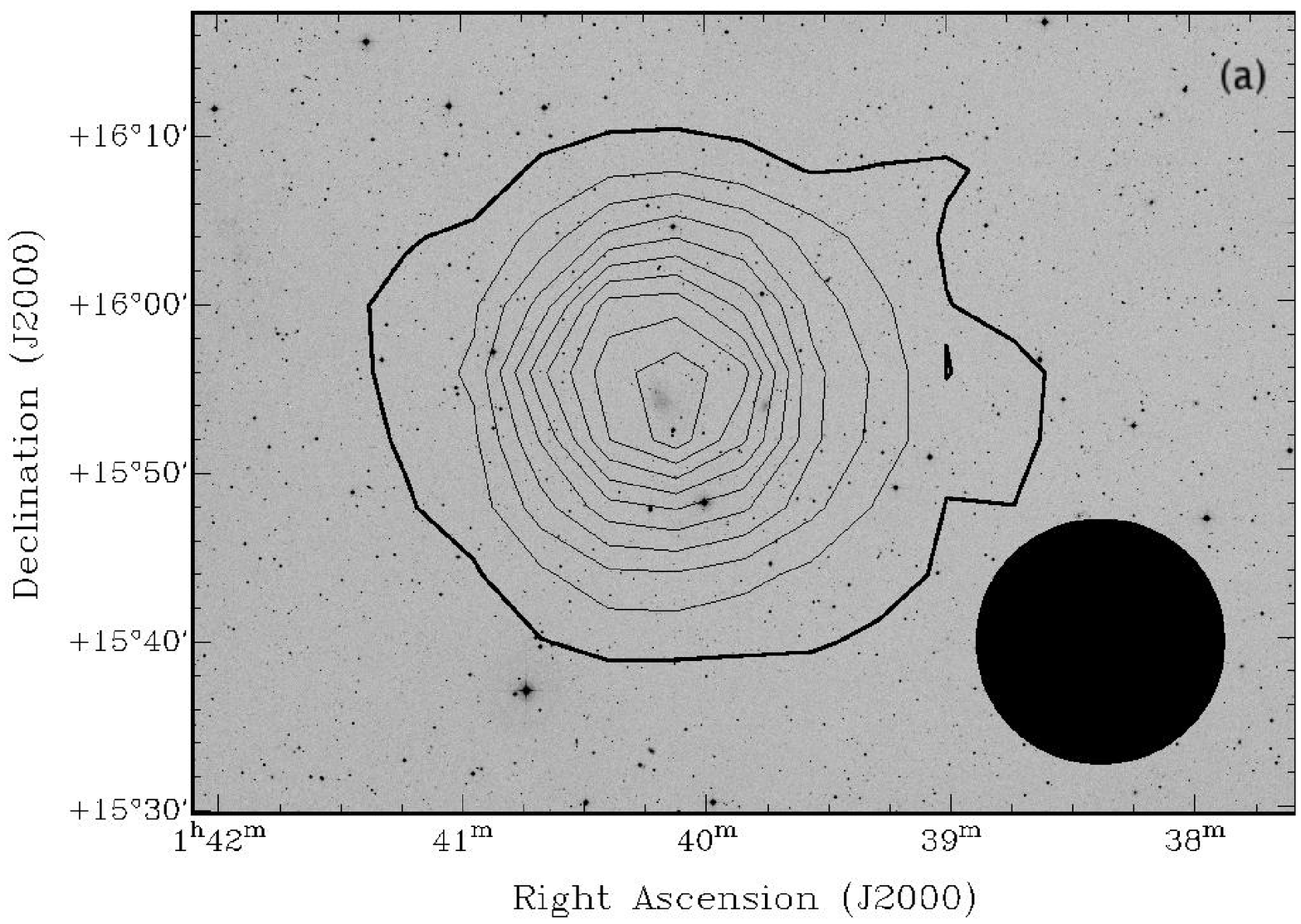}}
   \subfigure{
     \includegraphics[angle=0,width=0.45\textwidth]{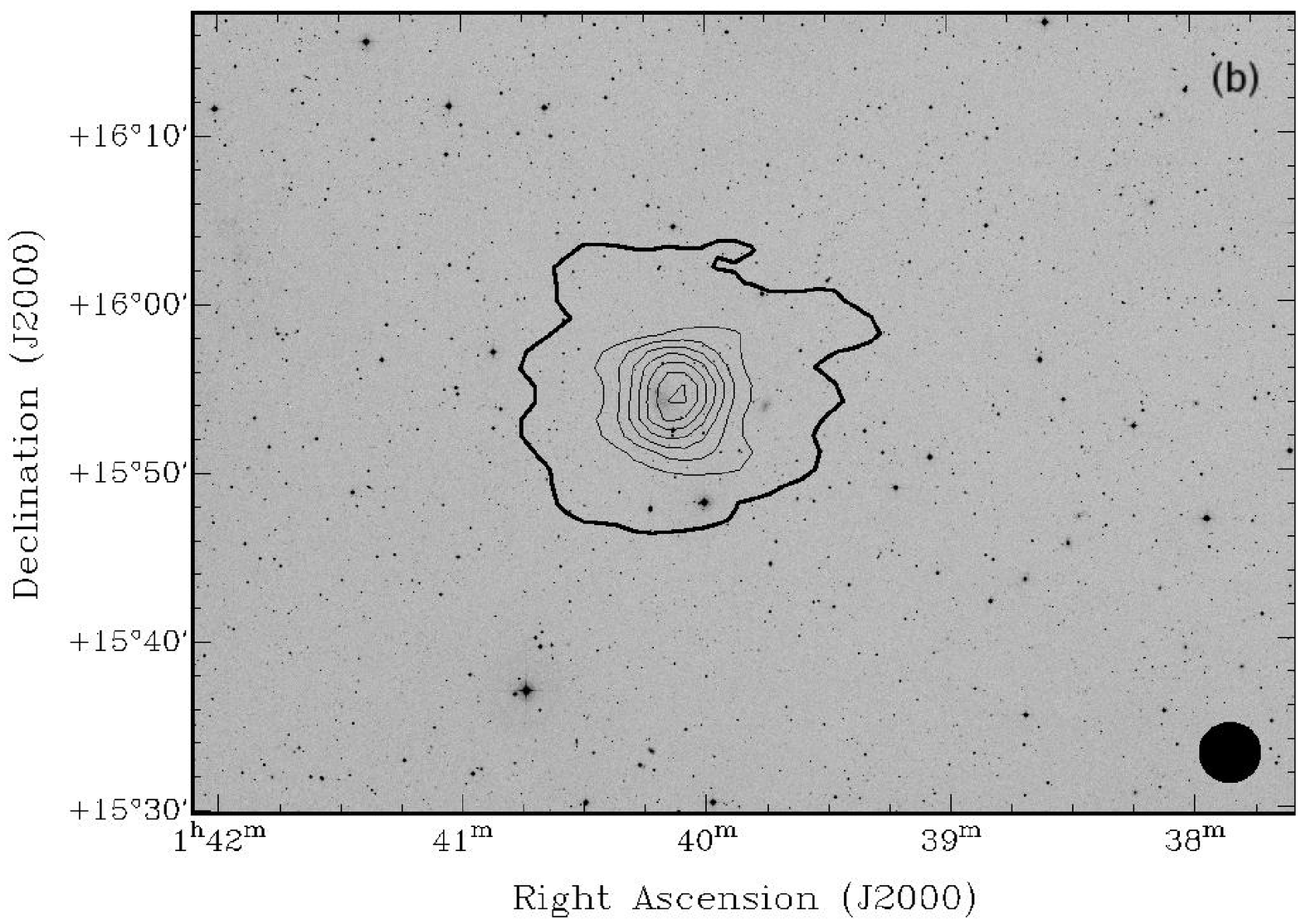}}
   \subfigure{
     \includegraphics[angle=0,width=0.45\textwidth]{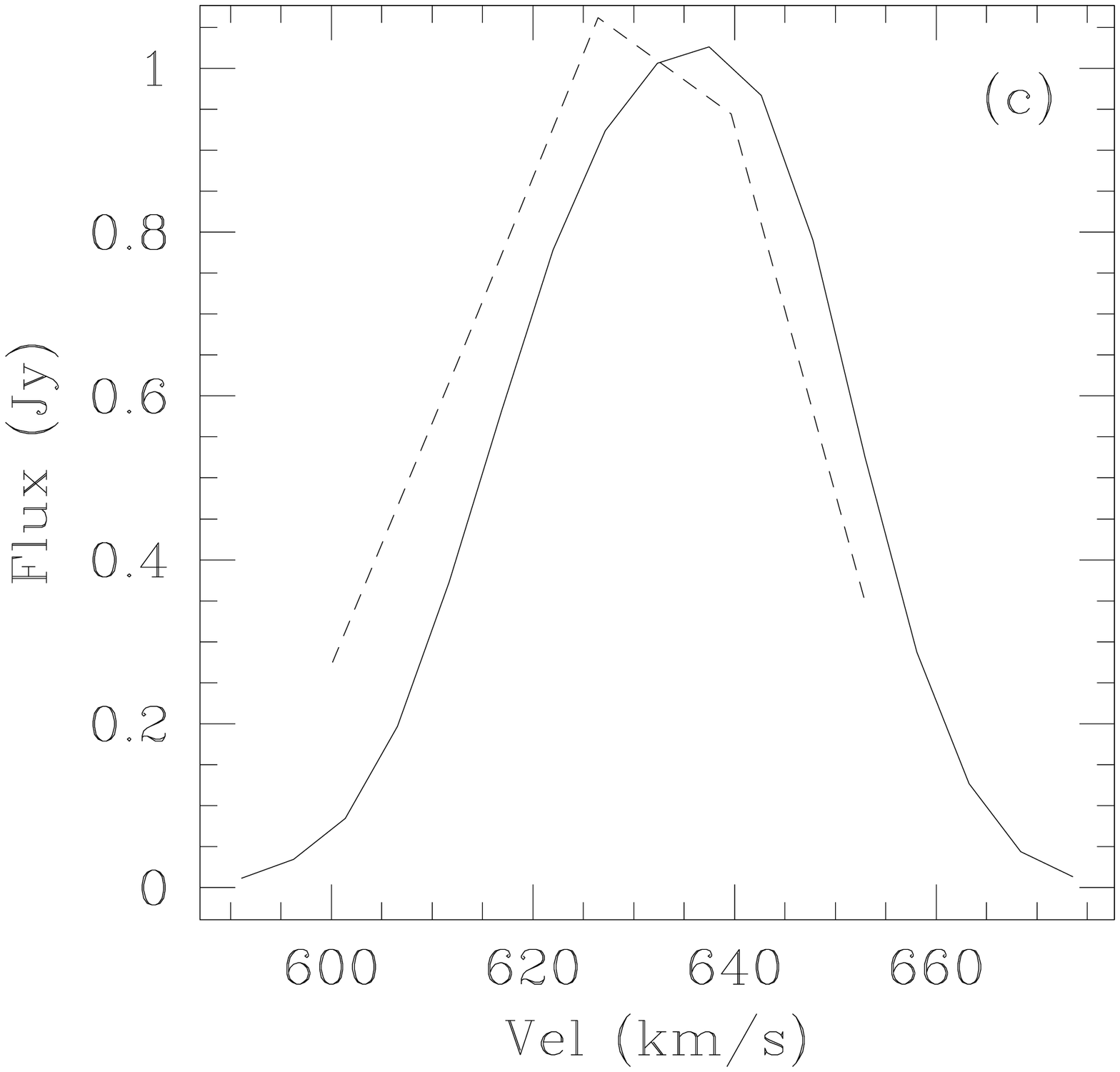}}
   \subfigure{
     \includegraphics[angle=0,width=0.45\textwidth]{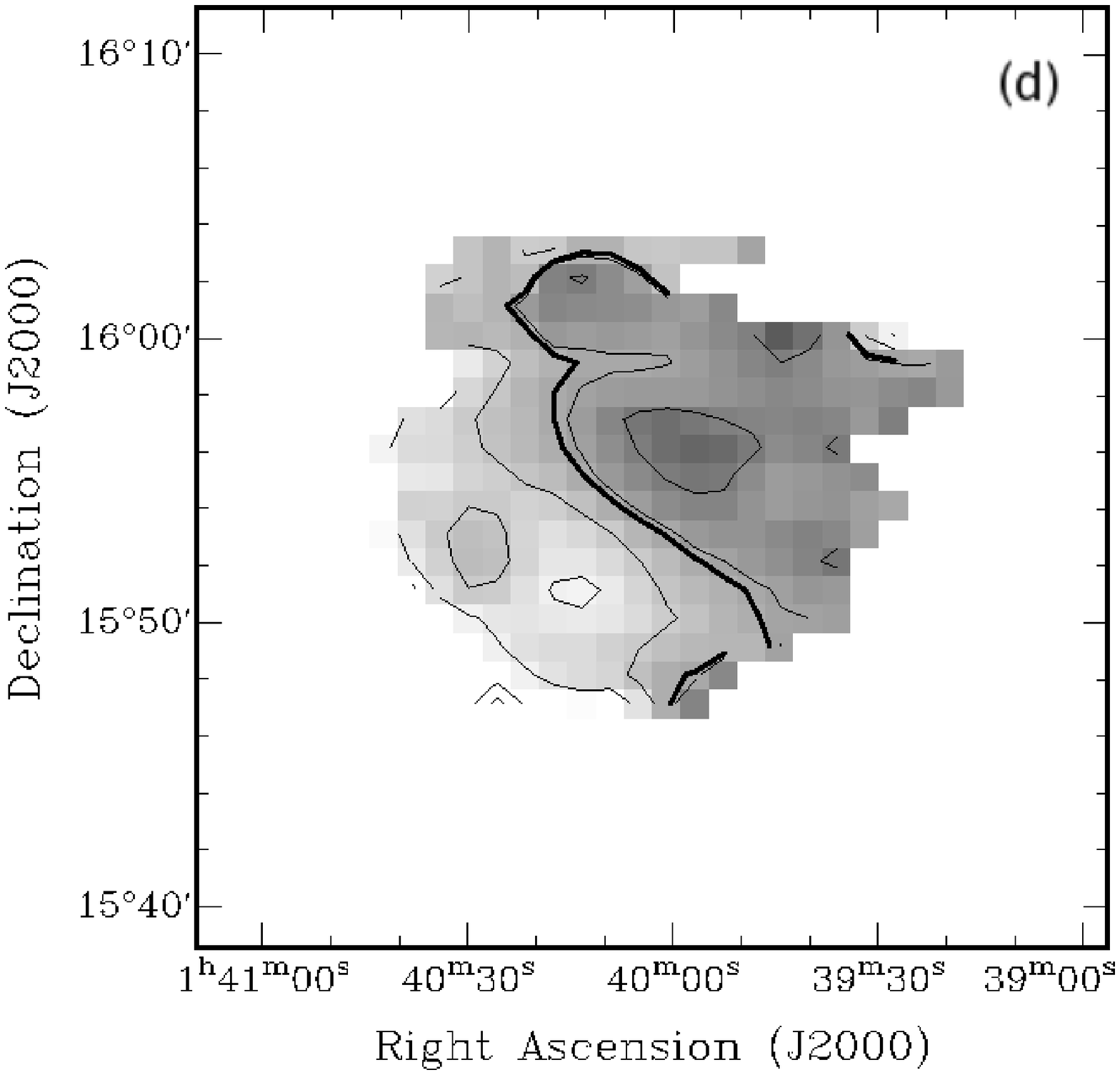}}
   \caption{{\bf UGC 1176} (a) HIPASS \hi\/ contours overlaid on a POSS
	     II image. Contours ($\times 10^{18}$ cm$^{-2}$): {\bf 2} (3$\sigma$)
  5, 10, 15, 20, 25, 30, 35, 40, 45. The
  solid circle illustrates the size of the HIPASS beam. (b) AGES \hi\/
  contours overlaid on a POSS II image. Contours ($\times 10^{18}$
  cm$^{-2}$): {\bf 2} (3$\sigma$), 60, 120, 180, 240, 300, 360, 420. The solid circle illustrates the size of the
  AGES beam. (c) \hi\/ velocity profiles: HIPASS is
  shown as the dashed line, AGES as the solid line. (d) Velocity
  field from AGES data. Isovelocity contours increase from light to
  dark: 620, 625, 630, {\bf   634}, 635, 640. \label{fig8}}
 \end{figure*}

\begin{figure*}
  \subfigure{
       \includegraphics[angle=0,width=0.45\textwidth]{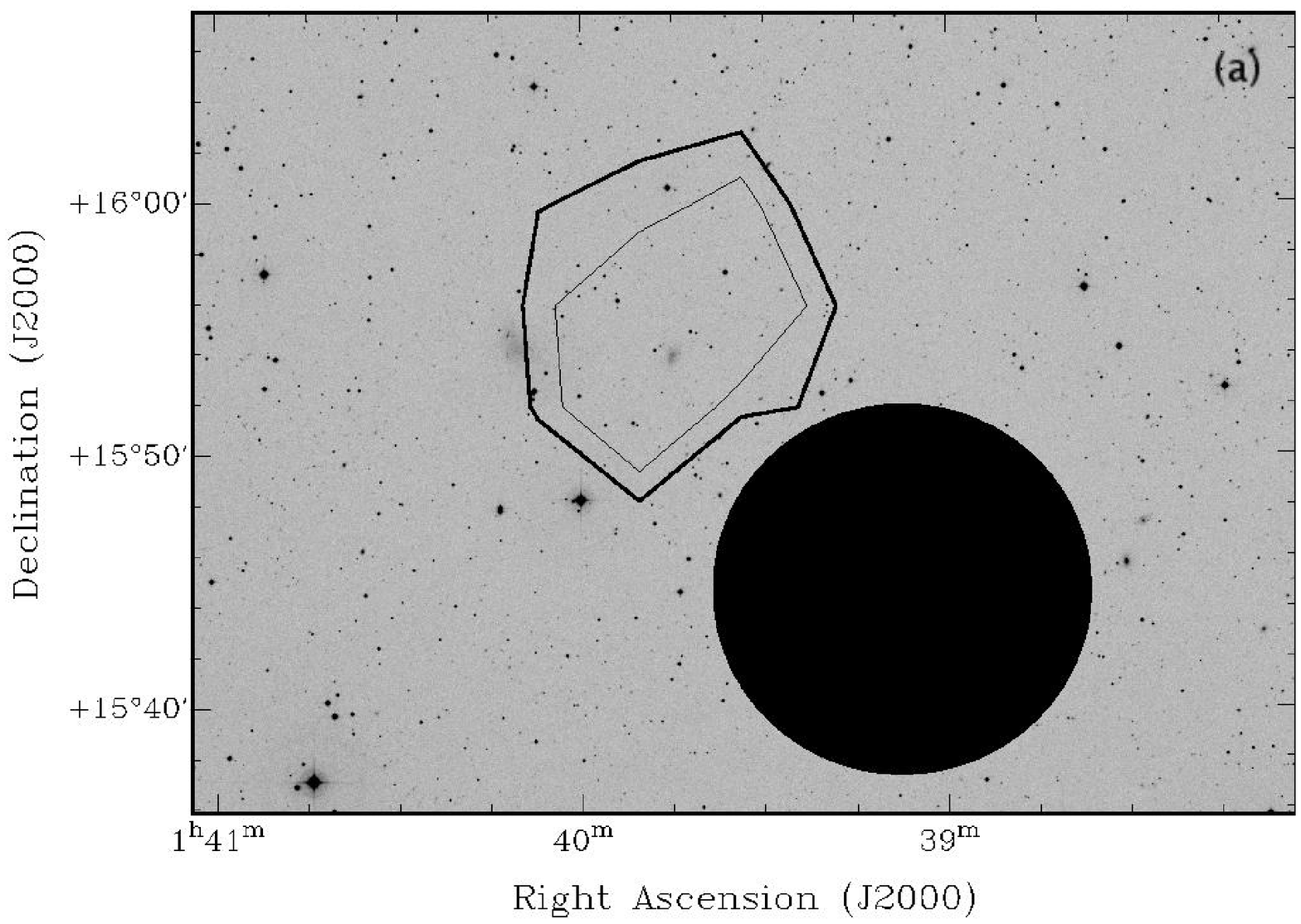}}
  \subfigure{
       \includegraphics[angle=0,width=0.45\textwidth]{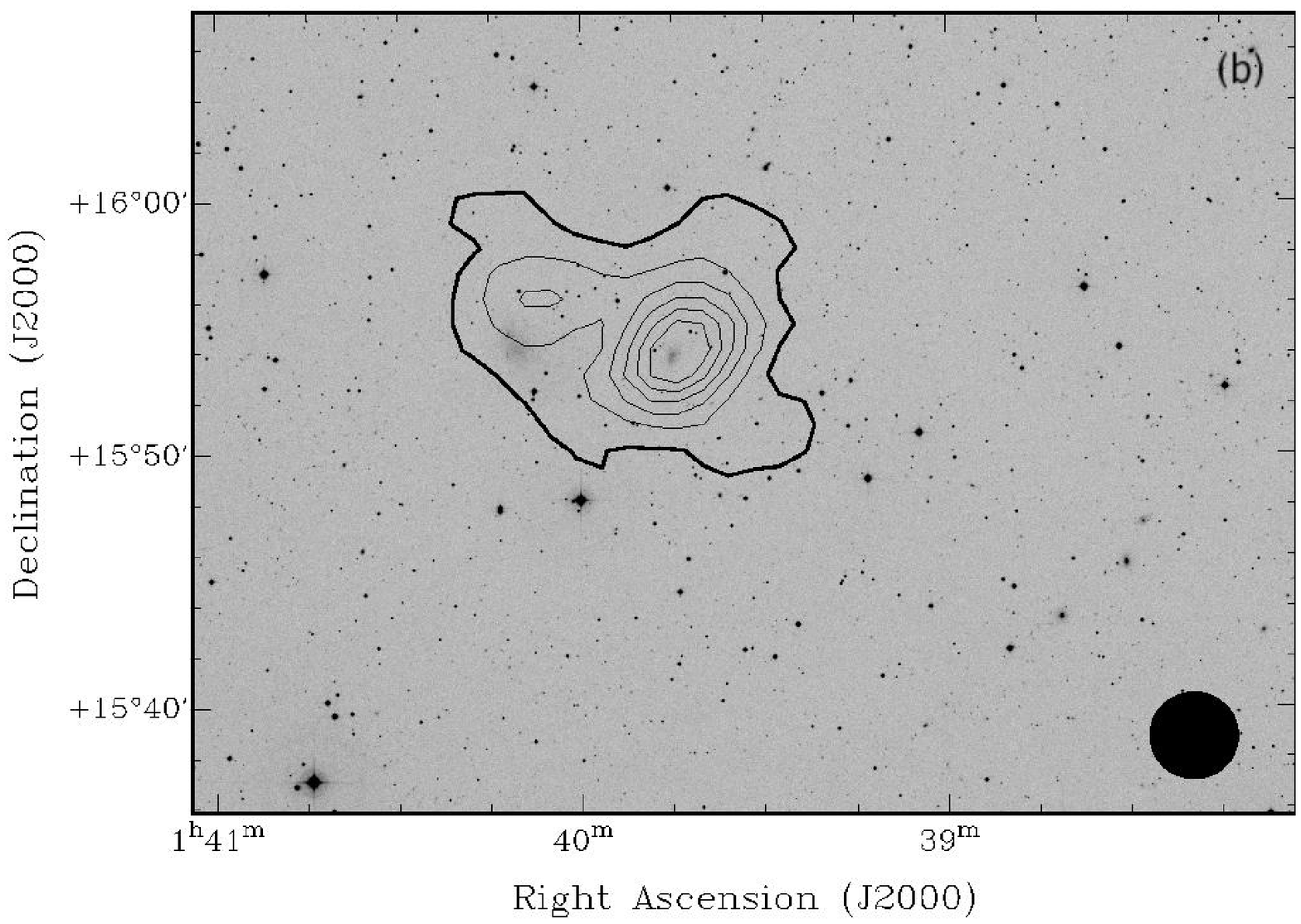}}
  \subfigure{
       \includegraphics[angle=0,width=0.5\textwidth]{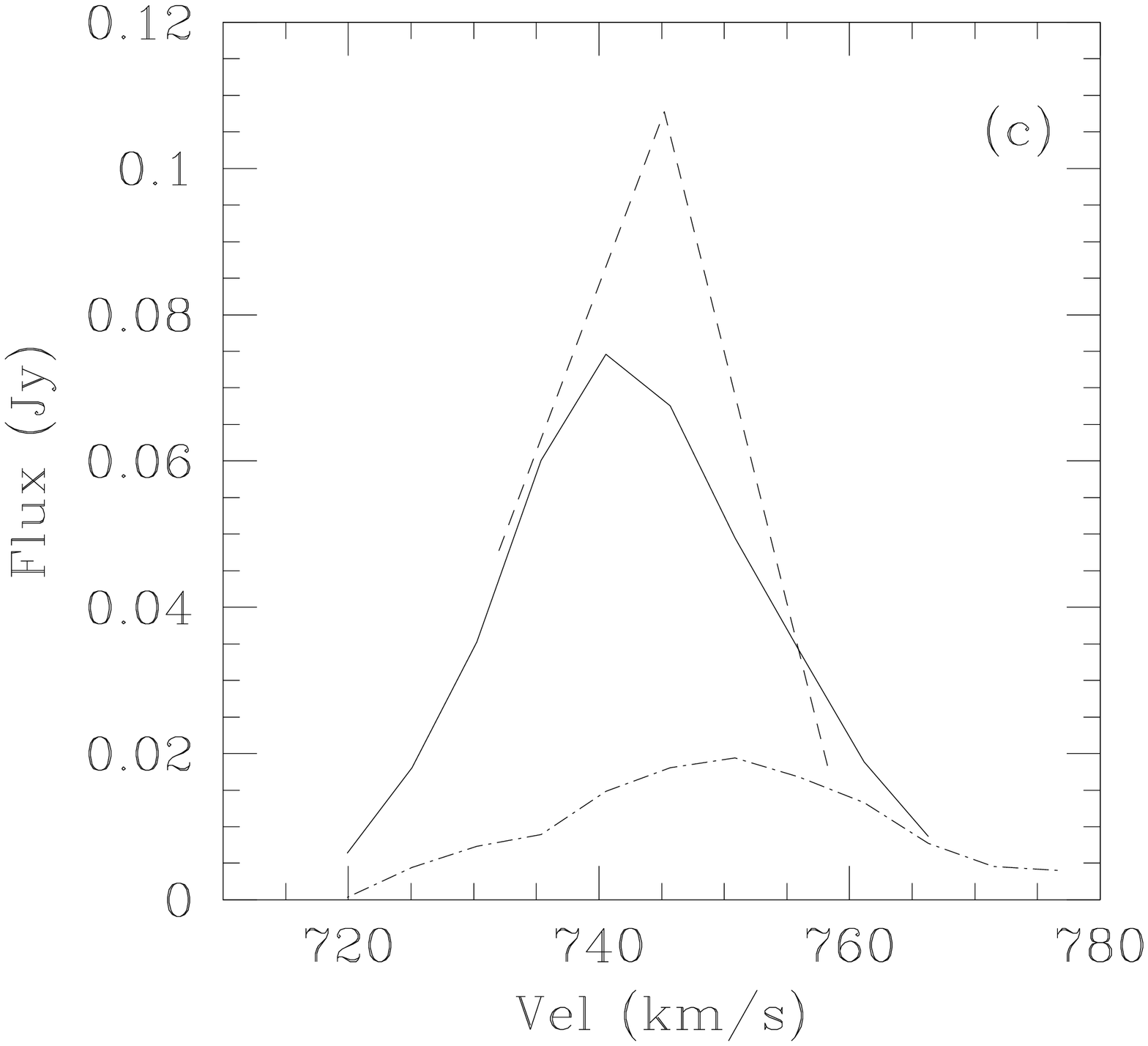}}
       \caption{{\bf UGC 1171 \& dw0137+1541} (a) HIPASS \hi\/
  contours overlaid on a POSS II image. Contours ($\times 10^{18}$
  cm$^{-2}$): {\bf 2} (3$\sigma$), 3. The solid circle illustrates the
  size of the HIPASS beam. (b) AGES \hi\/ contours overlaid on a POSS
  II image. Contours ($\times 10^{18}$ cm$^{-2}$): {\bf 2}
  (3$\sigma$), 5, 10, 20, 30. The solid circle illustrates the size of the
  AGES beam. (c) \hi\/ velocity profiles for UGC 1171 and dw0137+1541:
  HIPASS data for UGC 1171 is
  shown as the dashed line, AGES data for UGC 1171 is the solid
  line. Dw0137+1541 is shown by the dash-dot line.\label{fig9}}
 \end{figure*}

The reduced datacube was masked to isolate all the \hi\/ emission from
each member of the NGC 628 group and then moment maps were constructed for each
galaxy. Moment 0 maps were constructed to illustrate the \hi\/
distribution over the channel range occupied by each galaxy. Where the galaxy was
resolved, moment 1 maps (velocity fields) were also produced. These
moment maps are shown in Figs. \ref{fig6}--\ref{fig9} and also include 
the velocity profile of each galaxy. For comparison purposes we also
provide moment 0 maps produced from HIPASS data. None of the galaxies
were resolved by HIPASS and so there are no velocity fields from
HIPASS data. 

The 3$\sigma$ column density
limit is around 2$\times10^{18}$\/ cm$^{-2}$ in the integrated \hi\/ 
maps for both HIPASS and the AGES precursor. The benefit of our increased spectral
resolution is immediately clear in the velocity profiles. The
improvement in spatial resolution is also highlighted by the detection of
dw0137+1541. This dwarf is so close to UGC 1171 that it is
undetected in the HIPASS data but is clearly visible in the
AGES data (Fig. \ref{fig9}b) at position $(\alpha, \delta) = 01h\/
40m\/ 03s$, +15\degree\/ 56\arcmin\/ 00\arcsec. Fig. \ref{fig9}(b)
suggests that UGC 1171 \& dw0137+1541 are embedded in a neutral gas
envelope extending from northeast to southwest. This could be an
effect of beam smearing since there is no hint of
extended emission in Briggs \etal\/ (1986), but it is possible that it
is real and has been resolved out by the synthesised beam of the VLA. 

An important step is to compare the velocity profiles from AGES and
HIPASS to see if the flux calibration is being performed
correctly. From Figs. \ref{fig6}--\ref{fig9} the fluxes
calculated from the AGES data are generally in agreement with the
HIPASS data. The most probable reason for discrepancies is that the
gridding technique responds differently to sources of different size and
strength. The problem is that the gridding technique tends to
overestimate the flux of extended objects (objects comparable to or
larger than the beam size). This was an expected side effect of the
gridding process and Barnes \etal\/ (2001) simulated observations in
the HIPASS data using sources of known flux and size to quantify the
effect and calculate correction values. Another side effect of the
technique is to produce larger gridded beam sizes for stronger sources
(see Tables 2 \& 3 in Barnes \etal\/ 2001).

We have corrected for this as follows. First the ratio of the AGES
source size to the AGES beam was found. The source size was measured
from higher resolution \hi\/ data (Briggs, 1986; Briggs, 1992). This
ratio was multiplied by the HIPASS mean observing beam size
(14.4\arcmin) to find the size of a hypothetical HIPASS source that
would produce the equivalent source/beam ratio. Table 3 in Barnes
\etal\/ (2001) was then used to find the HIPASS flux-weighted beam
area for this hypothetical source. The ratio of the HIPASS
flux-weighted beam area to the mean observing beam area was then
calculated. Finally the AGES observing beam area was then multiplied
by this ratio to get the AGES flux-weighted beam area. The integrated
flux was then simply the summed AGES flux from the source multiplied
by the AGES pixel size divided by the AGES flux-weighted beam area.

There are other possible sources of inaccuracy in the AGES data. The
gain calibration did not take into account the telescope's response at
different zenith angles - beyond a zenith angle of $\sim$15\degree\/
the Arecibo 305m dish is known to lose sensitivity since only part of
the dish is illuminated. This effect was limited since very few scans
were taken at zenith angles greater than 15\degree\/.The system temperature is also known to vary
over the frequency range of the system which was not taken into
account during the data reduction. Steps are underway to document
these variations so that they may be included in future revisions of
{\sc livedata} and {\sc gridzilla}.

Table \ref{tbl1} shows the measured parameters from AGES, for each
galaxy, compared to those measured from HIPASS and also from other previous
surveys (Kamphuis \& Briggs, 1982; Briggs, 1986; Huchtmeier, 2003). Positions were found from applying the fitting routine {\sc imfit}
within {\sc miriad} to the flux map integrated over the galaxy's
velocity extent. Uncertainties are shown in brackets after each
measurement. Positional uncertainties for AGES data 
were estimated from fitting NRAO VLA Sky Survey (Condon \etal, 1998)
sources to a continuum map produced from our data. For AGES, positions of NVSS
sources were found to be accurate to within
$\pm$30\arcsec. Uncertainties in position from the HIPASS data were
taken from Zwaan \etal\/ (2003). The AGES
positions agree with HIPASS and previous measurements to within the
measured uncertainties.

The uncertainties in each of the \hi\/ properties depend on the S/N of
the source, the velocity width of the source, the velocity resolution
and the shape of the spectrum. An excellent summary of techniques for
estimating the uncertainties in these \hi\/ properties is given in
Koribalski \etal\/ (2004) and references therein. The uncertainty
in the integrated flux density is given by 
\begin{equation}
 \sigma(F_{HI}) =  4 \times SN^{-1}(S_{peak}F_{HI}\delta v)^{1/2}
\end{equation}
where $S_{peak}$ is the peak flux, SN is the ratio of $S_{peak}$ to $\sigma (S_{peak})$, F$_{HI}$ is
the integrated flux and $\delta v$ is the velocity resolution of the
data. $\sigma (S_{peak})$ is found to increase in extended sources,
regions of high 20-cm continuum emission and it also increases with
rising flux density values. We adopt the estimate of Koribalski
\etal\/ (2004) that \protect{($\sigma S_{peak})^2$ = rms$^2$ + (0.05 $S_{peak})^2$}.

The uncertainty in the systemic velocity is given by:
\begin{equation}
 \sigma(V_{sys}) = 3 \times SN^{-1}(P \delta v)^{1/2}
\end{equation}
where $P = 0.5 \times (W_{20} - W_{50})$, which is simply a measure of
the steepness of the profile edges. Errors in the widths are given
by $\sigma(W_{50}) = 2 \times\sigma(V_{sys})$ and $\sigma(W_{20}) = 3
\times\sigma(V_{sys})$. Distances are based on pure Hubble flow and
the associated errors arise from the errors in the integrated flux.

\begin{table*}
  \centering
  \begin{minipage}[c]{145mm}
    \caption{\hi\/ measurements of the NGC 628 group. \label{tbl1}}
    \begin{tabular}{lccccccc}
      Survey & RA & DEC & V$_{sys}$ &W$_{50}$ &W$_{20}$& F$_{HI}$  & M$_{HI}$ \\
      &(2000) &(2000) &(\kms) &(\kms) &(\kms) & (Jy \kms) & ($\times10^{8}$M$_{\odot}$)\\
      \hline
      {\bf NGC 628}&&&&&&\\
      AGES &  1 36 41.9 (2.0) & 15 47 44 (30) &659.2 (1) &
      56 (2)&73 (3)&352.6 (3.7) & 9.1 (0.1) \\
      HIPASS & 1 36 42.0 (4.5) & 15 47 34 (48) &654 (3)&
      50 (6) &75 (9) &431 (13) & 11.0 (0.3) \\
      Kamphuis& 1 36 42.0 (1.0)& 15 47 12 (14)& 657 (0.7)& 56 (1) &---& 470.1& 12\\
      \& Briggs  (1982)&&&&&&\\
      \\
      {\bf KDG010}&&&&&&\\
      AGES & 1 43 34.9 (2.0) & 15 41 48 (30) & 792 (1) & 34 (2) &49 (3) &3.13 (0.28) & 0.80 (0.07)\\
      HIPASS &  1 43 46.5 (4.6) & 15 42 48 (48) & 787 (5) & 27 (10) &45 (15) &2.0 (1.1)& 0.5 (0.3)\\
      Huchtmeier& 1 43 37.2 (4.6)& 15 41 43 (48) & 788.9 (0.4)& 32 (1.1)&---&2.35 & 0.6\\
       (2003)&&&&&&\\
      \\
      {\bf UGC 1176}&&&&&&\\
      AGES& 1 40 06.0 (2.0) & 15 54 45 (30) & 634 (1) & 36 (2) &54 (3) &27.4 (1.9)&
      7.0 (0.5)\\
      HIPASS& 1 40 08.4 (4.6) & 15 55 06 (48) & 628 (2) &
      40 (4) &58 (6)&28.2 (4.3)& 7.2 (1.8)\\
      Briggs  (1986)&  1 40 07.7 (1.0) & 15 53 55 (14) & 629 (1) & 38 (1)&---
      &30.17& 7.7\\
      \\
	{\bf UGC 1171}&&&&&&\\
	\\
      AGES&  1 39 38.6 (2.0) &15 55 19 (30) & 743 (1) & 29 (2) &41 (3) &1.46 (0.14)&
      0.37 (0.04)\\
      HIPASS& 1 40 00.2 (4.6) & 15 53 56 (48) & 743 (2) & 27 (4)& 45 (6)
      &2.2 (1.2)&0.56 (0.3)\\
      Briggs  (1986)& 1 39 44.2 (1.0)& 15 54 01 (14)& 740 (3)& 23 (4)&---&1.80& 0.5\\
      \\
	{\bf dw0137+1541}&&&&&&\\
	AGES&1 40 03.9 (2.0) & 15 55 45 (30) & 749 (1) & 28 (2) & 47
      (3) & 0.39 (0.21) &0.10 (0.05)\\
      HIPASS& --- & --- &   --- &  --- &  --- &  --- & ---\\
      Briggs  (1986)& 1 40 09.2 (1.0)& 15 56 16 (14)&750 (5)&
      23 (10)&---&0.27 & 0.07 \\
      \\
      \hline\\
	
    \end{tabular}
  \end{minipage}
\end{table*}
 
One of the aspects of AGES is the potential
to detect diffuse \hi\/ that is located in the intergroup or intracluster medium.
The high resolution of the AGES observations, however, has the advantage
of allowing differentiation between \hi\/ that may be associated with a hitherto
unknown LSB galaxy in the group (for instance, see Pisano \etal\/ 2004) and
a more extended, smooth, diffuse component or extended clump of emission within the
overall structure of the group. The nature of the distribution of baryonic material 
in this medium (e.g. White \etal\/ 2003) can provide interesting constraints 
on structure formation (see Mihos \etal\/ 2005; Osmond \etal\/ 2004). In
Fig. \ref{fig1}, the rectangular survey area has physical dimensions of
approximately 1 Mpc$\times$200 kpc. This provides an indication of how
spread out this group is. The overall physical radius of the NGC 628
extended group is approximately 1.5 Mpc with a relatively low velocity
dispersion of 100--150 \kms\/ depending on which choices are made for
group membership. Overall its physical extent is similar to our Local
Group, but it contains more than just 3 bright galaxies (all the
listed UGC and NGC galaxies are more massive than M33).

There
appears to be no evidence of intra-group gas down to a \hi\/ column density
of 2$\times10^{18}$\/ cm$^{-2}$ integrated over a velocity width,
$\Delta v$ = 50 \kms. In Table \ref{tbl1}, the sum of the detected
\hi\/ masses is approximately 2$\times 10^{10}$M$_{\odot}$, of which
1/2 belongs to NGC 628 and its extended \hi\/ emission.  Excising that
region from the search area and using our observed upper limit on
column density of any \hi\/ emission, leads to an upper limit of
5$\times 10^8$M$_{\odot}$ for the integrated smooth, intergroup \hi\/
mass that might be distributed between NGC 628 and KDG 010. This is
less than 3\% of the \hi\/ mass observed to exist in the bodies of the
galaxies and strongly suggests that tidal interactions in this
gas-rich group of galaxies have not been very robust in removing \hi\/
from the host galaxies. This result is consistent with the symmetry
and pronounced lack of tidal distortion of the extended \hi\/ emission
around NGC 628. Perhaps this suggests that loose galaxy structures,
like the NGC 628 group and the Local Group are too dynamically young
for sufficient interactions to have occurred.

\subsection{Objects beyond the NGC 628 group}

Due to the large observing bandwidth offered by the WAPPs (100 MHz) it
was expected that a number of background objects would also be
detected. The datacube was examined by eye for more distant 
objects and 22 other objects were detected out to a velocity of
$cz$=17500 \kms. Including the NGC 628 group this gives a detection rate of 5.4 objects deg$^{-2}$. This implies
that the total number of galaxies detected by AGES will be
$\sim$1500. The ALFALFA
survey (Giovanelli \etal\/ 2005) has predicted that they will detect a
20000 objects over an area of 7000 deg$^2$, giving an average
detection rate of 2.9 deg$^{-2}$. 

Another important aspect of this project is to identify potential optical
counterparts. Positional data is not enough for optical
associations, it is also necessary to obtain optical spectral data to
compare the redshift of the gas with that of the visible galaxies in
the image  field. If the positional or spectral data of the radio
observations don't correlate with that of the optical observations,
then an optical object cannot be considered the optical counterpart to
the \hi\/ emission. 

NED was used to search for objects within 
a 3\arcmin\/ radius of each \hi\/ detection. 6\arcmin\/$\times$6\arcmin\/
POSS II images, taken from the SuperCOSMOS database, were also
examined for possible optical counterparts. Of the 22 detections, 9
appear to be previously uncatalogued and none of the objects are
listed in NED as having previous \hi\/ measurements. SDSS redshift
data is available for only 3 objects, but these compare favourably
with the \hi\/ redshifts as described below. The \hi\/ detections are
listed in Table \ref{tbl6} along with any objects listed in NED within
a 3\arcmin\/ radius. The final column in Table \ref{tbl6} shows the
angular distance between the centre of the \hi\/ detection and the NED
object's position. Fig. \ref{fig10} shows the SuperCOSMOS
images for each object, centred on the \hi\/ detection, together with
the \hi\/ spectra. The Arecibo beam is represented by the circle and
is to scale. Optical detections within the Arecibo beam have
been identified with an arrow and previously catalogued objects have been
labelled. Descriptions of the images and the corresponding spectra are
given below. Although some objects coincide with the centre of the
\hi\/ emission, one should bear in mind that without optical redshift
data, it is impossible to say for certain whether or not these objects are
associated with the gas. 

\begin{table*}
    
    \begin{sideways}
      \begin{tabular}{lccc@{}c@{}ccc@{}c@{}ccc}
      \hline
      & RA & DEC & V$_{sys}$  & Distance  & W$_{50}$& W$_{20}$
      & F$_{HI}$  & M$_{HI}$  &
      Previously catalogued&Angular & M$_{HI}/$L$_{H}$\\
      AGES ID &(2000)  &(2000)  &(\kms) &(Mpc) &(\kms)  &(\kms) & (Jy \kms) & ($\times
      10^{8}$M$_{\odot}$)&objects&separation\\&&&&&&&&&& (arcmin)& M$_{\odot}/$L$_{\odot}$\\
      \hline
      J013149+152353& 01 31 48.9 & 15 23 53 & 10705 (3)& 146 & 178
      (6)&220 (9)& 1.44 (0.11)& 72.2
      &SDSS J01314980+1523265&0.5&0.9\\
      J013204+152947& 01 32 04.1 & 15 29 47 & 13196 (7) & 178 & 326
      (14) &400 (21)& 1.01 (0.11)& 75.5
      &SDSS J013204.6+153001.2&0.3&0.1\\
      &&&&&&&&&2MASX J01320599+1529298&0.5&0.5\\
      J013313+160139& 01 33 12.9 & 16 01 39 & 8895 (3) & 122 & 138
      (6)&194 (9)& 1.69 (0.11)& 59.0
      &2MASXi J0133132+160125&0.2&---\\
      J013323+160222& 01 33 22.8 & 16 00 21 & 8587 (1)& 118 & 69 (2)
      &81 (3) & 0.07 (0.02)& 2.1 & New & ---&---\\
      J013538+154850& 01 35 38.0 & 15 48 50 & 8012 (3)& 110 & 132
      (6)&153 (9) & 0.61 (0.08)& 17.5 &
      New&---&---\\ 
      J013718+153635& 01 37 18.0 & 15 36 35 & 8174 (4)& 112 & 144 (8)&
      169 (12) & 0.46 (0.08)& 13.7 &
      NVSS J013714+153722&1.2&0.3\\
      J013807+154328& 01 38 07.0 & 15 43 28 & 13237 (2)& 179 & 162 (4)
      &169 (6)& 0.60 (0.08) & 45.4 &New&---&---\\
      J013827+154728& 01 38 26.8 & 15 47 28 & 8339 (2)& 114 & 182 (3)
      &201 (5)& 1.74 (0.11)& 53.4 &
      New&---&--- \\
      J013917+154613& 01 39 17.1 & 15 46 13 & 17080 (5) & 228 & 51 (9)
      &98 (14)& 0.27 (0.05)& 33.5 &New&---&---\\
      J013953+151955& 01 39 52.7 & 15 19 55 & 16806 (6) & 224 & 231
      (12) &284 (18)& 0.43 (0.07) & 51.1
      &SDSS J013955.02+152036.7&0.9&---\\
      J013956+153135& 01 39 56.2 & 15 31 35 & 17343 (7) & 231 & 105
      (14) &214 (21)& 0.42 (0.07) & 53.7 & New&---&---\\
      J014013+153319& 01 40 13.0 & 15 33 19 & 17052 (7) & 227 & 535
      (14) &566 (21)& 1.33 (0.13) & 163.0
      &2MASX J01401331+1533345&0.3&0.2\\
      J014025+151903& 01 40 25.2 & 15 19 03 & 16848 (5) & 225 & 44 (9)
      &101 (14)& 0.31 (0.06) & 36.7
      &2MASX J01402402+1518240&0.7&0.2\\
      J014033+160513 & 01 40 32.6 & 16 05 13 & 13017 (4)& 176 & 101
      (8) &152 (12)& 0.59 (0.07)& 42.6
      &2MASXi J0140360+160414&1.3&---\\
      J014430+161502& 01 44 30.3 & 16 15 02 & 5020 (2)& 70 & 107 (3)
      &121 (5)& 0.61 (0.07)& 7.0 &
      New&---&---\\
      J014524+155923& 01 45 23.7 & 15 59 23 & 5050 (4)& 70 & 34 (7) &
      60 (11)&
      0.18 (0.04) & 2.1 & New&---&---\\
      J014630+154332& 01 46 30.4 & 15 43 32 & 17495 (3)& 232 & 132
      (6)&167 (9)& 0.73 (0.09)& 92.7 &
      2MASX J01463012+1543250&0.1&0.4\\
      J014644+155622& 01 46 44.0 & 15 56 22 & 7063 (5)& 97 & 27 (10)
      &51 (15)&
      0.10 (0.05) & 2.2 &2MASX J01465179+1557192&2.1&0.07\\
      J014719+155603& 01 47 19.1 & 15 56 03 & 5191 (3)& 72 & 86(7)
      &104 (11)& 0.35 (0.07)& 4.3 &NVSS
      J014722+155737& 1.8&---\\
      J014742+161317 & 01 47 41.7 & 16 13 17 & 4890 (3)& 68 & 138 (6)
      & 177 (9)& 1.02 (0.09)& 11.1
      &2MASXi J0147459+161318&1.0&---\\
      J014752+155855& 01 47 51.7 & 15 58 55 & 17514 (11) & 233 & 450
      (22) &578 (33)& 1.59 (0.13) & 200.0 &
      2MASX J01475424+1558247&0.8&1.0\\
      &&&&&&&&& 2MASX J01475424+1559387&1.0&---\\
      J014834+152756& 01 48 33.7 & 15 27 56 & 12985 (4) & 175 & 114
      (8) &129 (12)& 0.50 (0.07)& 36.3 &New&---&---\\
      \hline\\
      \end{tabular}
      \end{sideways}
    \caption{\hi\/ properties of objects detected beyond the NGC 628
    group. \label{tbl6}}
\end{table*}
    {\bf AGES J013149+152353}
This signal coincides with the SDSS galaxy, SDSS J013149.81+152326.5. There
is an optical redshift available for this galaxy placing it at $cz=$10717
\kms. The radio data place the galaxy at a recessional velocity of $cz=$10705
\kms. SDSS J013149.81+152326.5 is a spiral galaxy and this morphology
is reflected in the \hi\/ profile. This galaxy appears to be somewhat
isolated since there are no galaxies reported in this region within
1700 \kms.

{\bf AGES J013204+152947}
The SDSS galaxy, SDSS J013204.6+153001.2, falls almost in the centre
of the Arecibo beam. It has an optically measured redshift of
$cz=$13195 \kms\/ from the SDSS. Our radio measurement of the recessional
velocity is $cz=$13196 \kms. The good agreement between the positional and
redshift data make it almost certain that this is the source of the
\hi\/. This galaxy has a red optical colour ($g-r\sim$0.75 mag) 
typically observed in early type galaxies (Bernardi \etal\/ 2003). 
Moreover the SDSS nuclear spectrum shows a red continuum and no sign 
of recent star formation (i.e. no emission lines). 
A similar red colour ($g-r\sim$0.84 mag) is associated with the
companion, 2MASX J01320599+1529298, which lies within the beam.

{\bf AGES J013313+160139}
Spiral structure is clearly seen in the \hi\/ profile and is
probably coming from 2MASXi J0133132+160125 which lies in the
centre of the beam. 

{\bf AGES J013323+160222}
The \hi\/ profile for this object is not the large S/N signal centred
at $cz\sim$8650 \kms\/ in Fig. \ref{fig14}, but that located at
$cz\sim 8350$ \kms. The large signal is
coming from AGES J013313+160139 (Fig. \ref{fig12}). AGES J013323+160222
is admittedly a marginal detection ($\sim3\sigma$) and will no doubt
require deeper \hi\/ observations to confirm. There is a faint smudge
visible in {\it B} and {\it R} (not shown) located at 01\rah\/ 33\ram\/ 14.8\ras,
16\degree\/ 02\arcmin\/ 56\arcsec.

{\bf AGES J013538+154850}
The \hi\/ position
is likely to be coming from the spiral located near the centre of
the image. This is supported by the \hi\/ profile which is a typical
spiral, but only optical redshift data would really tie down the
\hi\/ emission to this galaxy.

{\bf AGES J013718+153635}
The \hi\/ profile for this object is typical of a spiral galaxy. Tying
it down to a single galaxy is difficult though. There is a 21cm
continuum source (NVSS J013714+153722) located at a distance of
1.2\arcmin. This frame is rather densely populated with at least five
galaxies present, two of which fall within the Arecibo beam. Neither
of these sources have any redshift information listed in NED for
comparison. The continuum source could be associated with the smudge
at 01\rah\/ 37\ram\/ 17.7\ras, 15\degree\/ 38\arcmin\/ 00\arcsec and this just
falls within the edge of the \hi\/ beam. The \hi\/ signal could also
be coming from the object at 01\rah\/ 37\ram\/ 17.0\ras, 15\degree\/
39\arcmin\/ 50\arcsec. 

{\bf AGES J013807+154328}
The \hi\/ profile of this galaxy suggests a spiral galaxy of mass
$\sim$4.5 $\times10^{9}$M$_{\odot}$. The optical counterpart is probably
 the object located near the centre of the POSS II frame
 (Fig. \ref{fig11}) but this object is not listed in NED. There is a
 hint of spiral structure in the optical which does lend support to
 this being the counterpart. Optical spectroscopy would be needed to
 help confirm this.

{\bf AGES J013827+154728}
While the bright galaxy to the south of the frame in Fig. \ref{fig13} is
identified by NED as 2MASX J01383039+1545455, the object in the centre of the
frame is not listed at all. The double horn \hi\/ profile indicates a
large, spiral galaxy. The optical image, however, shows a galaxy with a
rather clumpy structure, although it may be exaggerated by the
presence of foreground stars. Without optical redshift data one can't
say for certain that this galaxy is the source of the \hi\/
emission. However the correlation between the positions of the radio and
optical data is very good. 

{\bf AGES J013917+154613}
There is a faint galaxy that can been seen
nestling amongst the cluster of stars just south of the centre of the
image. Although quite dim compared with the surrounding stars, the
galaxy's neutral gas content (3.3$\times10^{9}$M$_{\odot}$) suggests we
are looking at a Milky Way sized galaxy that is merely far away
($\sim$ 230 Mpc). This is unusual since this galaxy also has
approximately the same mass as object 2 and is at the same distance,
yet nothing as obvious as 2MASX J01463012+1543250 lies within the
Arecibo beam.

{\bf AGES J013953+151955}
This galaxy is probably the edge-on galaxy located in Fig. \ref{fig11}
at 01\rah\/ 39\ram\/ 55.0\ras, 15\degree\/ 20\arcmin\/ 34\arcsec. This galaxy
has also been identified by SDSS with an {\it optical} recessional
velocity of $cz=$16693 \kms. Our redshift of $cz$=16806 \kms\/
compares reasonably well with this. The shape of the spectrum is
fairly Gaussian and has a narrow width (W$_{50}$ = 49 \kms). It would
be highly unusual for a galaxy as massive as this one
(5.1$\times10^{9}$M$_{\odot}$) to have such a narrow profile unless we
were viewing it face-on. SDSS J013955.02+152036.7 looks more like an
edge-on galaxy, so either some of the \hi\/ flux is hidden in the
noise or this galaxy is not responsible for the \hi\/ emission. More
sensitive radio observations would show for certain if we are just
measuring the peak of one of the double-horns that one would expect
from such an edge-on system. 

{\bf AGES J013956+153135}
The \hi\/ signal
is about a 5-$\sigma$ detection with a peak of around 5.5 mJy and W$_{50}$=105
\kms. The optical counterpart is a little more tricky. There is a
small smudge in Fig. \ref{fig10} (object 4) at 01\rah\/ 39\ram\/ 55.0\ras, 15\degree\/
30\arcmin\/ 22\arcsec. However, the \hi\/ mass is $\sim 5\times
10^{9}$H$_{\odot}$, which is similar to object 2. Therefore we might
expect to see a galaxy of a similar size to 2MASX
J01463012+1543250. This is a very interesting detection. There are no
obvious optical counterparts but the POSS II image is relatively
shallow so it is possible this \hi\/ detection has a
low surface brightness optical counterpart. Deep optical follow-up
observations are essential to try and determine whether this object
has an optical association. 

{\bf AGES J014013+153319}
Two objects have been identified within the beam that could be
contributing to the \hi\/ emission. The \hi\/ profile shows an excess
of neutral gas in the approaching half of the spectrum. This could be
due to superposition of individual profiles or due to asymmetry in
the gas distribution of one of the galaxies, if these galaxies are
indeed responsible for the \hi\/ emission. 

{\bf AGES J014025+151903}
This object has a narrow velocity width (W$_{50}$ = 44 \kms) but
is a 6-$\sigma$ detection. 2 objects exist in NED in this
area, but one of them is a distant quasar (SDSS
J014026.6+151813.5). 2MASX J01402402+1518240 falls within the Arecibo
beam and shows up in the POSS II image as a faint smudge
(Fig. \ref{fig11}). Follow-up optical spectroscopy and radio
interferometry would enable us to determine if 2MASX J01402402+1518240
is the source of the \hi\/ emission.

{\bf AGES J014033+160513}
The POSS II plate presents a particularly interesting object, 2MASXi
J0140360+160414, which lies just on the edge of the beam. At first
glance the optical image looks like a spiral galaxy with a spiral arm
extending to the south. A closer inspection of the POSS II images has
revealed that the central part of the
galaxy can almost be resolved into two objects and the trailing spiral
arm looks like a tidal stream linking a third object to the south. It
looks as if there is an interesting merger going on here. Further
investigation would be required to ascertain the nature of this
merger, and to determine whether or not these objects are responsible
for the \hi\/ emission.

{\bf AGES J01430+161502}
The \hi\/ signal
is unmistakable but locating the optical source is not as
straightforward. 2MASX J01443758+1615501 is present in the optical frame
in Fig. \ref{fig15} but its position would place it at the very
edge of the beam. The \hi\/ position is more likely linked to the
smudge nearer the centre of the beam at 01\rah\/ 44\ram\/ 33.5\ras,
16\degree\/ 14\arcmin\/ 48\arcsec. Its elongated structure is pointing
towards 2MASX J01443758+1615501 which may be the result of an
interaction with this galaxy. 

{\bf AGES J014524+155923}
Although the \hi\/ detection is narrow
(W$_{50}$ = 30 \kms) the peak flux is a 6-$\sigma$ detection and the
source of the \hi\/ is more or less centred on the galaxy in
Fig. \ref{fig15} at 01\rah\/ 45\ram\/ 22.5\ras, 15\degree\/ 59\arcmin\/
00\arcsec. This is the least massive of the detected galaxies, with a
\hi\/ mass of $\sim$2.1 $\times10^{8}$ M$_{\odot}$.

{\bf AGES 014630+154332}
The \hi\/ profile is double-horned with a width of 132 \kms,
suggesting this object is a spiral galaxy but the galaxy located at
the centre of optical image does not reflect this. It is a very
circular object, so if this object is the optical counterpart to the
\hi\/ it must have a very bright bulge or very diffuse arms. It's
circular nature also suggests we are looking at the galaxy face-on,
but if this was the case the \hi\/ profile would look more Gaussian. 

{\bf AGES J014644+155622}
Although the \hi\/ signal of this object is very narrow (26 \kms) it
persists over several channels suggesting that it is indeed real even
though the peak is only a 4-$\sigma$ detection. There are two objects
visible on the optical image (Fig. \ref{fig13}) that are on the
edge of the beam. The first object, 2MASX J01465179+1557192, is very bright and
large although it is 2.1\arcmin\/ away from the centre point of the
\hi\/ emission. The second object is only visible as a faint smudge
in the {\it B}-band and {\it R}-band images but is closer in separation
($\sim$1.5\arcmin). The second object is not listed in NED.

{\bf AGES J014719+155603}
The optical frame shows only one possible optical counterpart. The
object is located at 01\rah\/ 47\ram\/ 21.5\ras, 15\degree\/ 57\arcmin\/
00\arcsec. NVSS J014722+155737 is listed in NED less than 30 \arcsec
away from the optical detection and is probably the source of the
radio emission. Whether or not it is the source of the \hi\/ emission
is difficult to say. The positions do not match that well but optical
spectra would help make this distinction.

{\bf AGES J014742+161317}
The \hi\/ profile is fairly Gaussian with a velocity width of
W$_{50}$ = 138 \kms. The optical image shows two possible counterparts at
01\rah\/ 47\ram\/ 39.0\ras, 16\degree\/ 13\arcmin\/ 56\arcsec\/ (most likely)
and 01\rah\/ 47\ram\/ 46.0\ras, 16\degree\/ 13\arcmin\/ 18\arcsec\/ (2MASXi
0147459+161318). Further investigation will reveal the true source of
the \hi\/ emission.

{\bf AGES J014752+155855}
There are 2 galaxies in the {\it B}-band image that fall within the
beam. Both are listed as 2MASS objects and the \hi\/ emission may be coming
from one, both, or neither of these objects. Without higher resolution \hi\/
images and optical spectra it is impossible to distinguish between
them. 

{\bf AGES J014834+152756}
The double-horned \hi\/ velocity profile and the optical image both
indicate the this galaxy is a spiral. On closer inspection of the
optical data it seems that the galaxy may be a one-armed spiral. Again
this object is not listed in NED. It would be
interesting to acquire higher resolution radio data of
this object to see if there is tidal debris linked to this perturbed
spiral.

\begin{figure*}
\centering
 \subfigure{
    \includegraphics[width=0.32\textwidth]{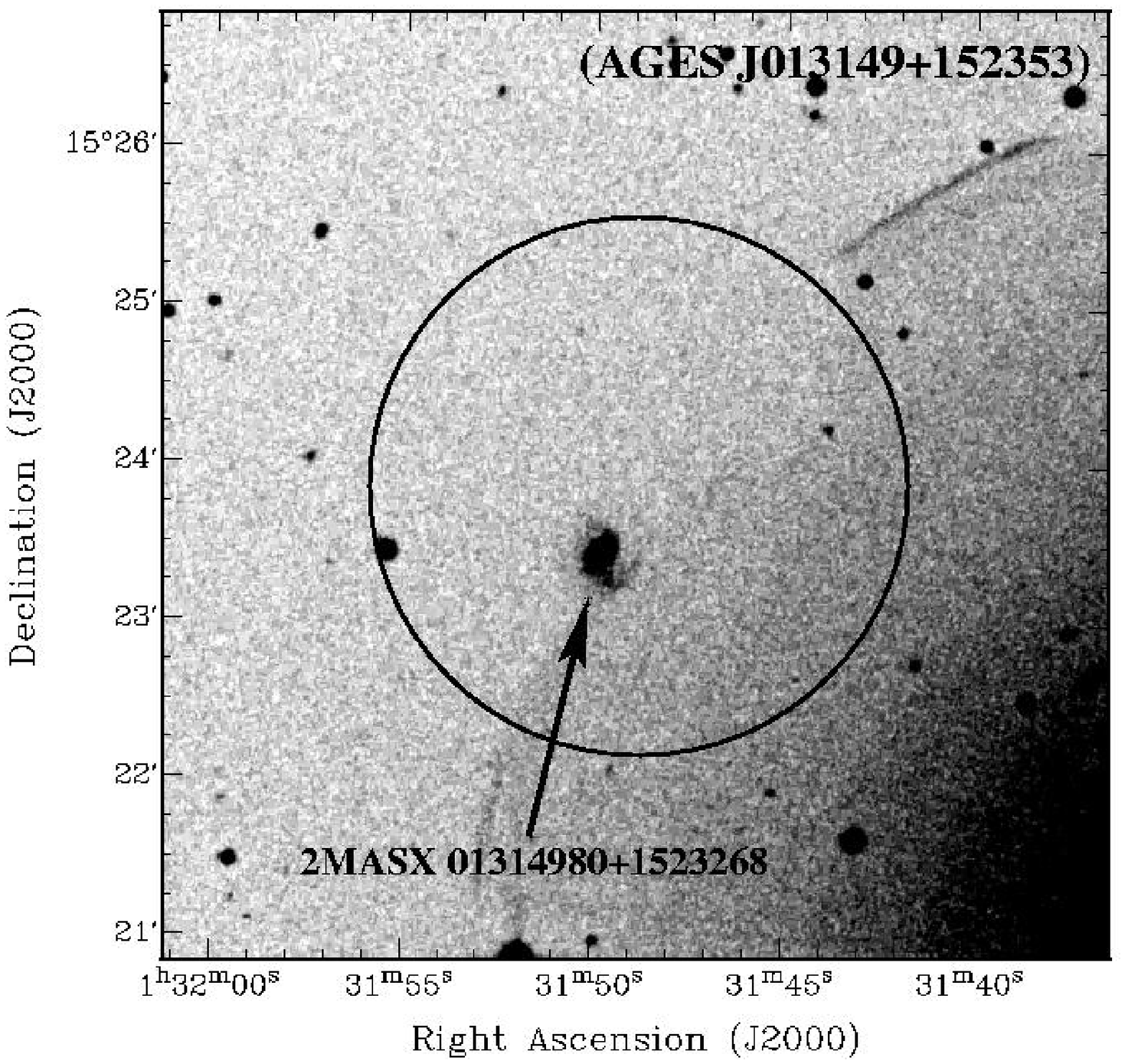}}
  \subfigure{
    \includegraphics[width=0.45\textwidth]{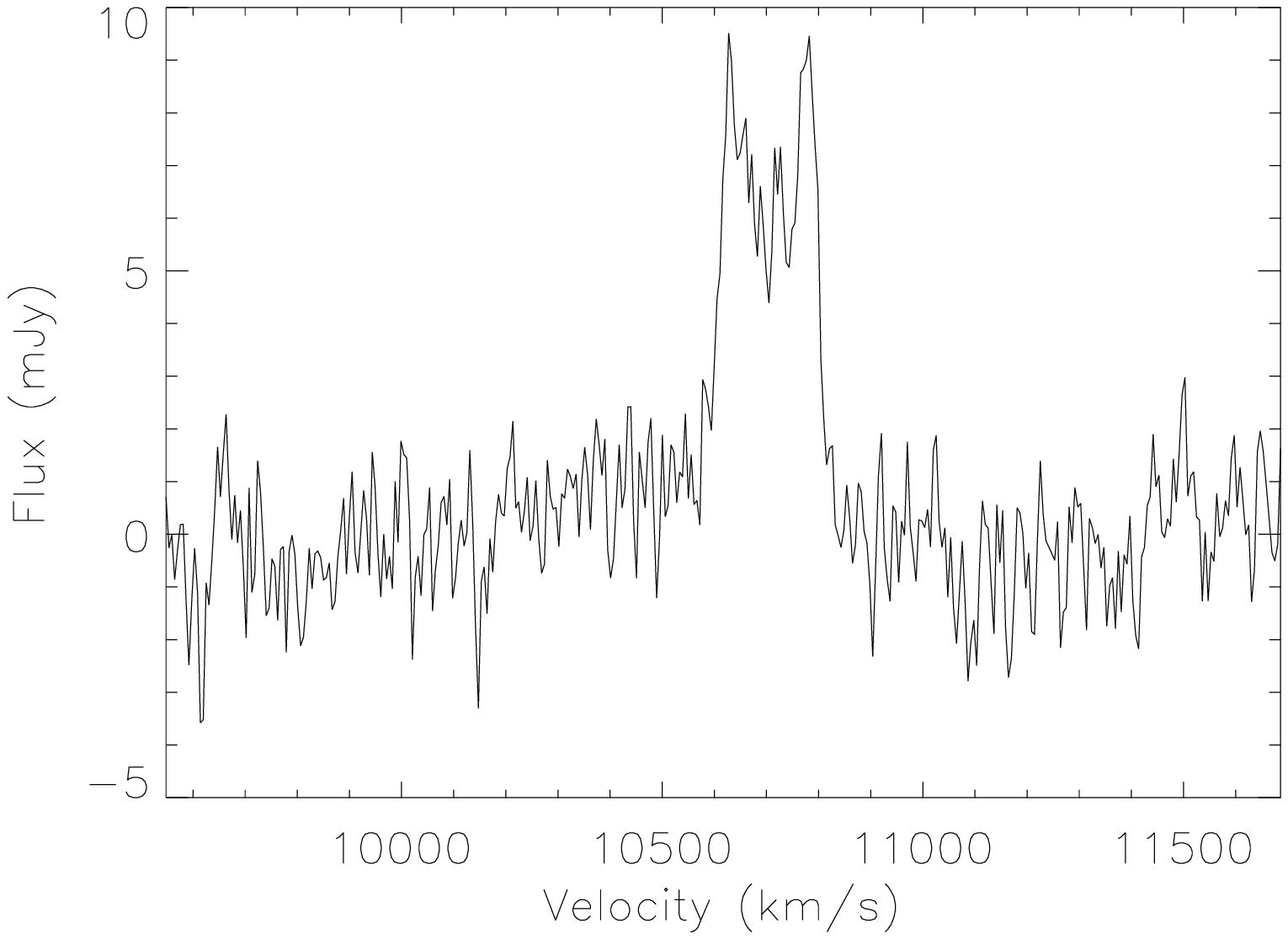}}
  \subfigure{
    \includegraphics[width=0.32\textwidth]{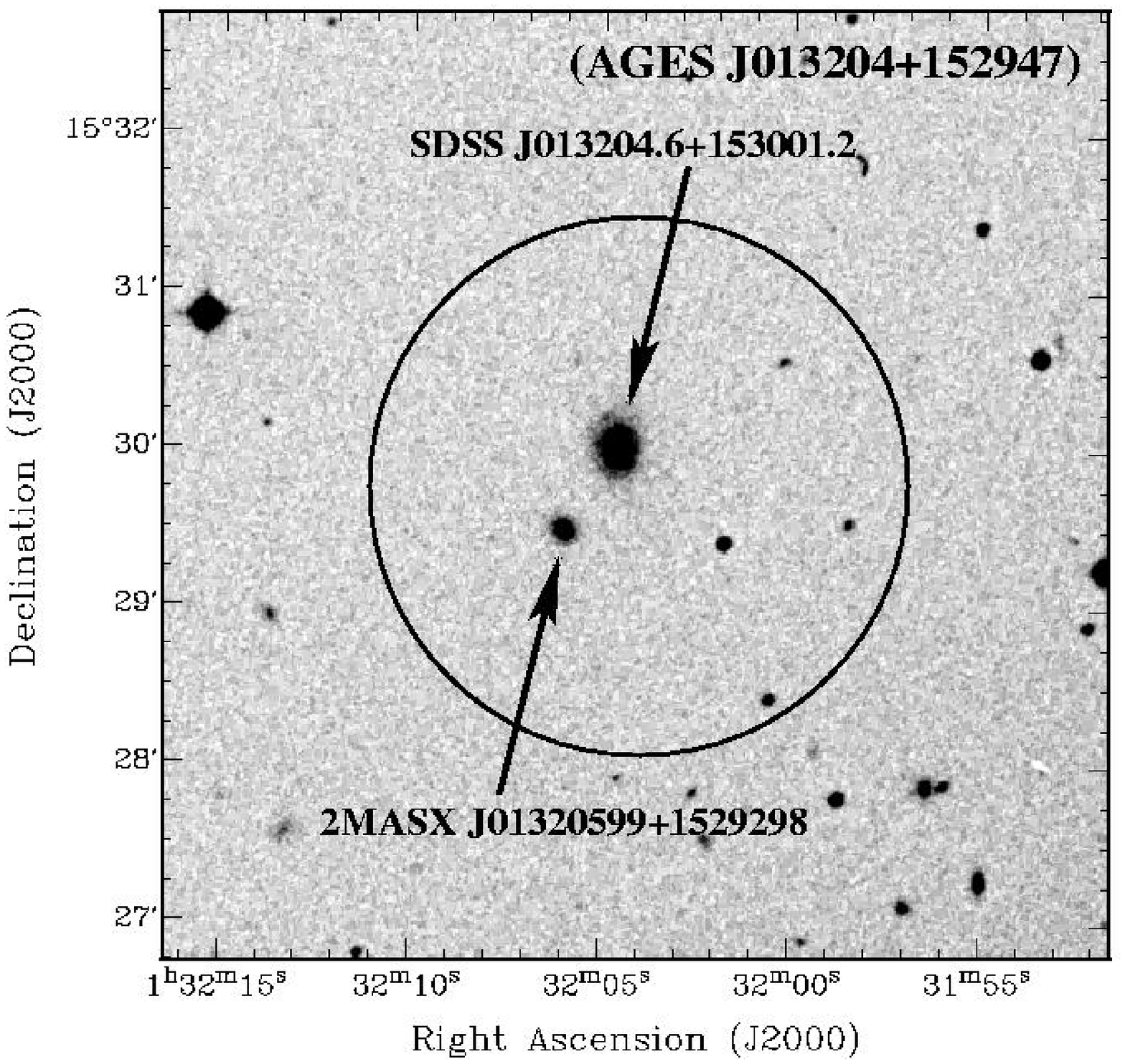}}
  \subfigure{
    \includegraphics[width=0.45\textwidth]{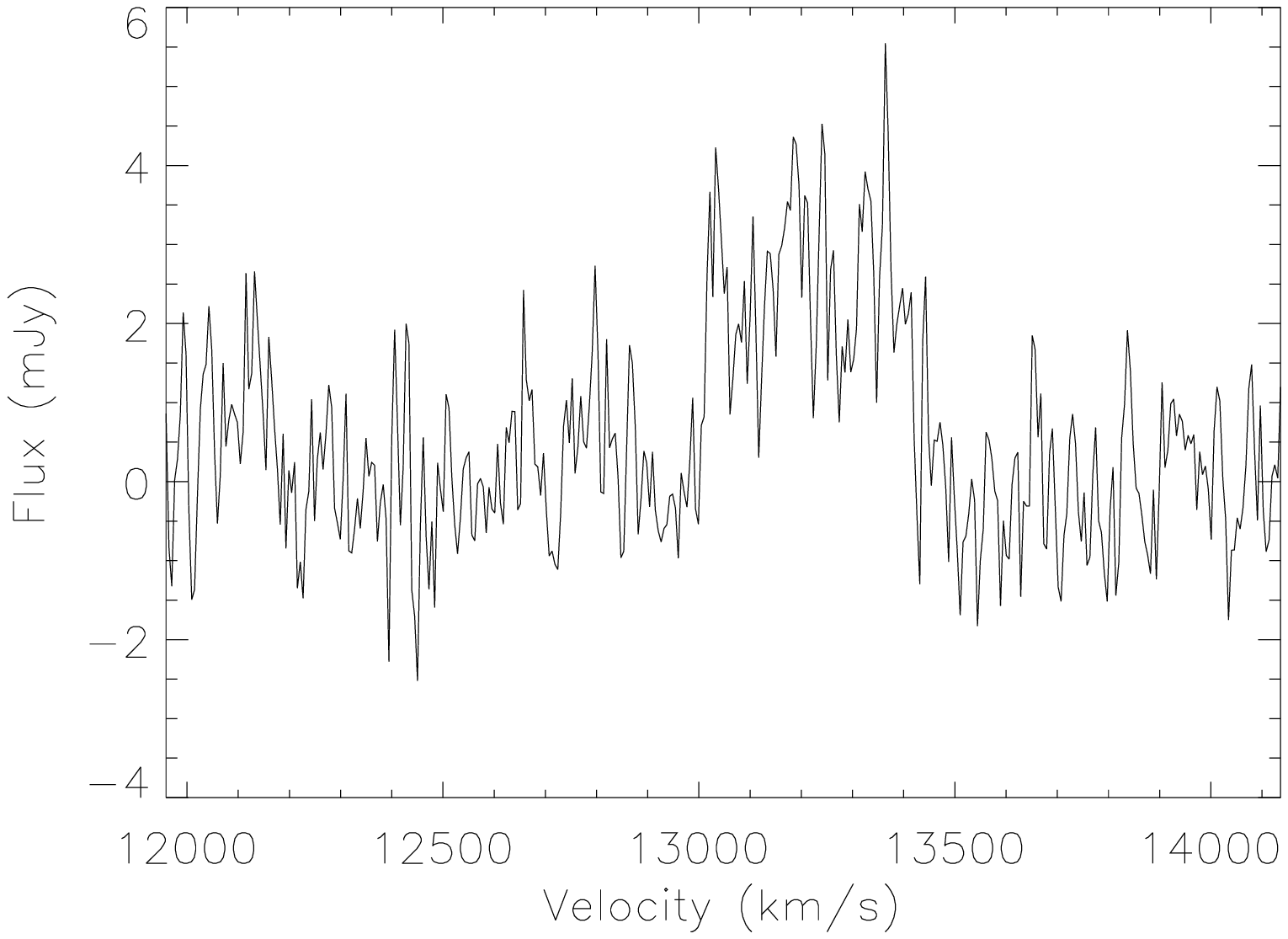}}
  \subfigure{
    \includegraphics[width=0.32\textwidth]{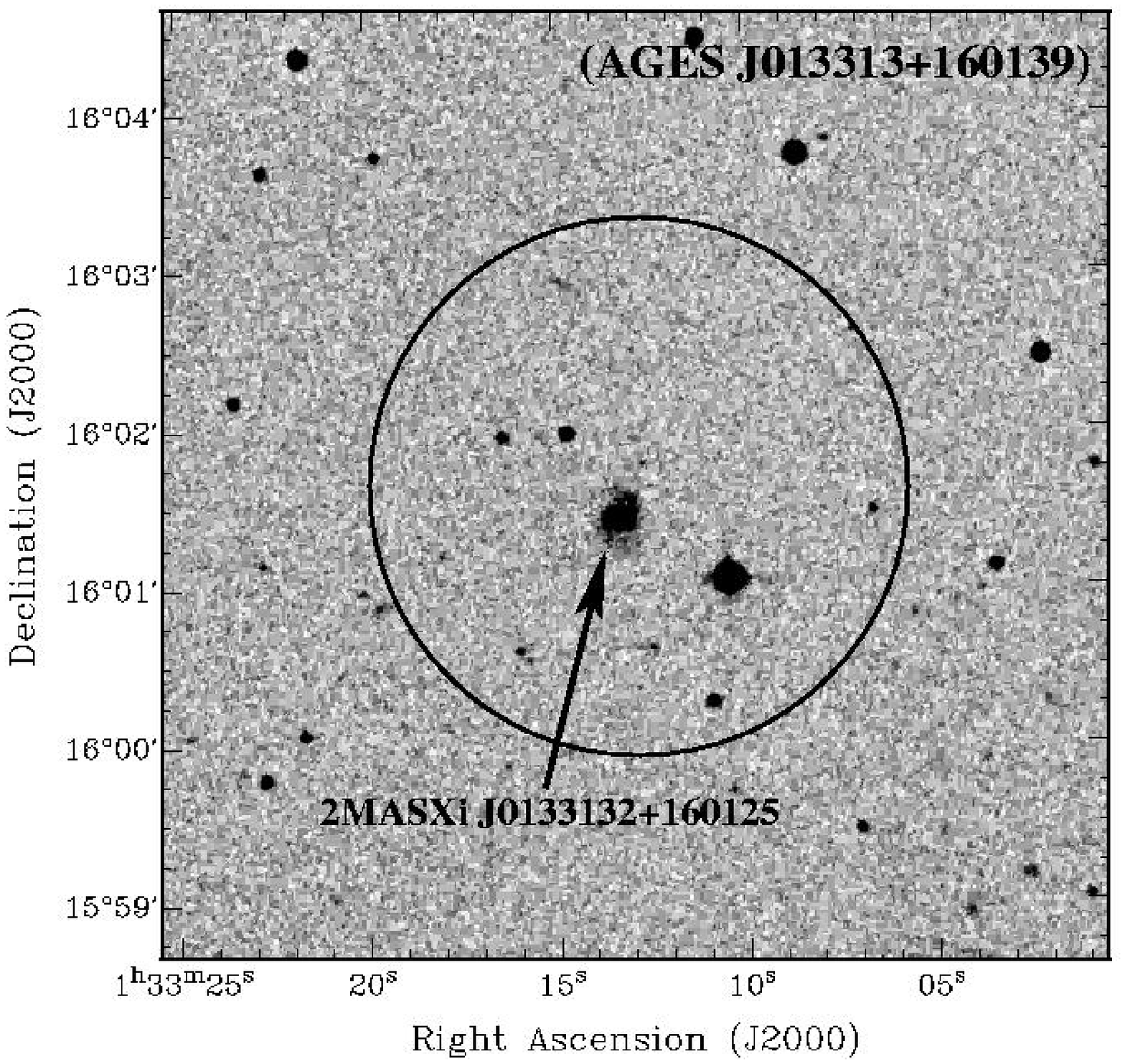}}
  \subfigure{
    \includegraphics[width=0.45\textwidth]{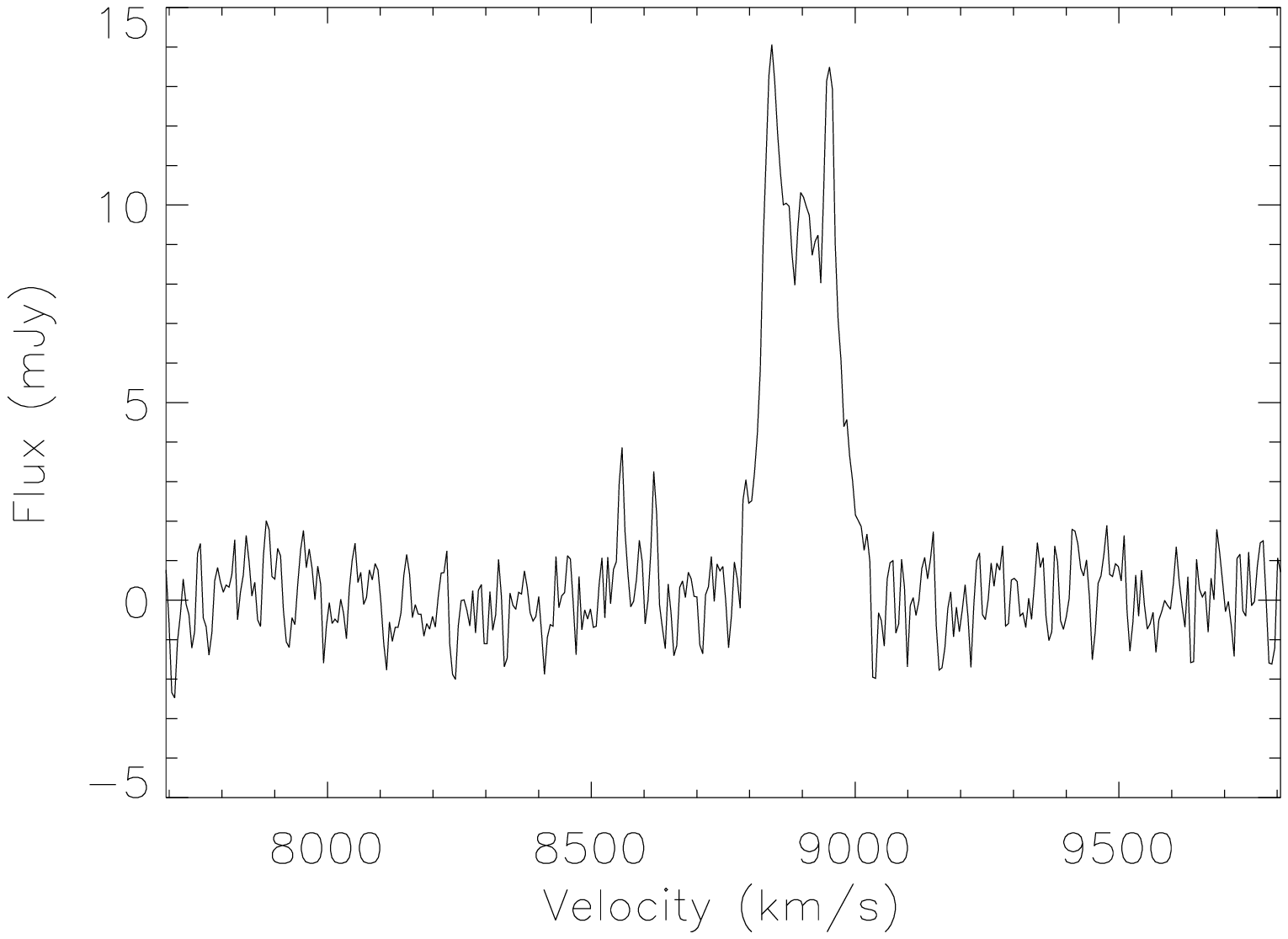}}
  \subfigure{
    \includegraphics[width=0.32\textwidth]{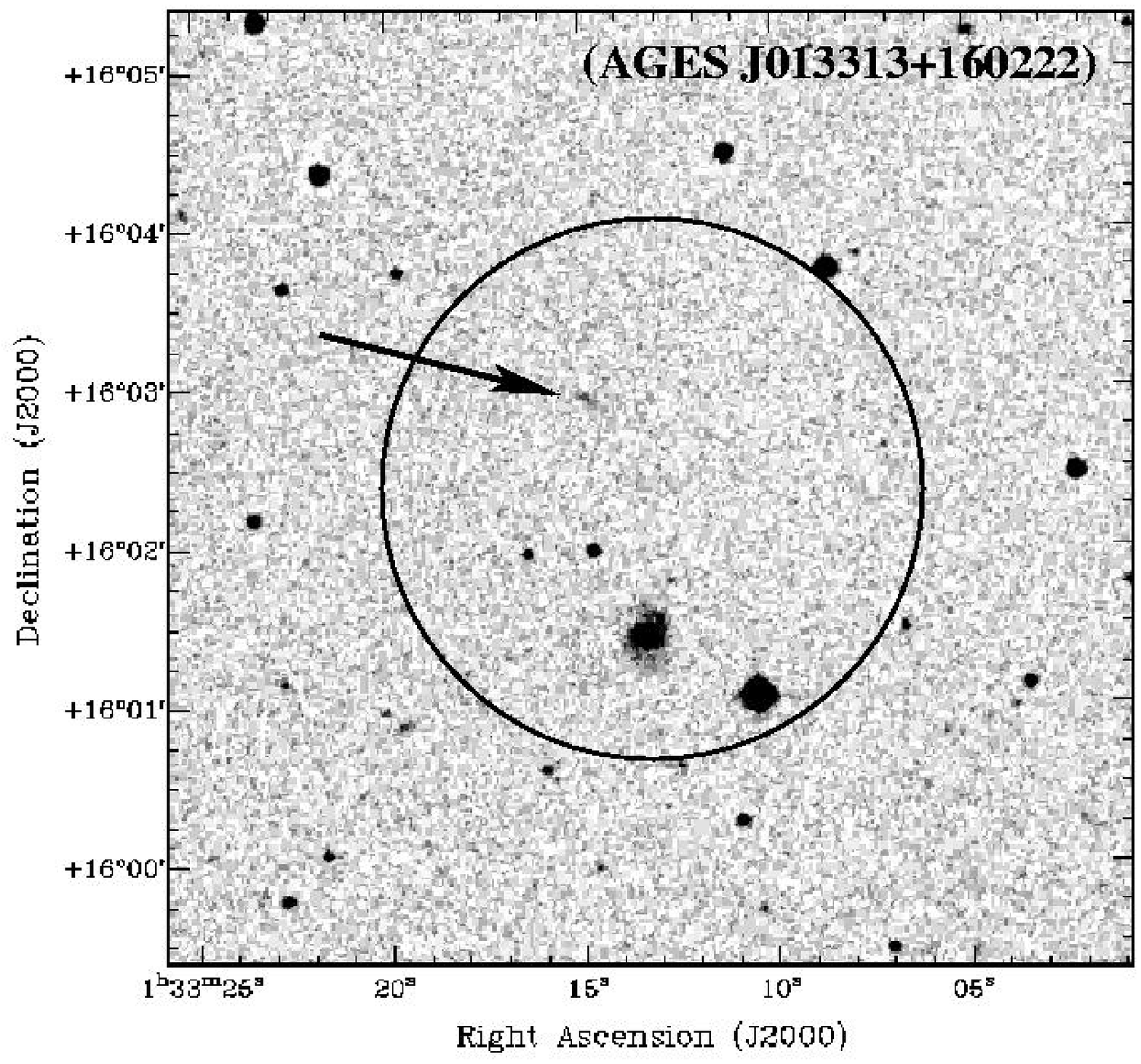}}
  \subfigure{
    \includegraphics[width=0.45\textwidth]{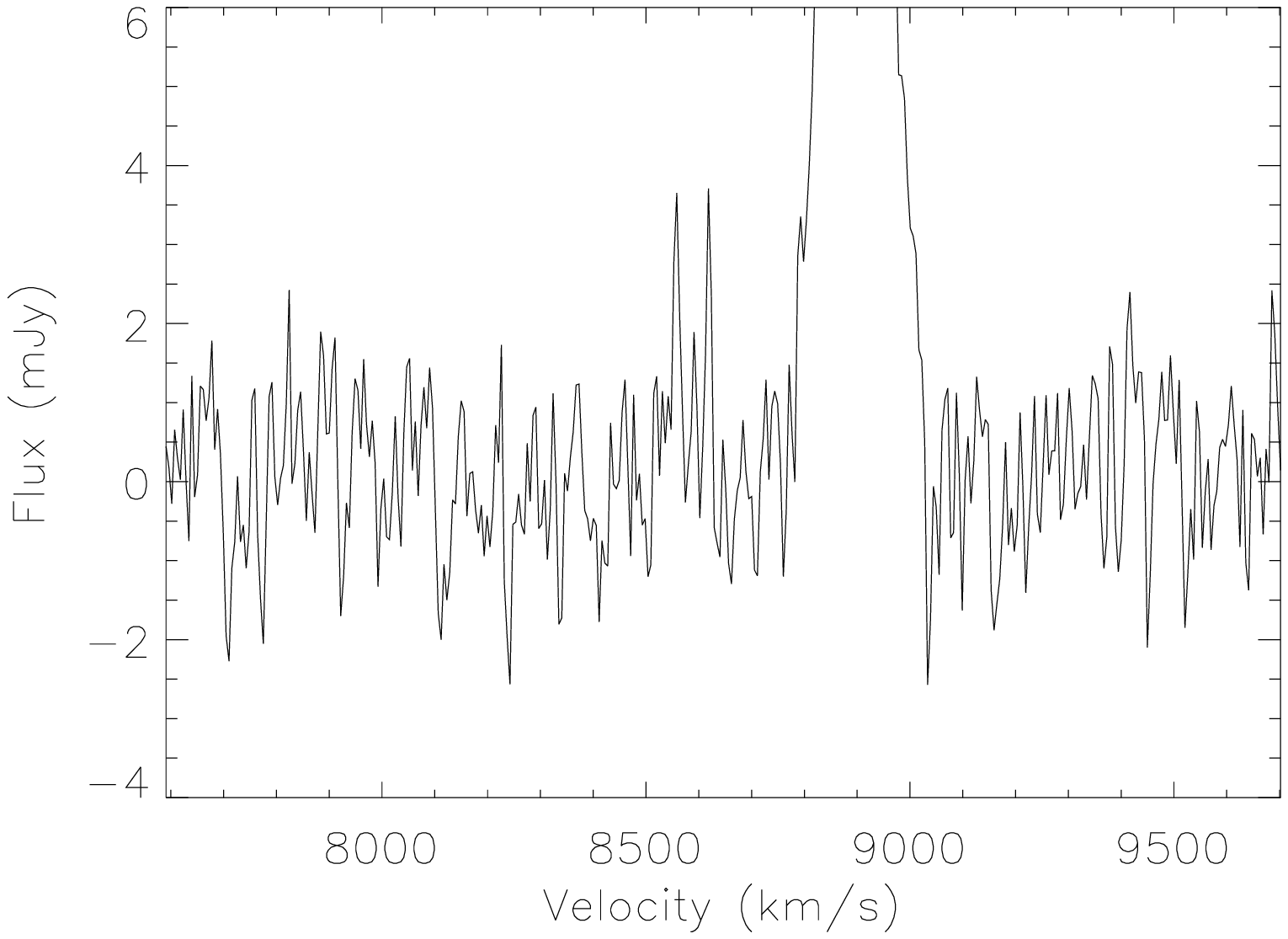}}
  \caption{Top to bottom: {\it B} band images and accompanying \hi\/
  spectra for AGES objects J013149+152353, J013204+152947,
  J013313+160139, J013313+160222\label{fig10}}
\end{figure*}
\begin{figure*}
\centering
  \subfigure{
    \includegraphics[width=0.32\textwidth]{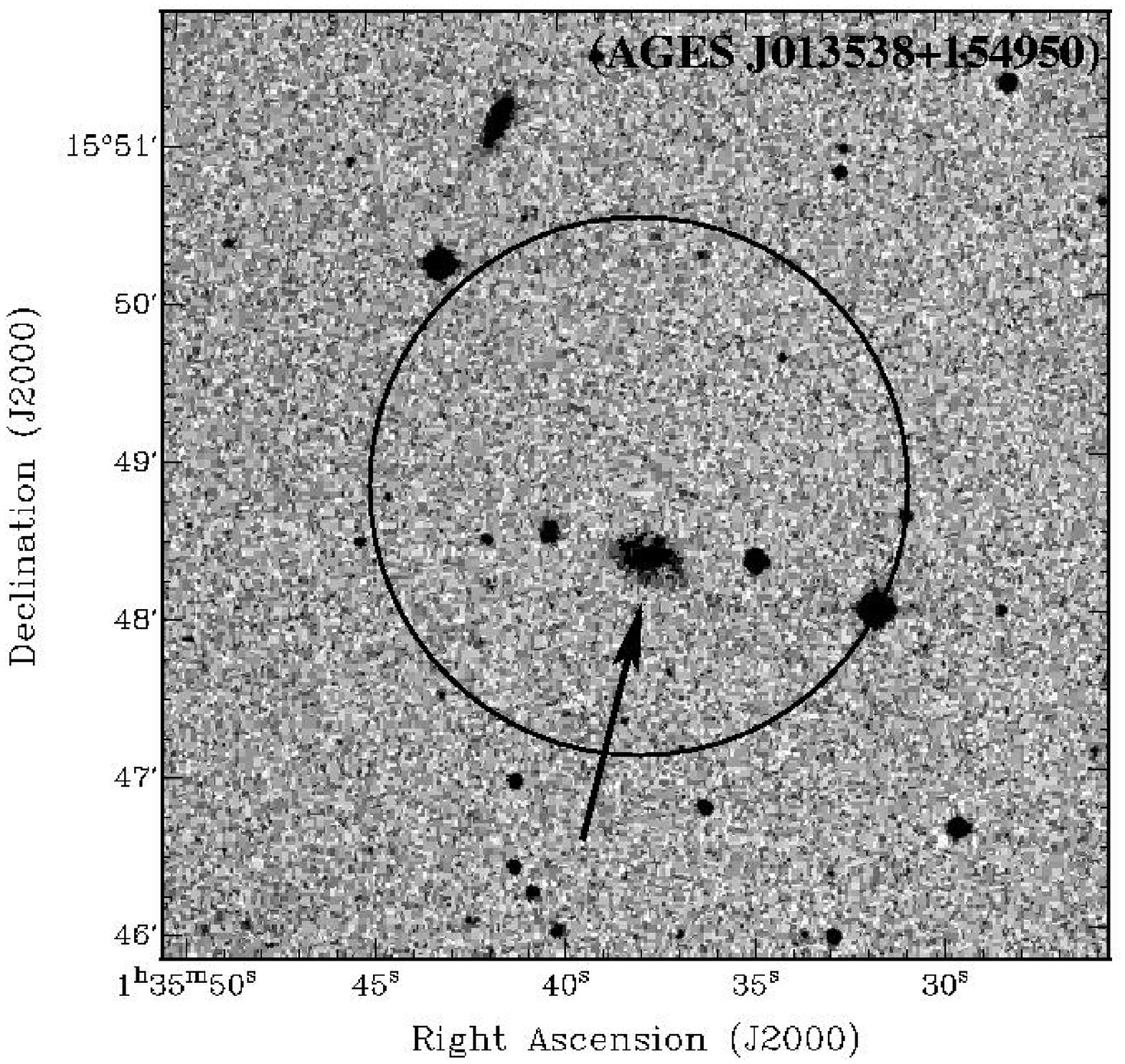}}
  \subfigure{
    \includegraphics[width=0.45\textwidth]{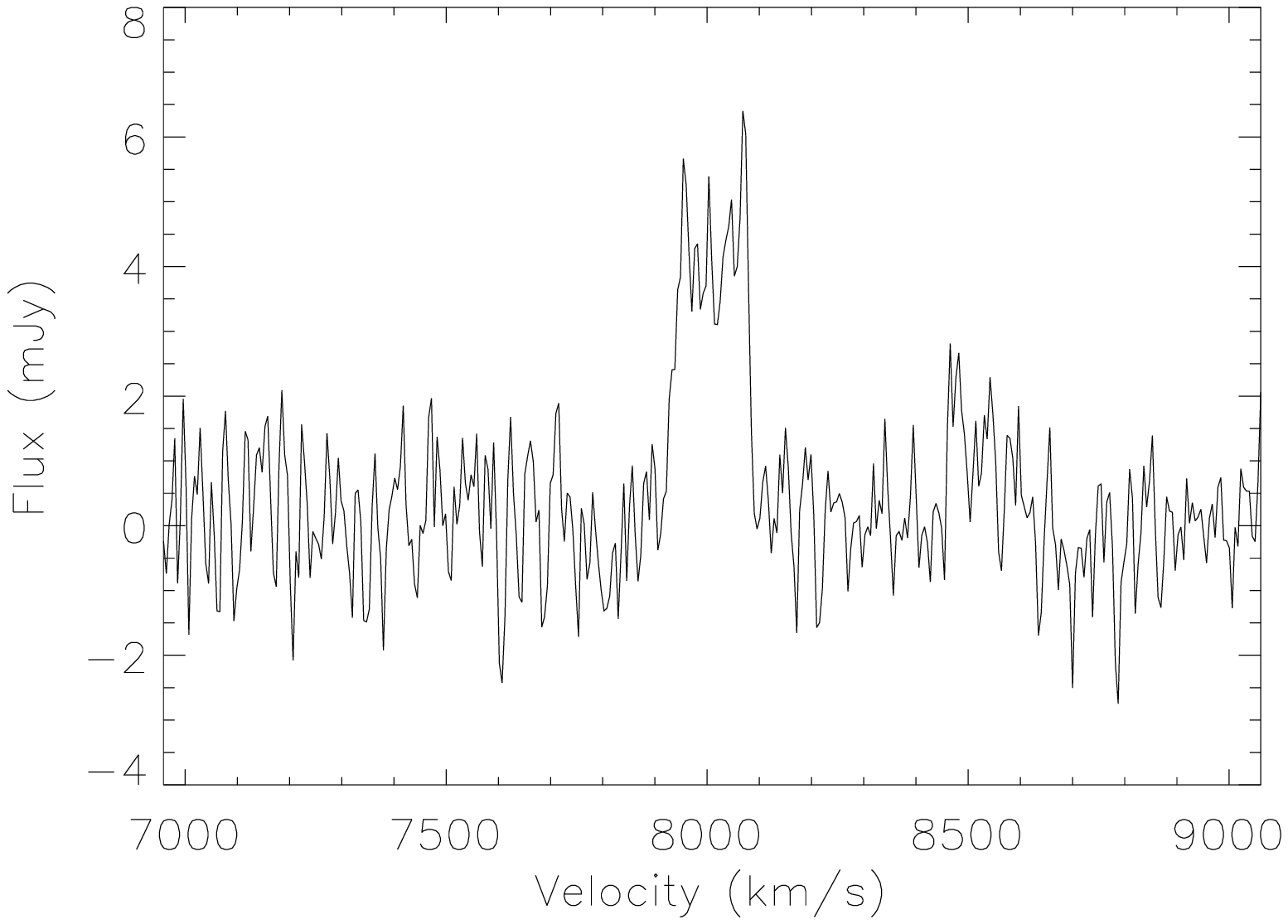}}
  \subfigure{
    \includegraphics[width=0.32\textwidth]{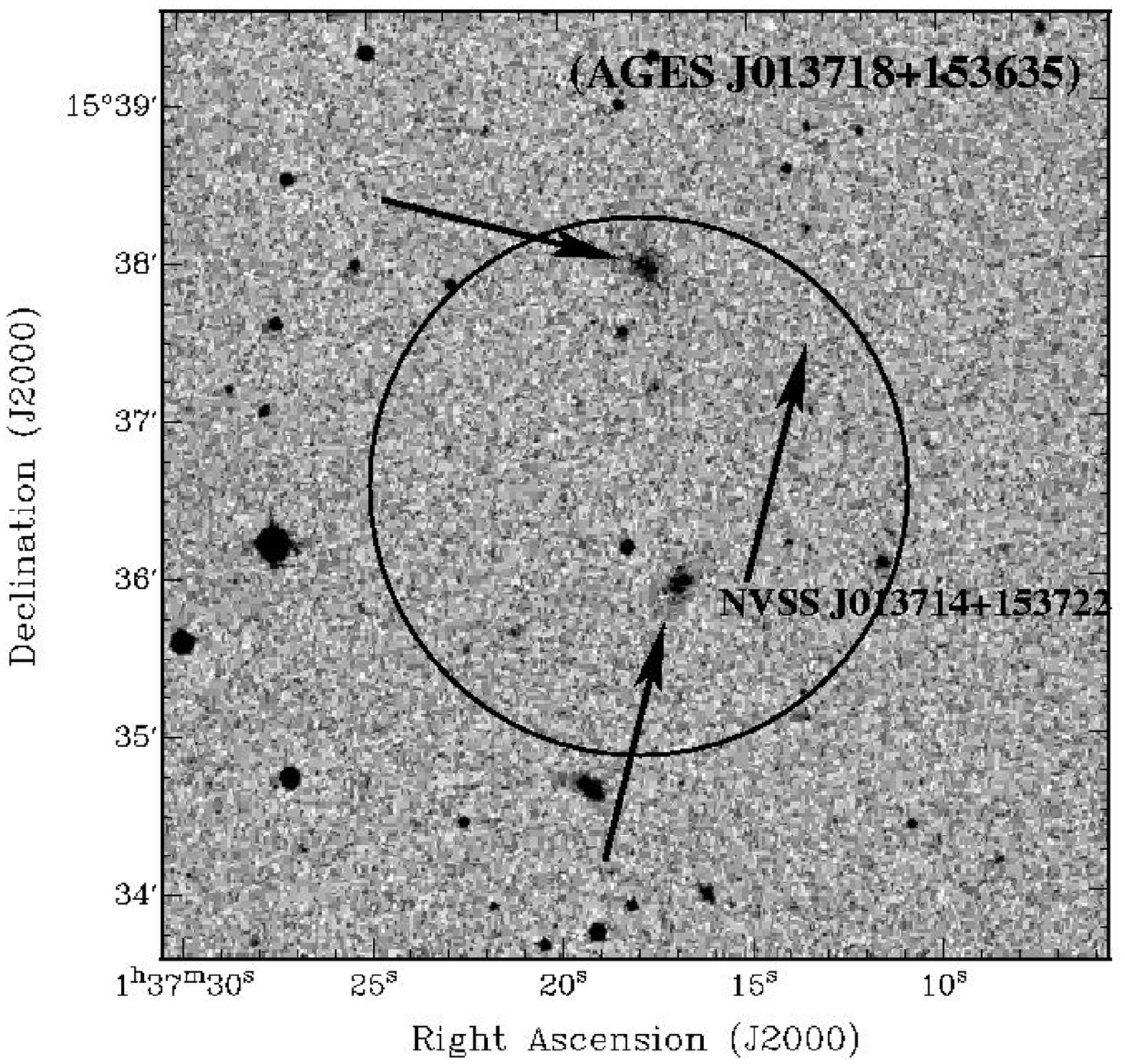}}
  \subfigure{
    \includegraphics[width=0.45\textwidth]{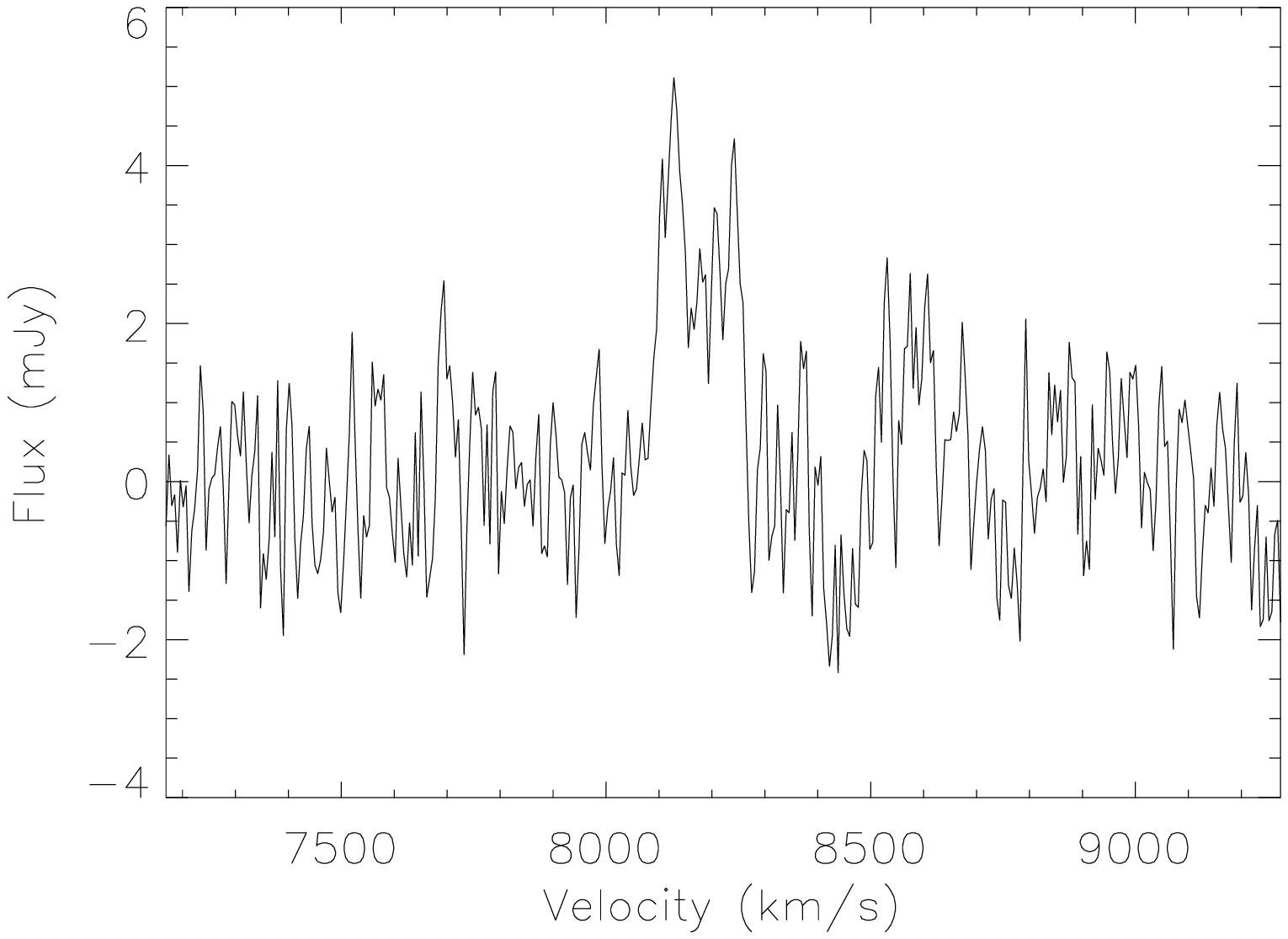}}
  \subfigure{
    \includegraphics[width=0.32\textwidth]{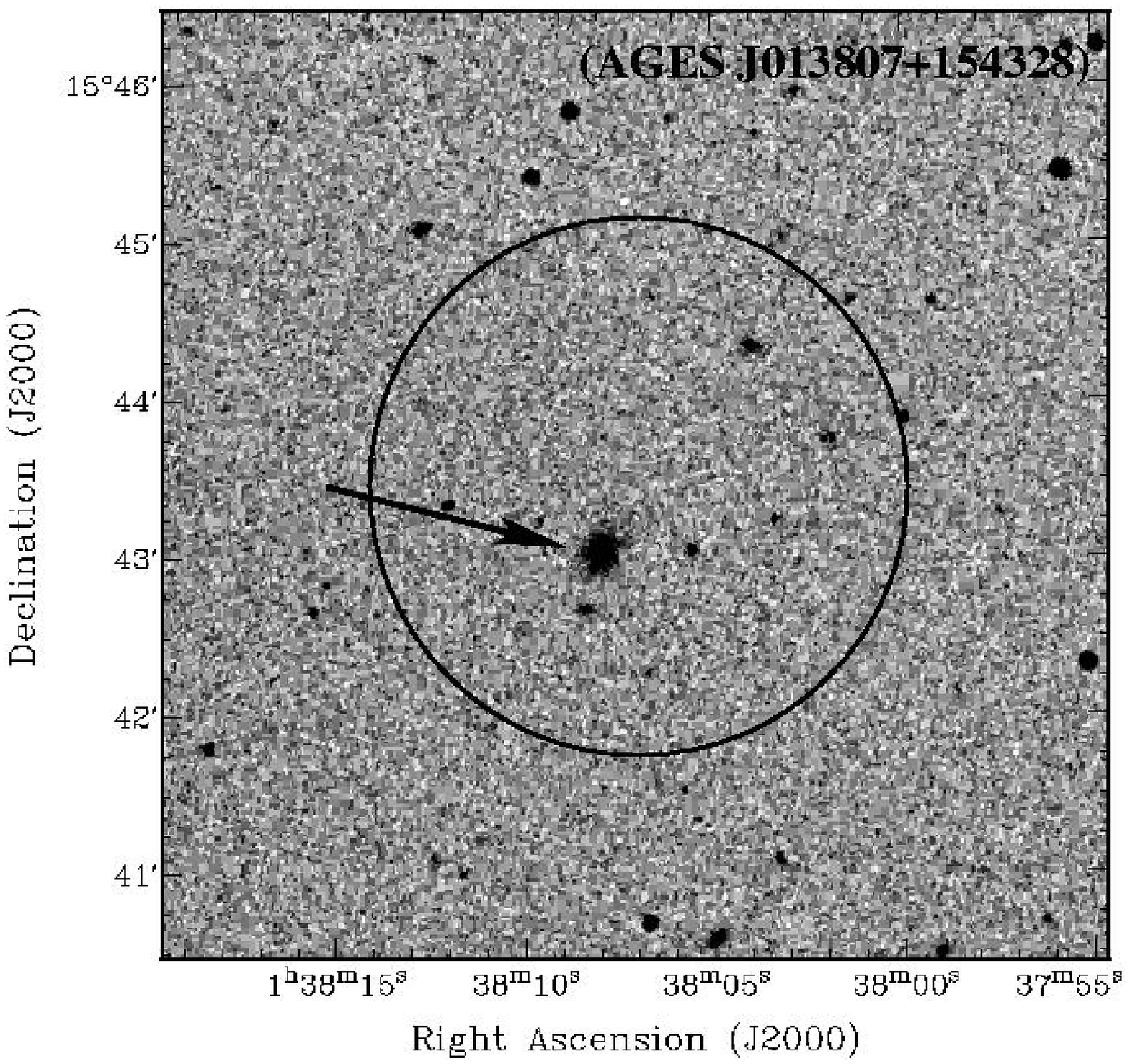}}
  \subfigure{
    \includegraphics[width=0.45\textwidth]{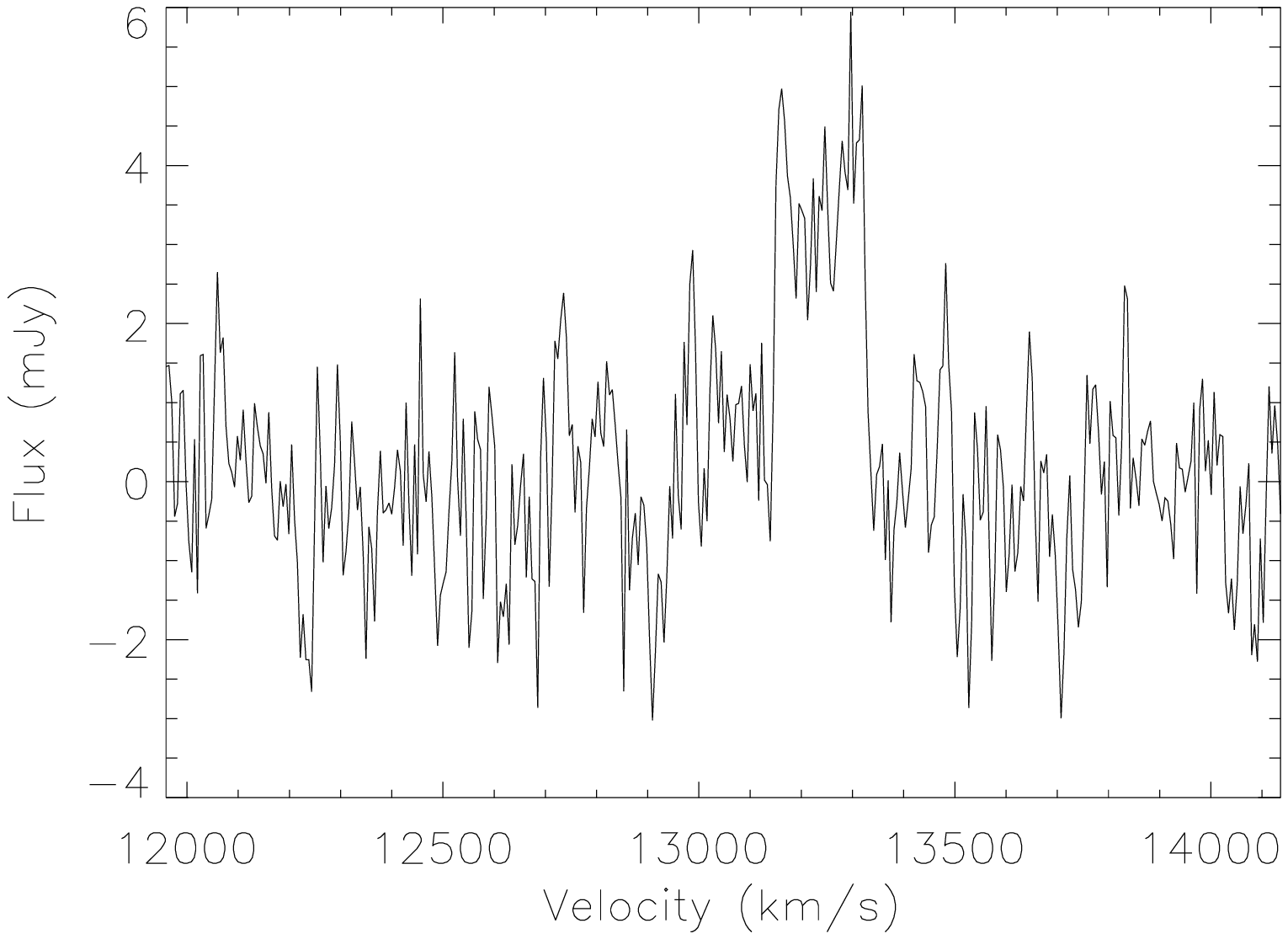}}
  \subfigure{
    \includegraphics[width=0.32\textwidth]{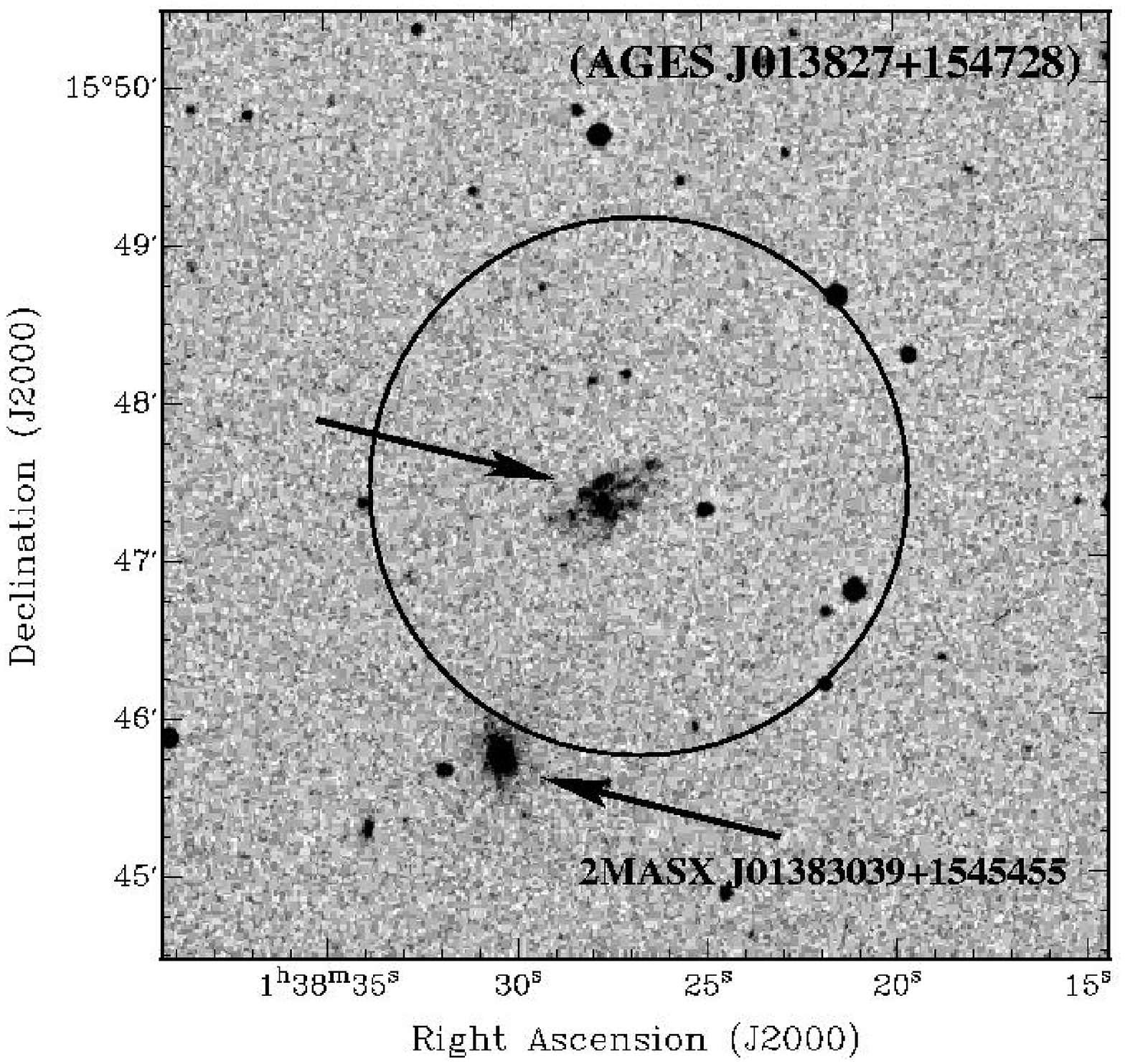}}
  \subfigure{
    \includegraphics[width=0.45\textwidth]{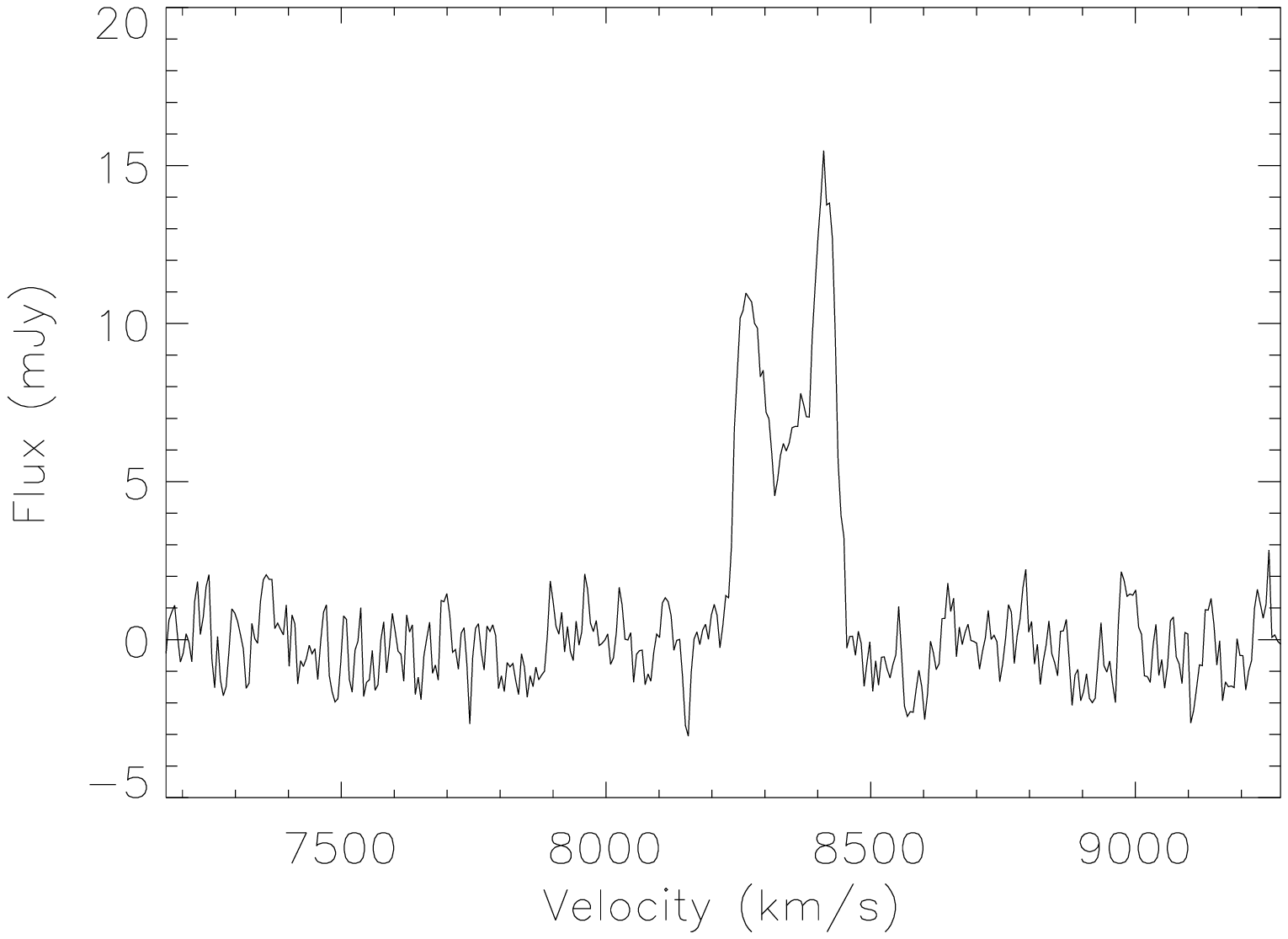}}

  \contcaption{Top to bottom: {\it B} band images and accompanying \hi\/
  spectra for objects J013538+154850, J013718+153635, J013807+154328, and J013827+154728\label{fig11}}
\end{figure*}
\begin{figure*}
\centering
  \subfigure{
    \includegraphics[width=0.32\textwidth]{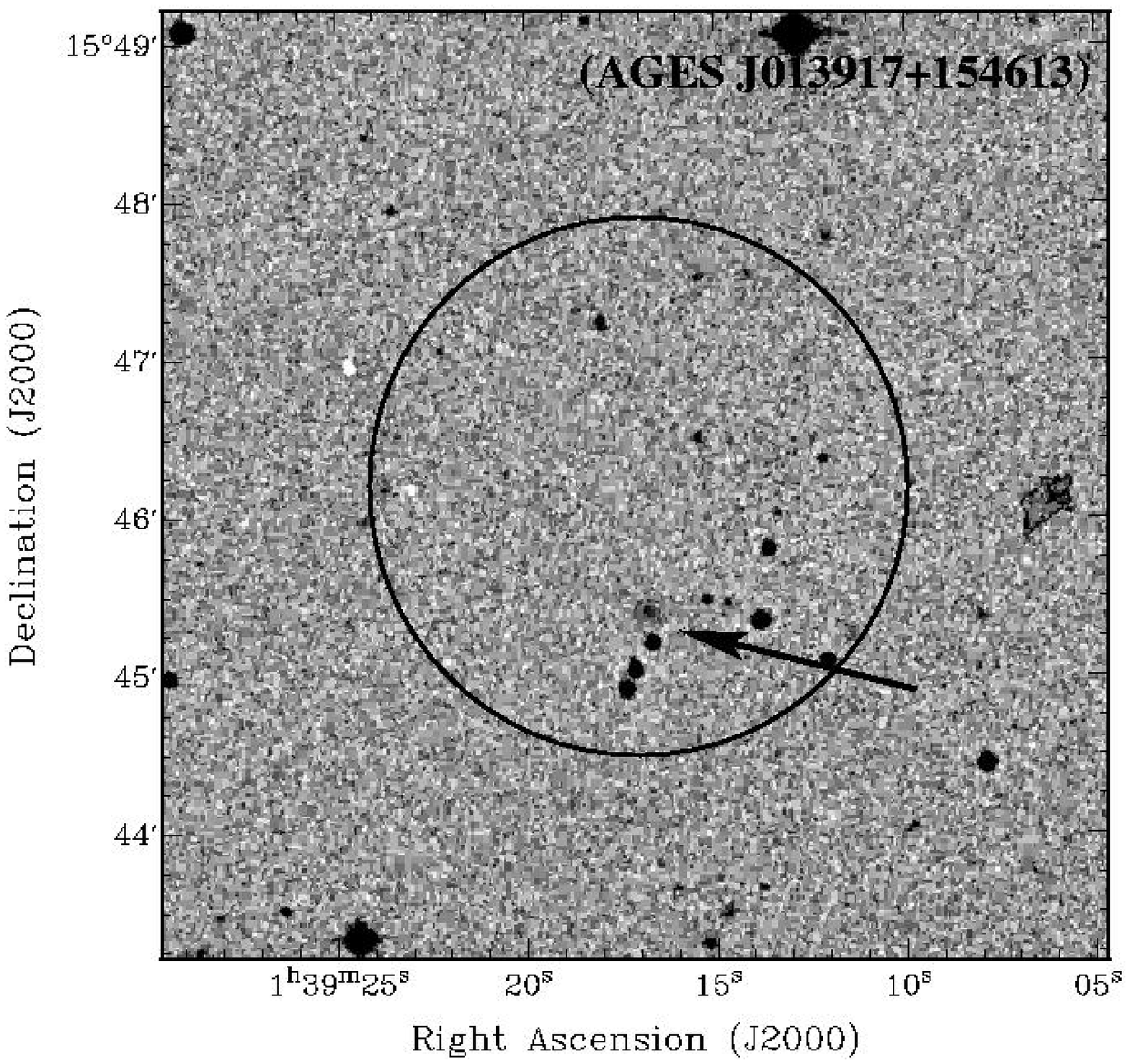}}
  \subfigure{
    \includegraphics[width=0.45\textwidth]{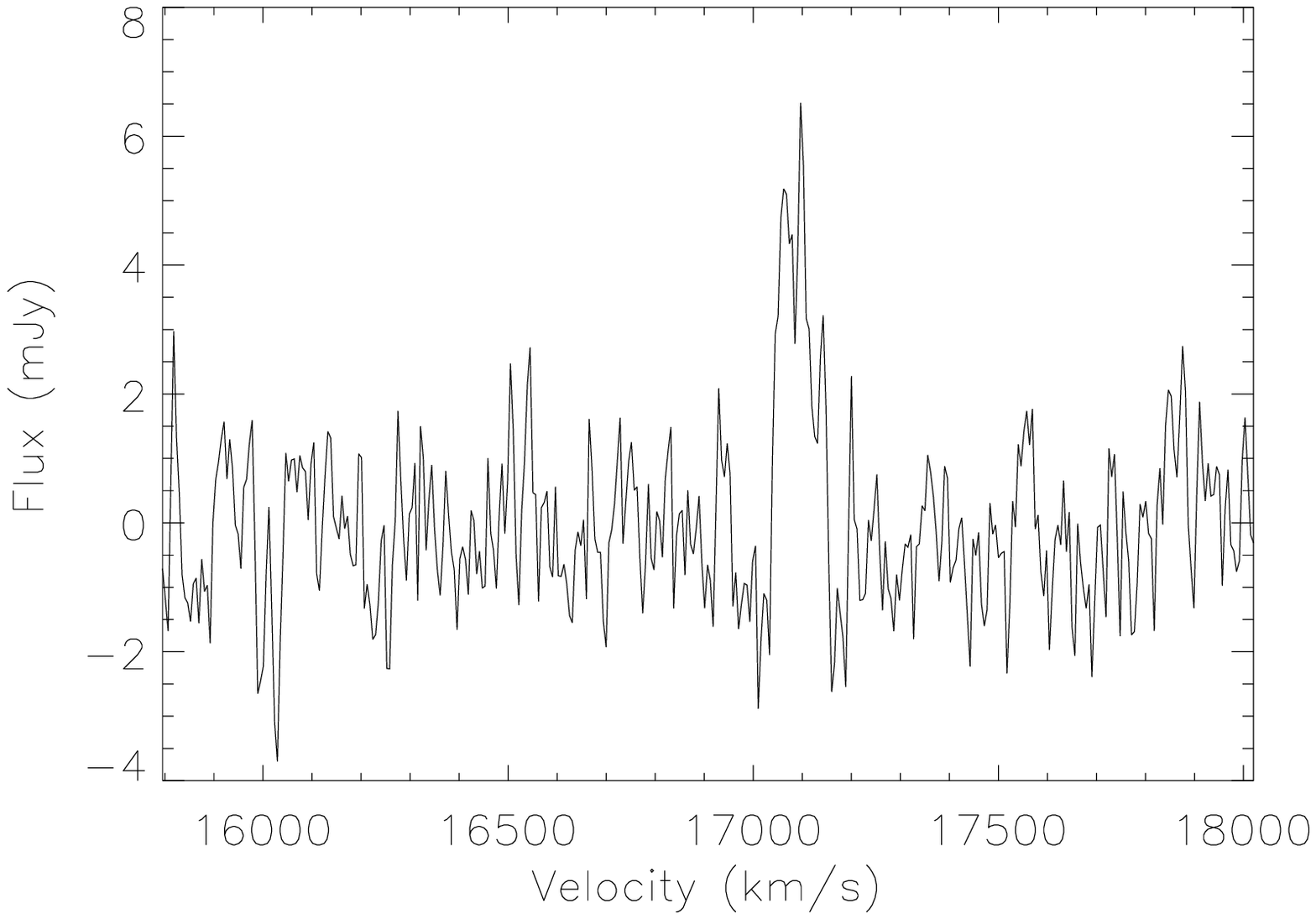}}
  \subfigure{
    \includegraphics[width=0.32\textwidth]{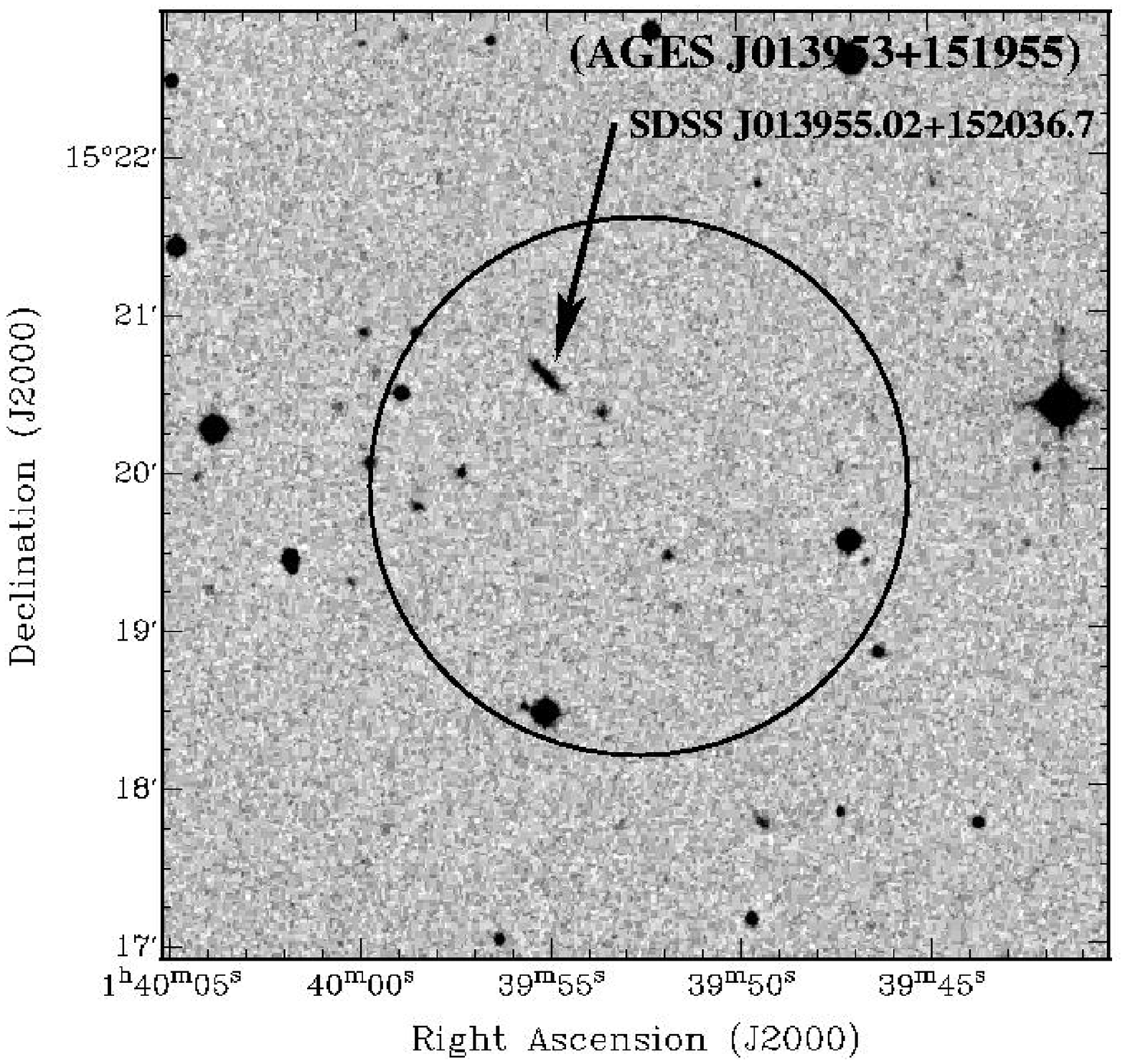}}
  \subfigure{
    \includegraphics[width=0.45\textwidth]{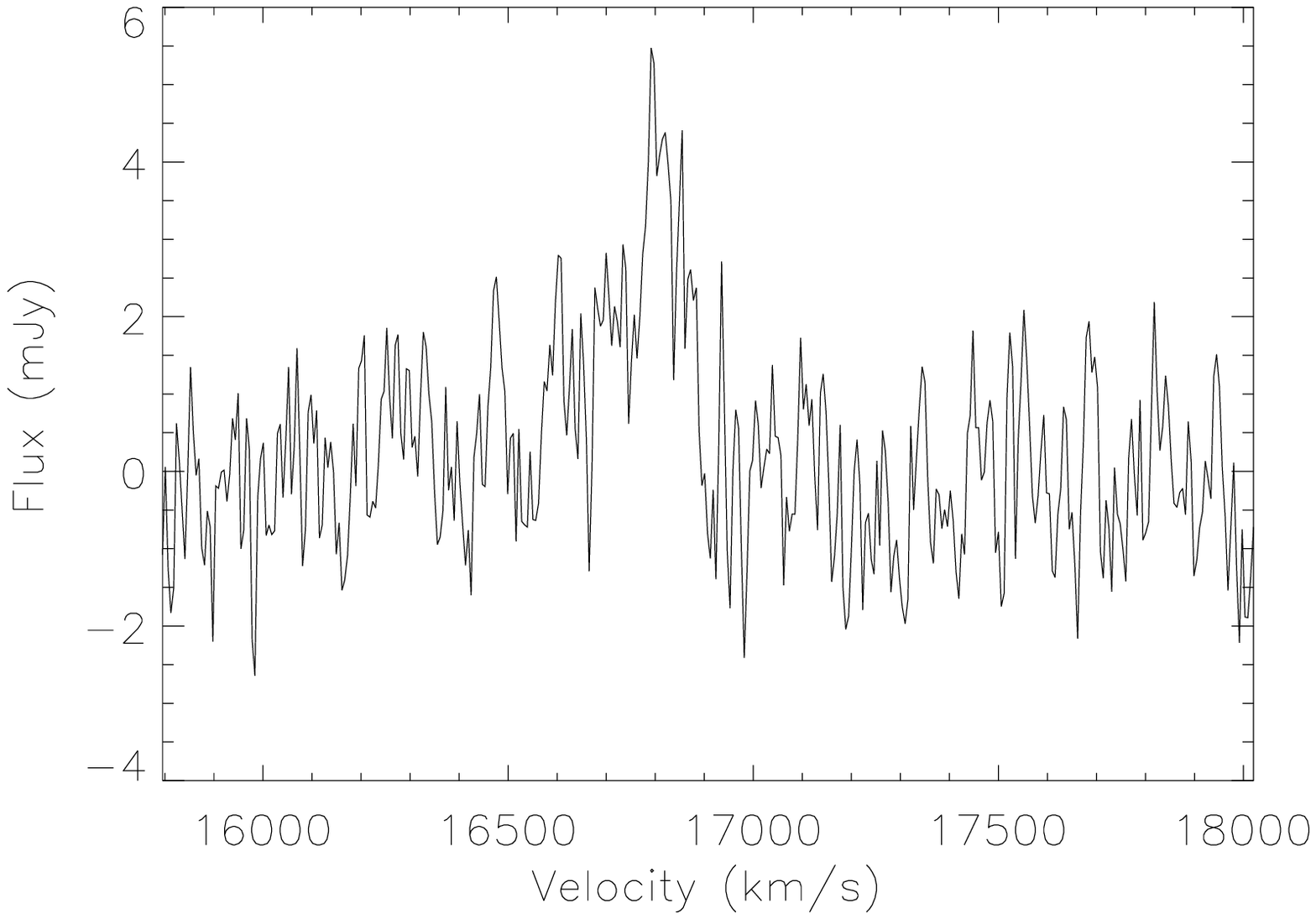}}
  \subfigure{
    \includegraphics[width=0.32\textwidth]{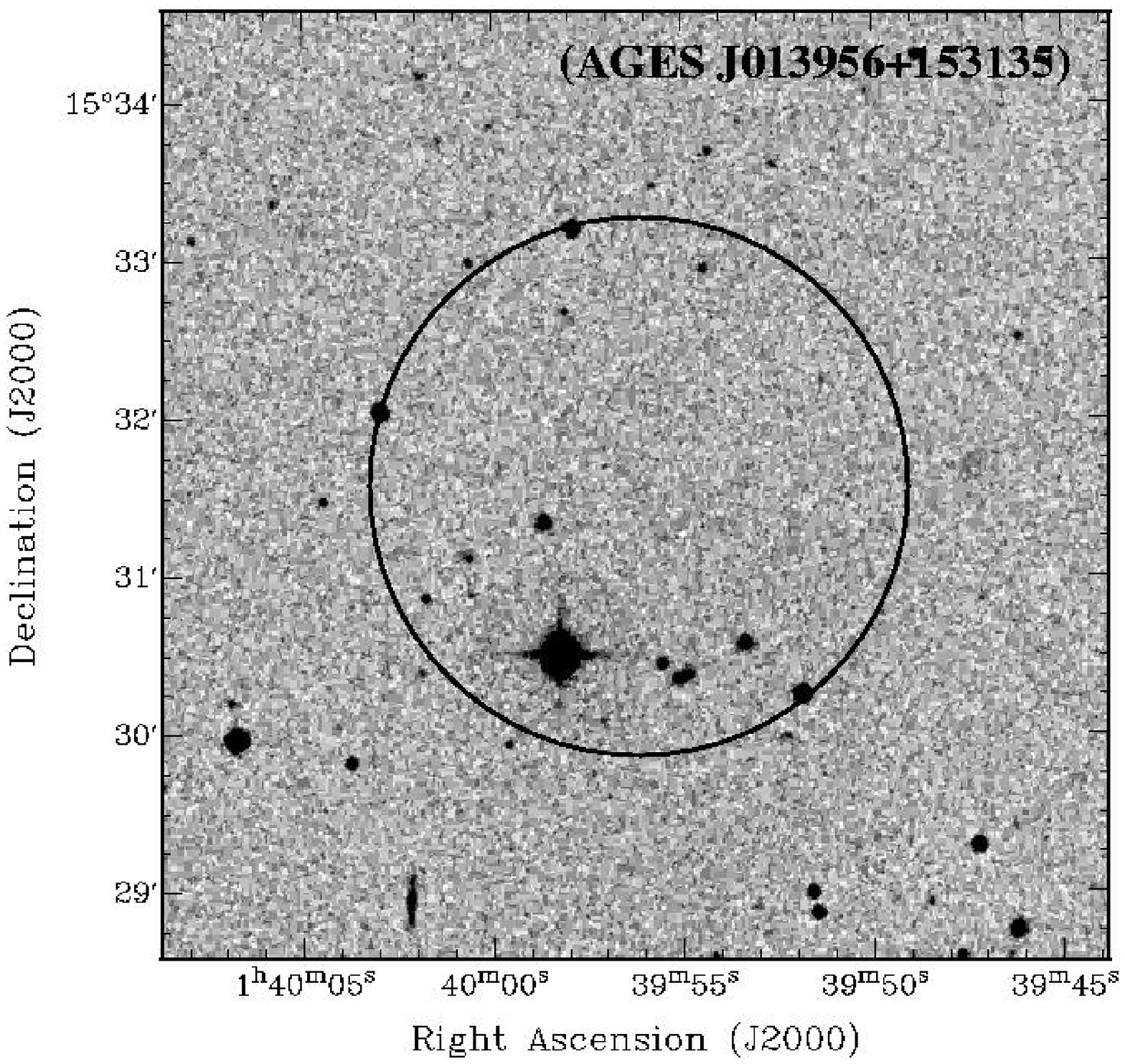}}
  \subfigure{
    \includegraphics[width=0.45\textwidth]{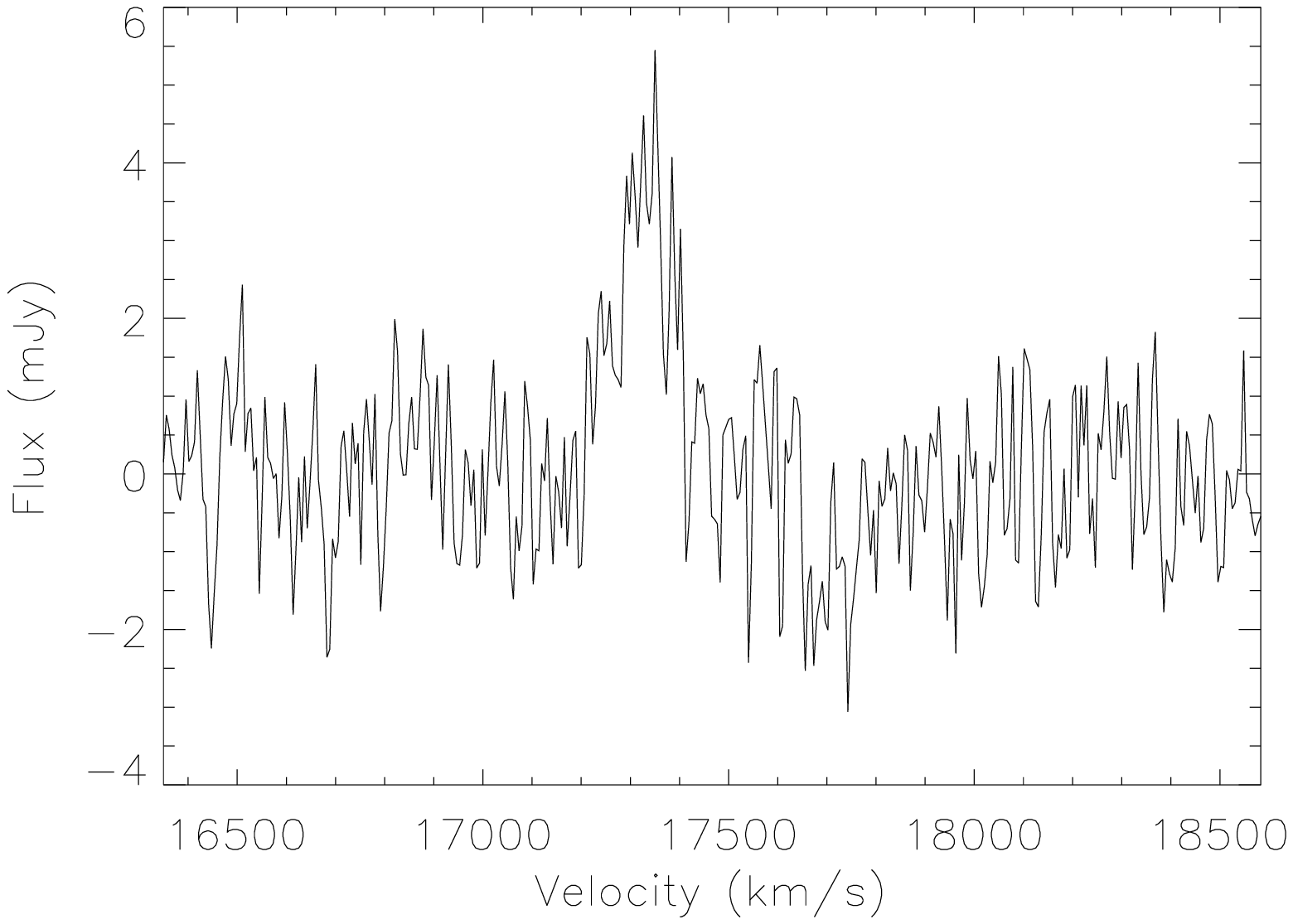}}
  \subfigure{
    \includegraphics[width=0.32\textwidth]{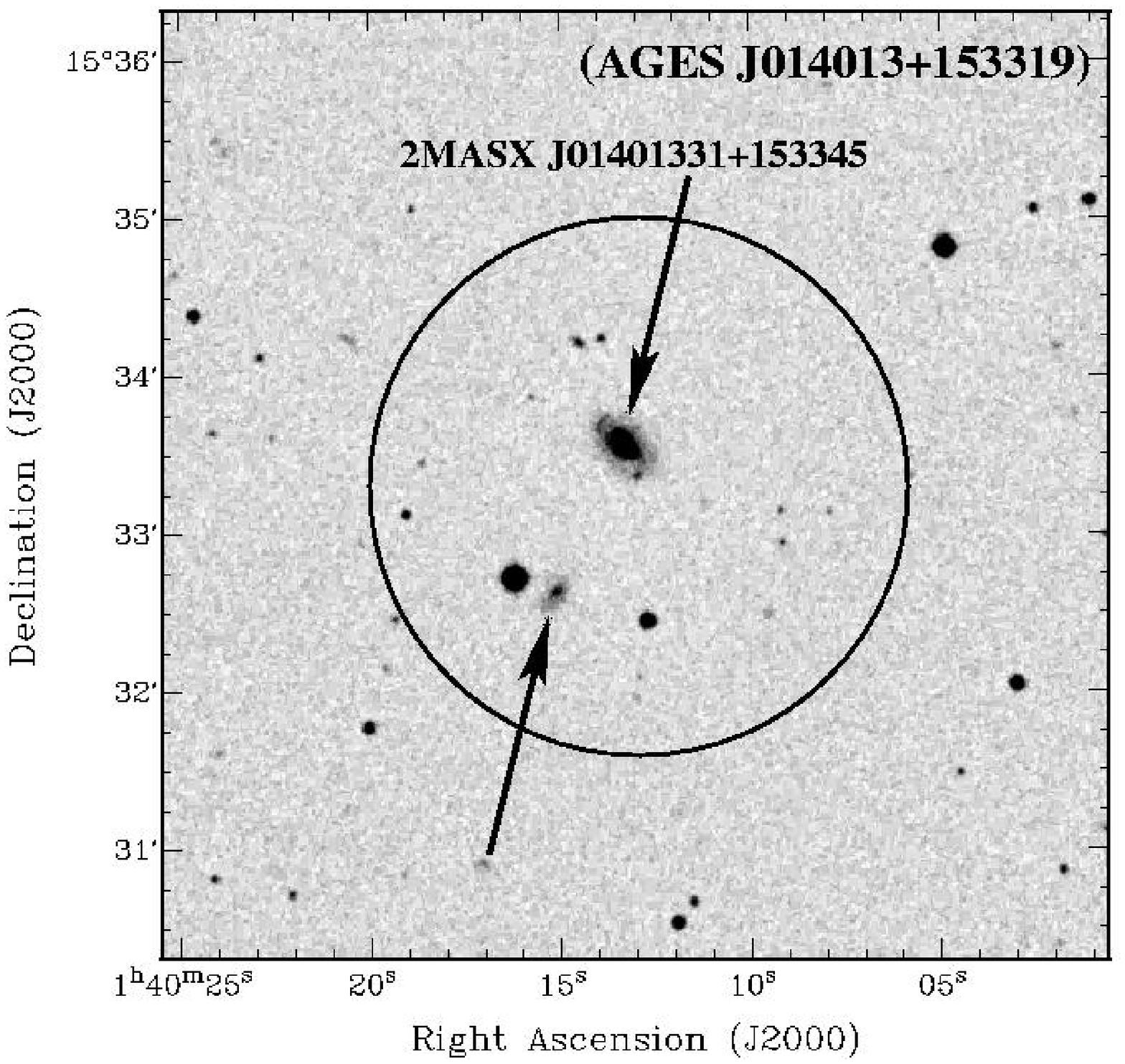}}
  \subfigure{
    \includegraphics[width=0.45\textwidth]{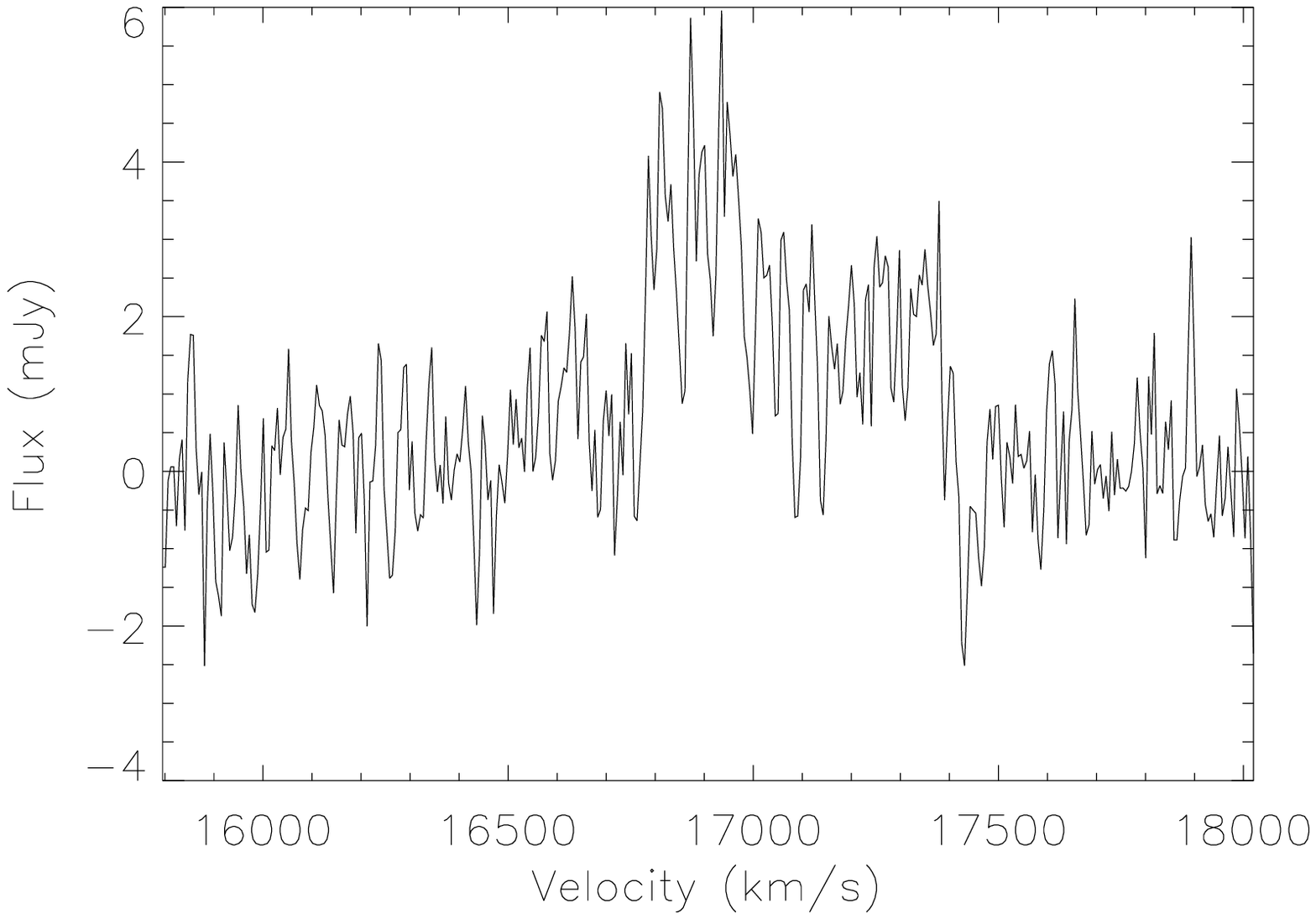}}
  
  \contcaption{Top to bottom: {\it B} band images and accompanying \hi\/
  spectra for objects J013917+154613, J013953+151955, J013956+153135 and J014013+153319\label{fig12}}
\end{figure*}
\begin{figure*}
\centering
  \subfigure{
    \includegraphics[width=0.32\textwidth]{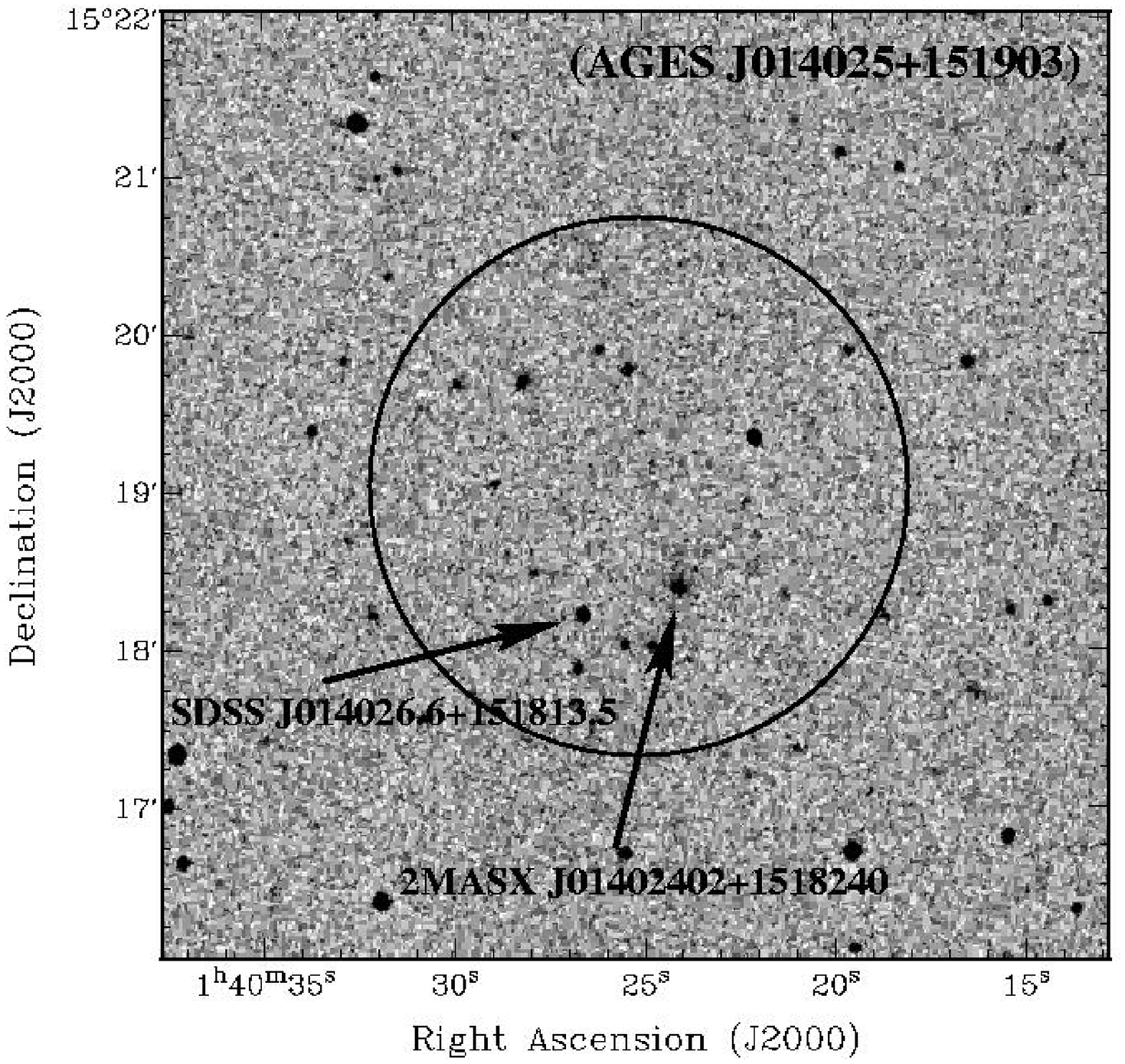}}
  \subfigure{
    \includegraphics[width=0.45\textwidth]{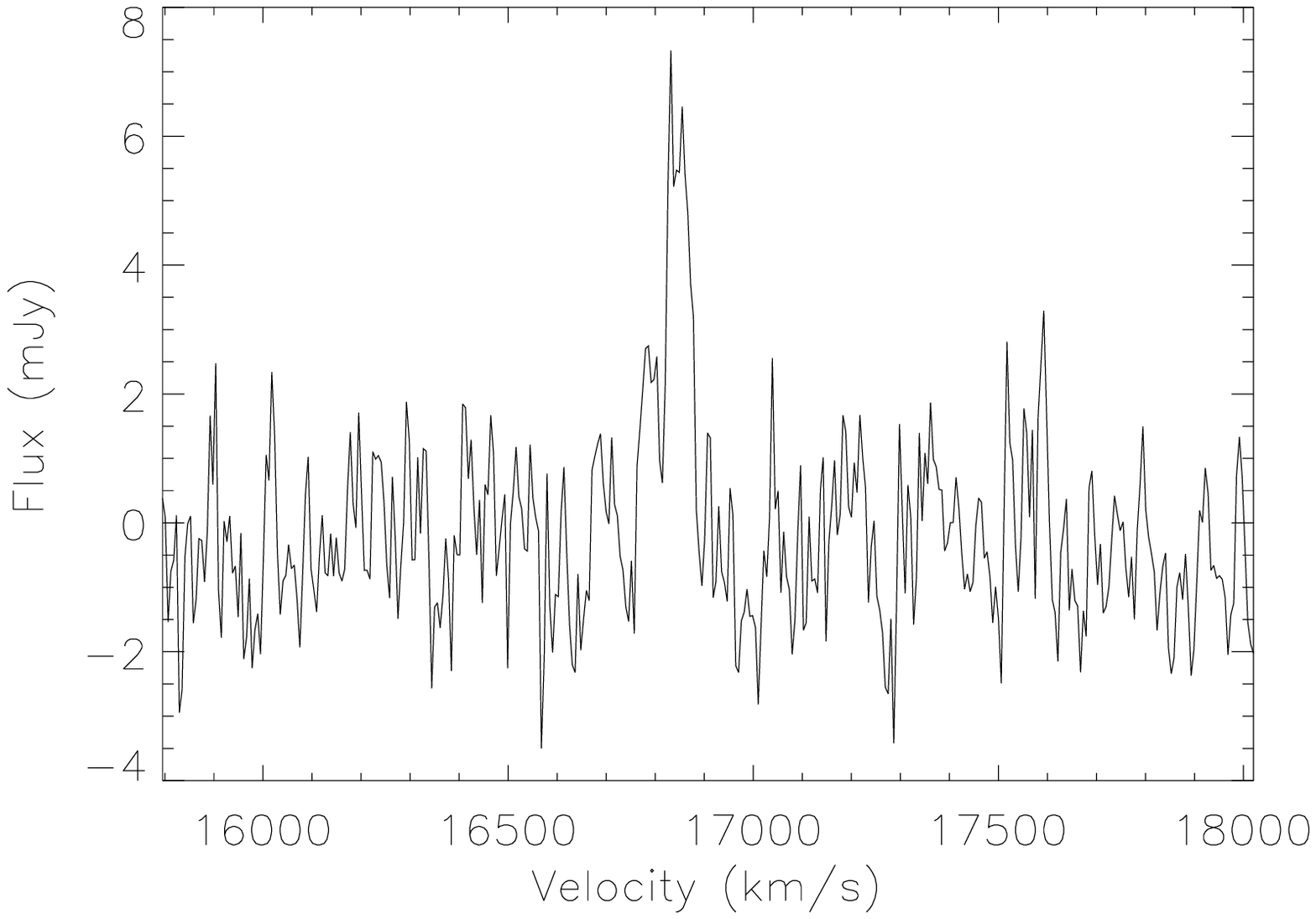}}
  \subfigure{
    \includegraphics[width=0.32\textwidth]{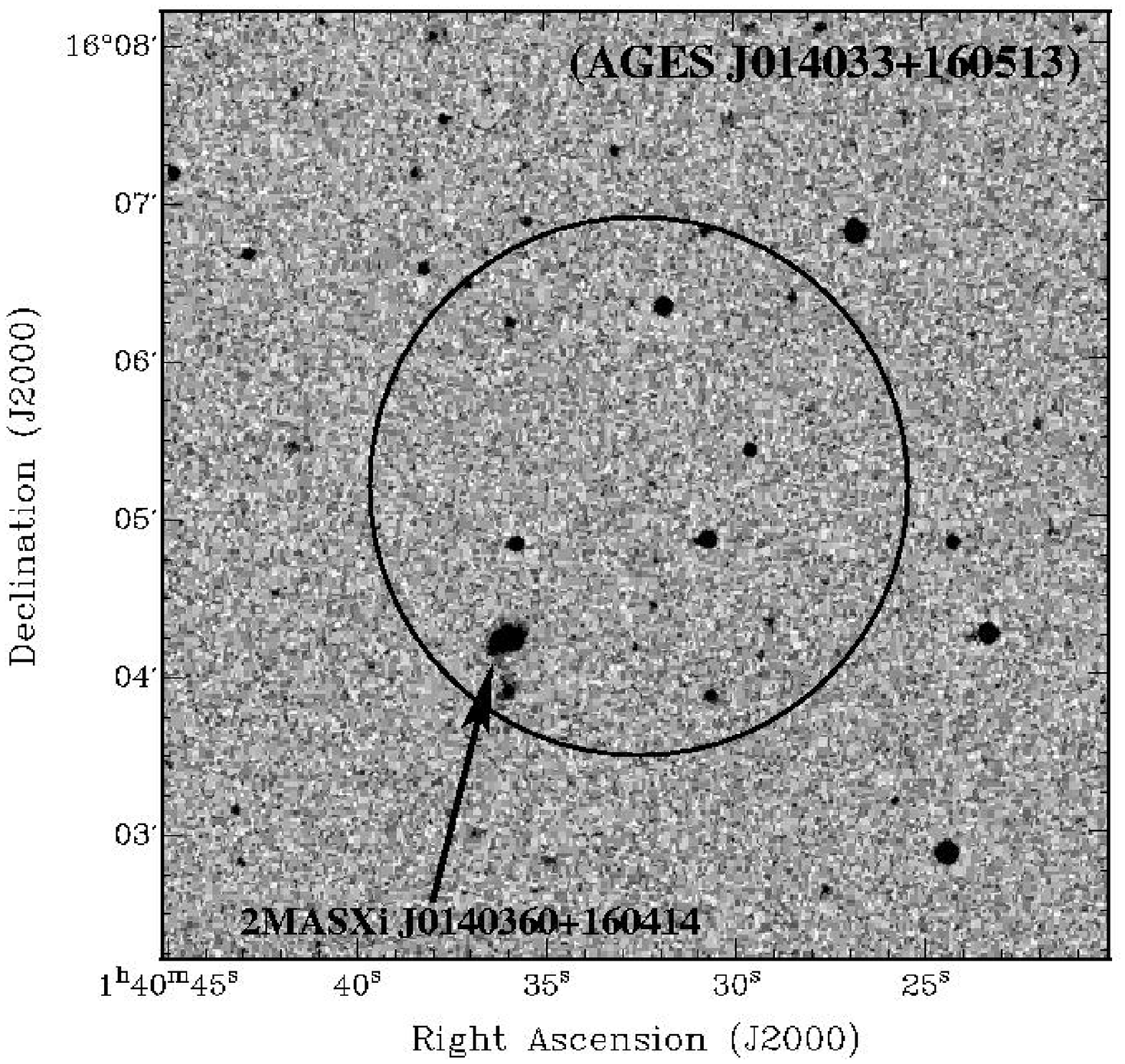}}
  \subfigure{
    \includegraphics[width=0.45\textwidth]{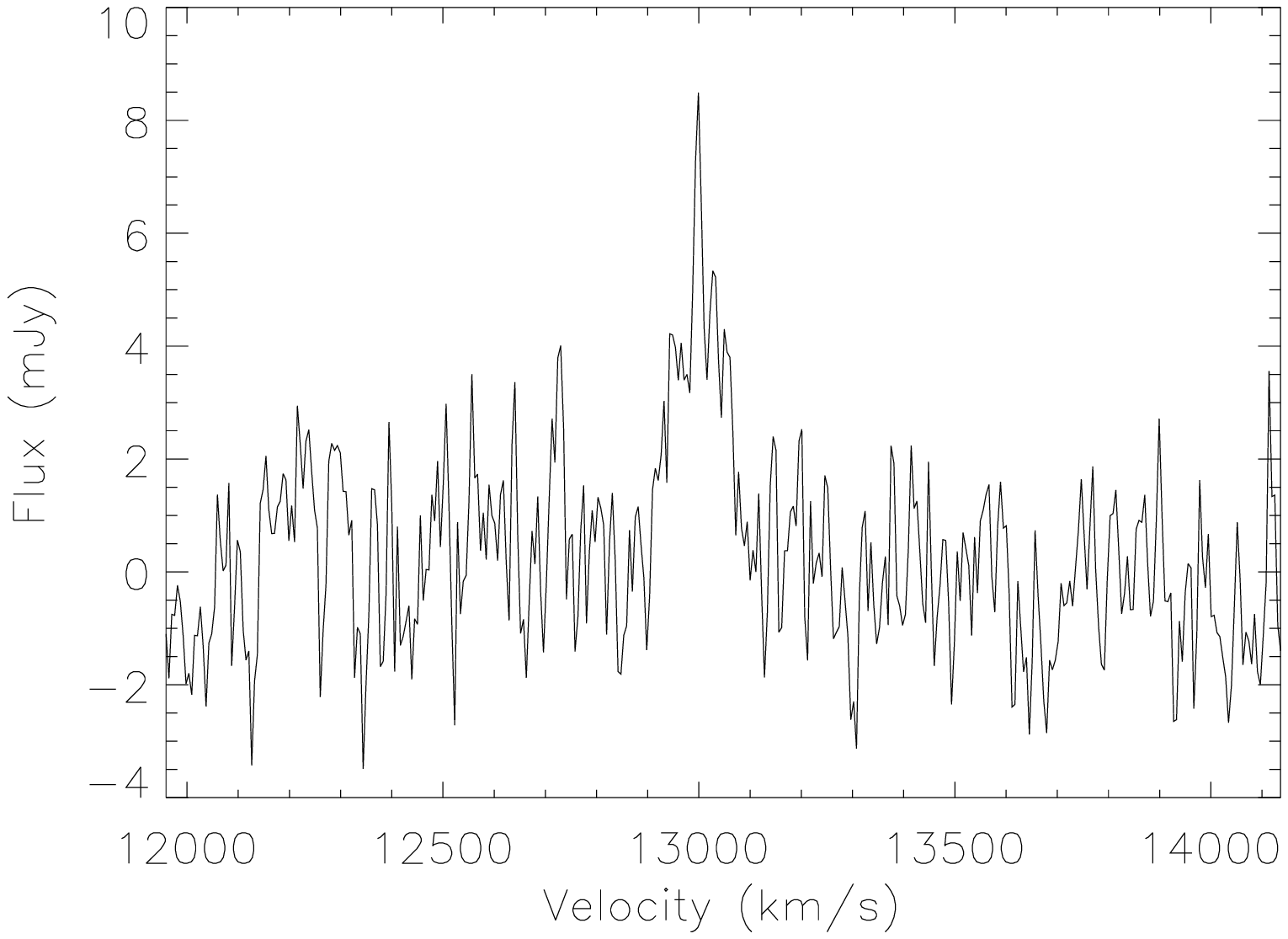}}
  \subfigure{
    \includegraphics[width=0.32\textwidth]{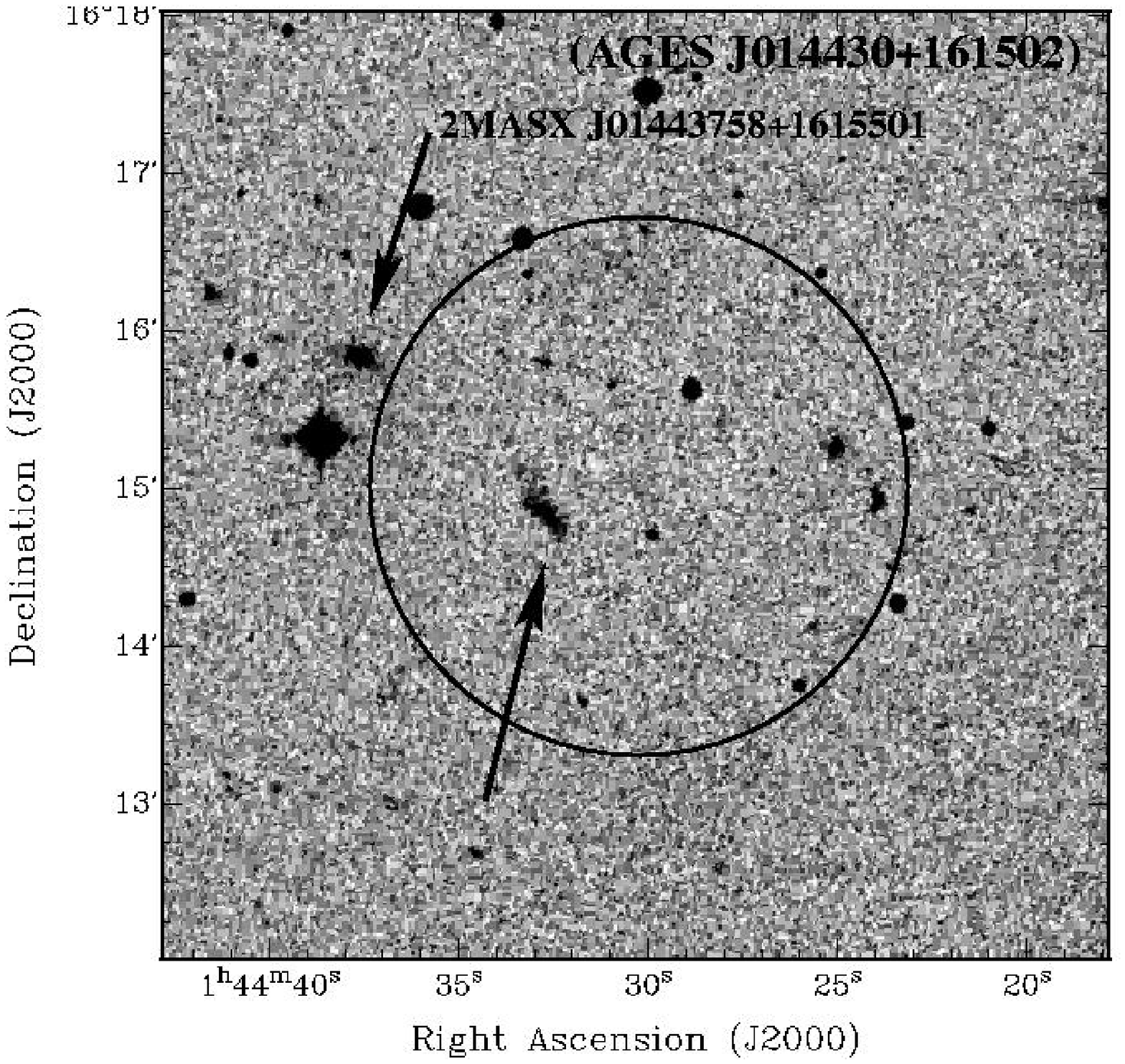}}
  \subfigure{
    \includegraphics[width=0.45\textwidth]{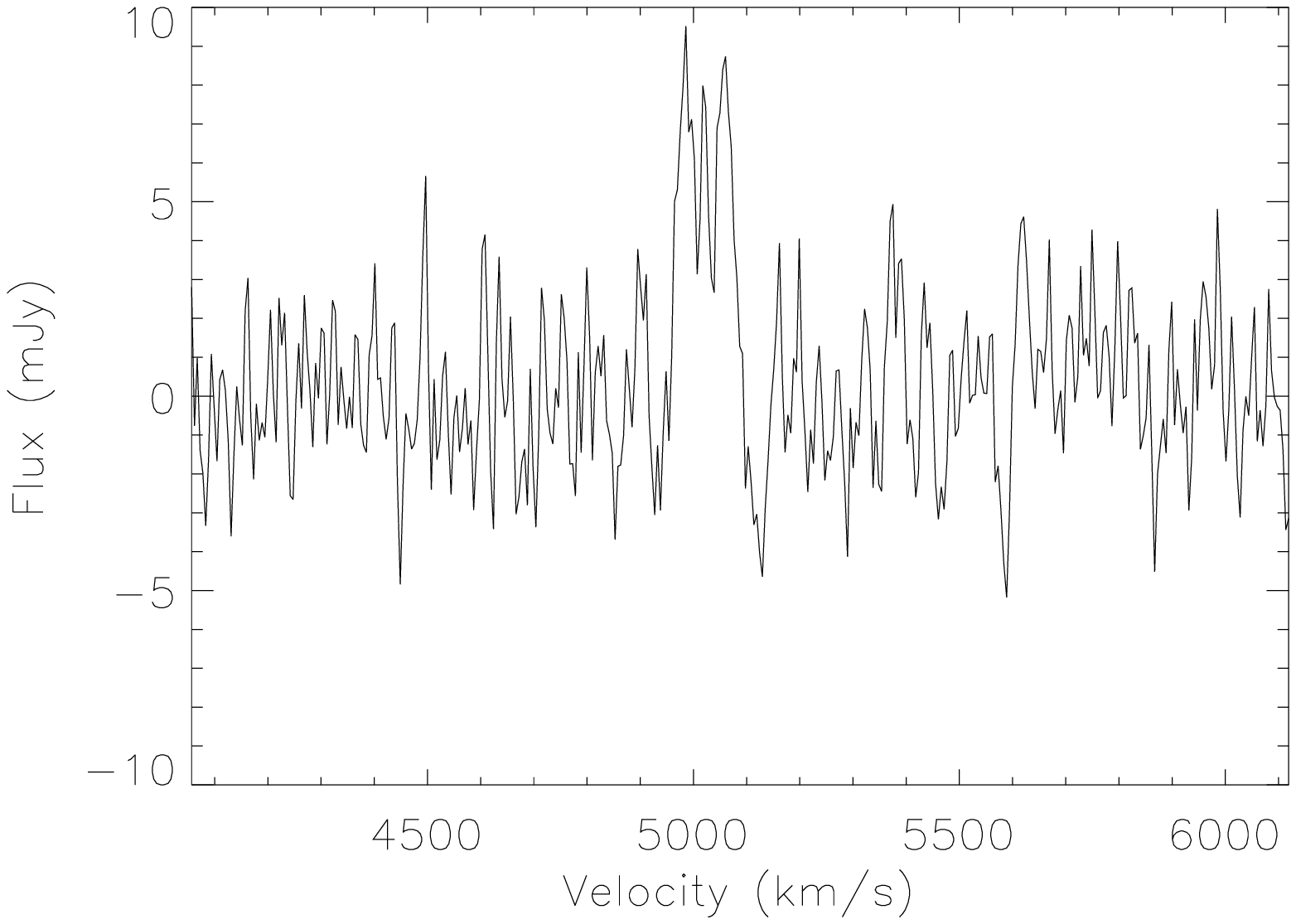}}
  \subfigure{
    \includegraphics[width=0.32\textwidth]{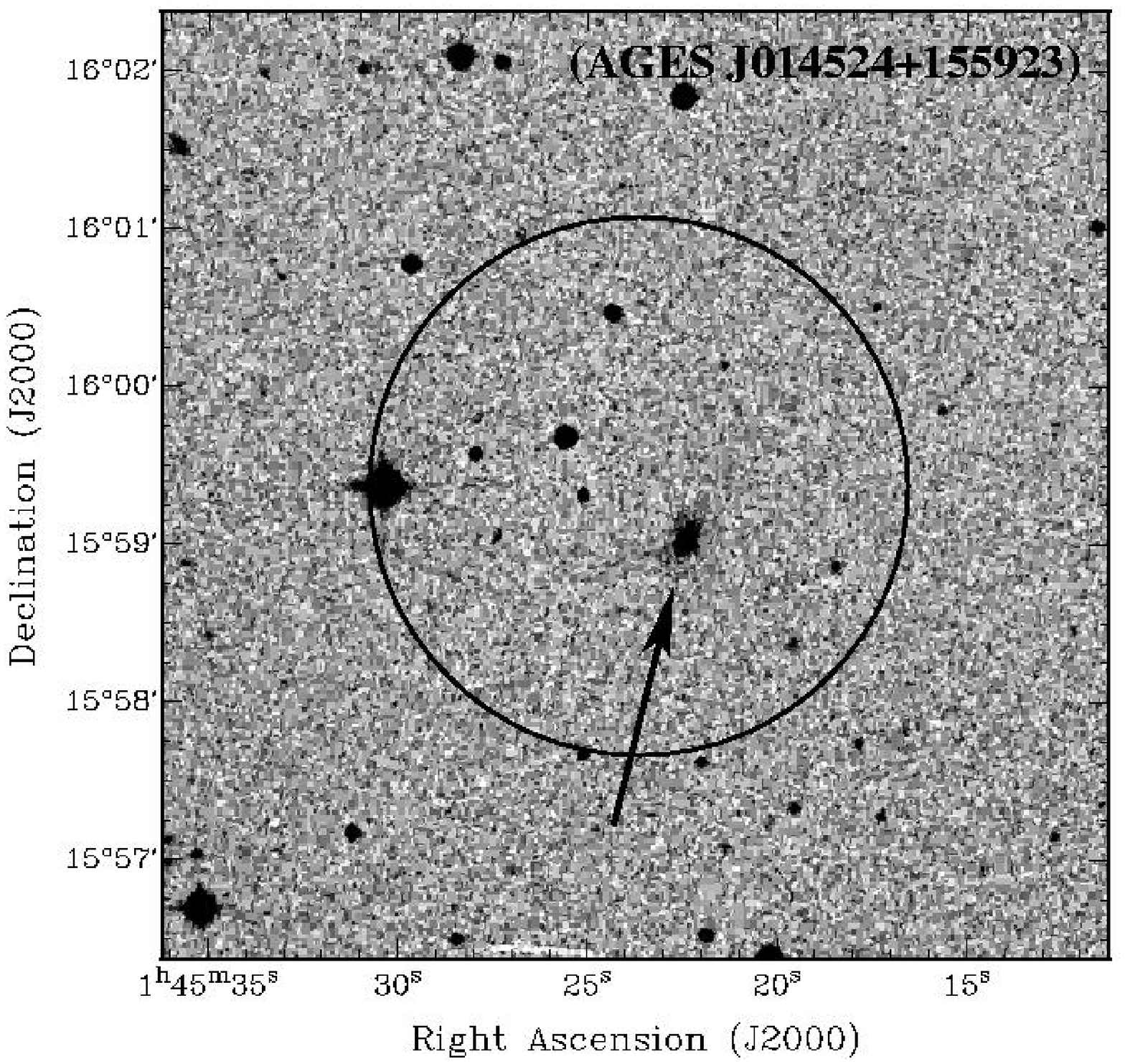}}
  \subfigure{
    \includegraphics[width=0.45\textwidth]{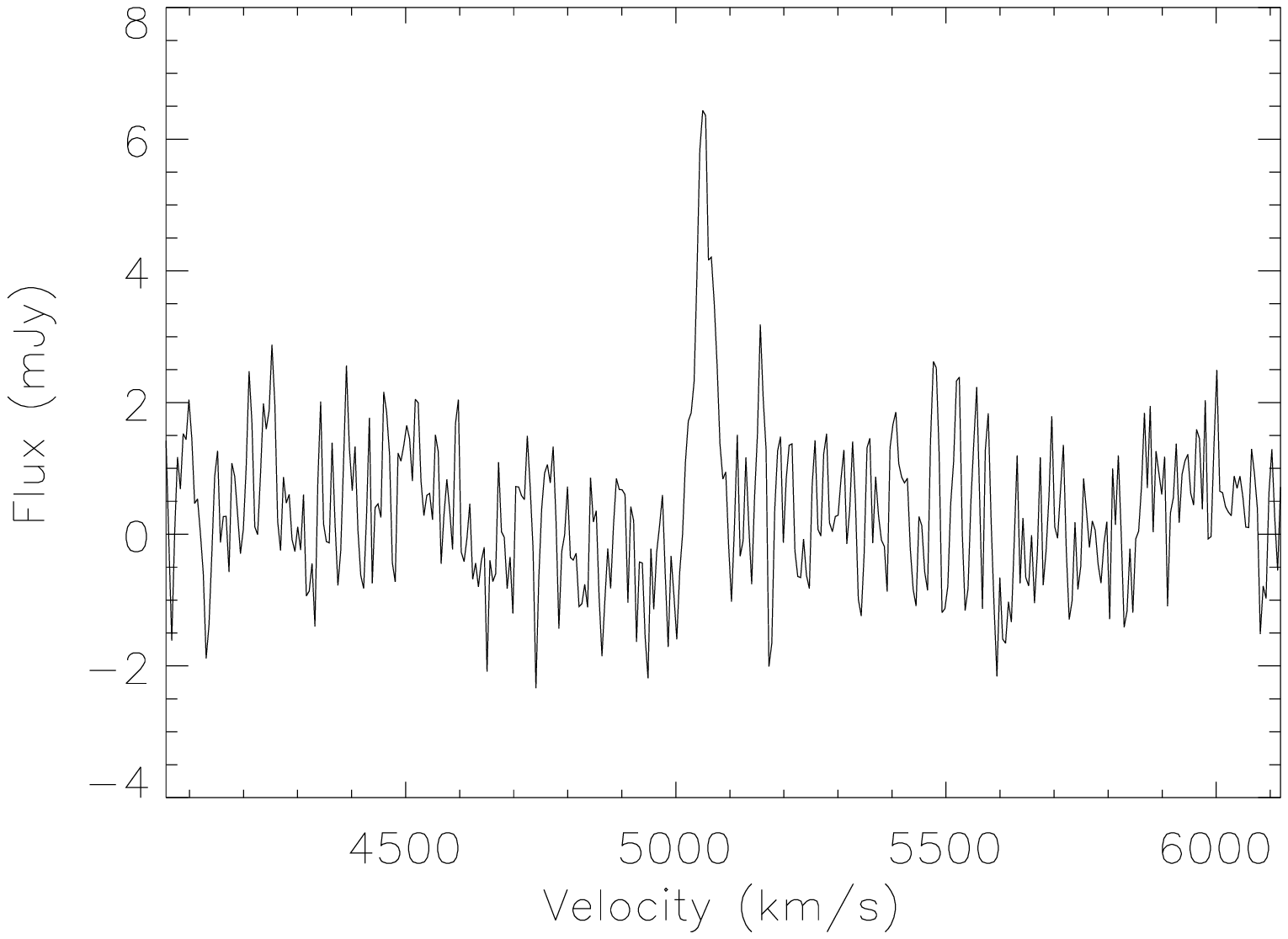}}
  
  \contcaption{Top to bottom: {\it B} band images and accompanying \hi\/
  spectra for objects J014025+151903, J014033+160513, J014430+161502 and J014524+155923\label{fig13}}
\end{figure*}
\begin{figure*}
\centering
  \subfigure{
  \includegraphics[width=0.32\textwidth]{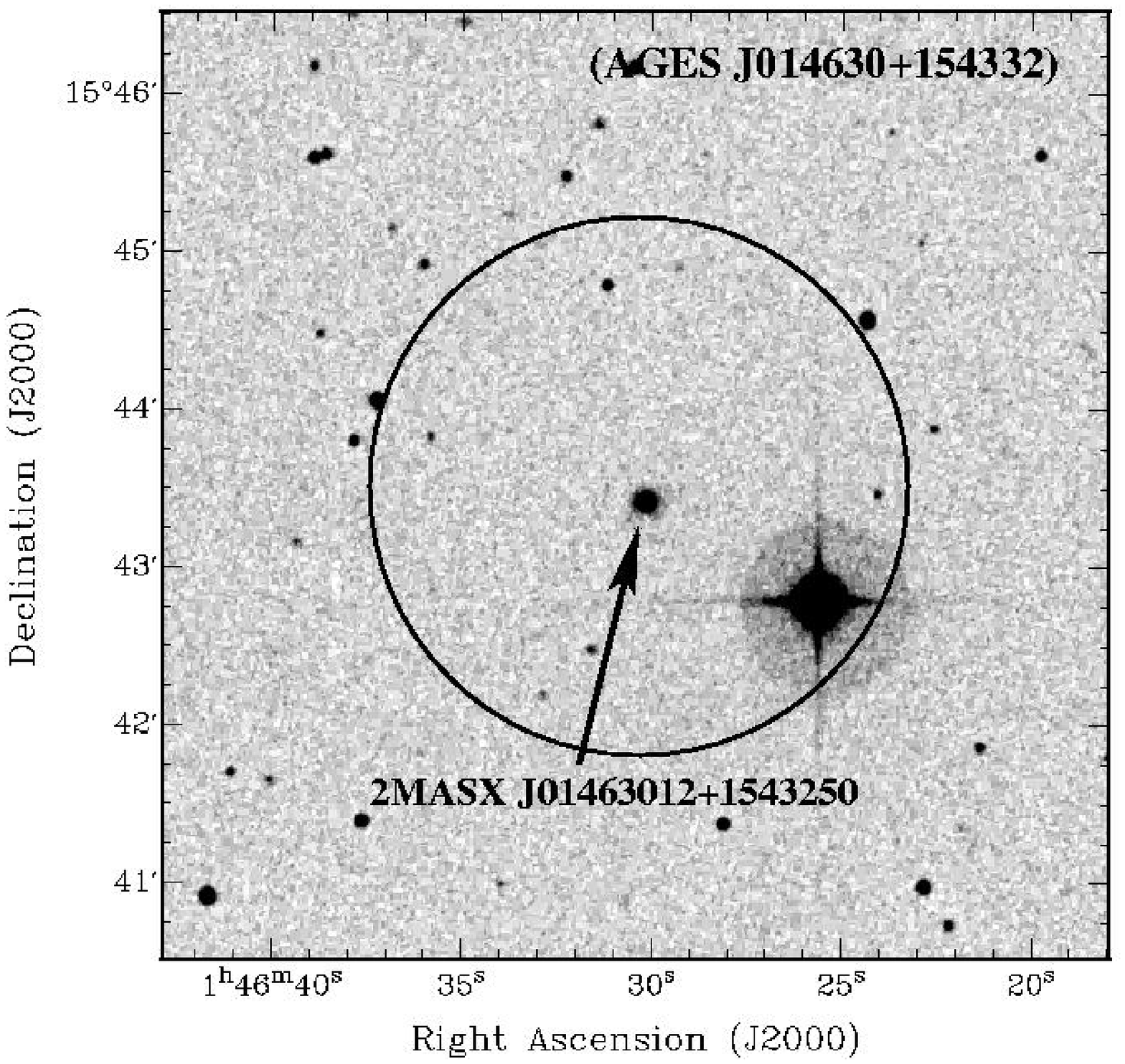}}
\subfigure{
  \includegraphics[width=0.45\textwidth]{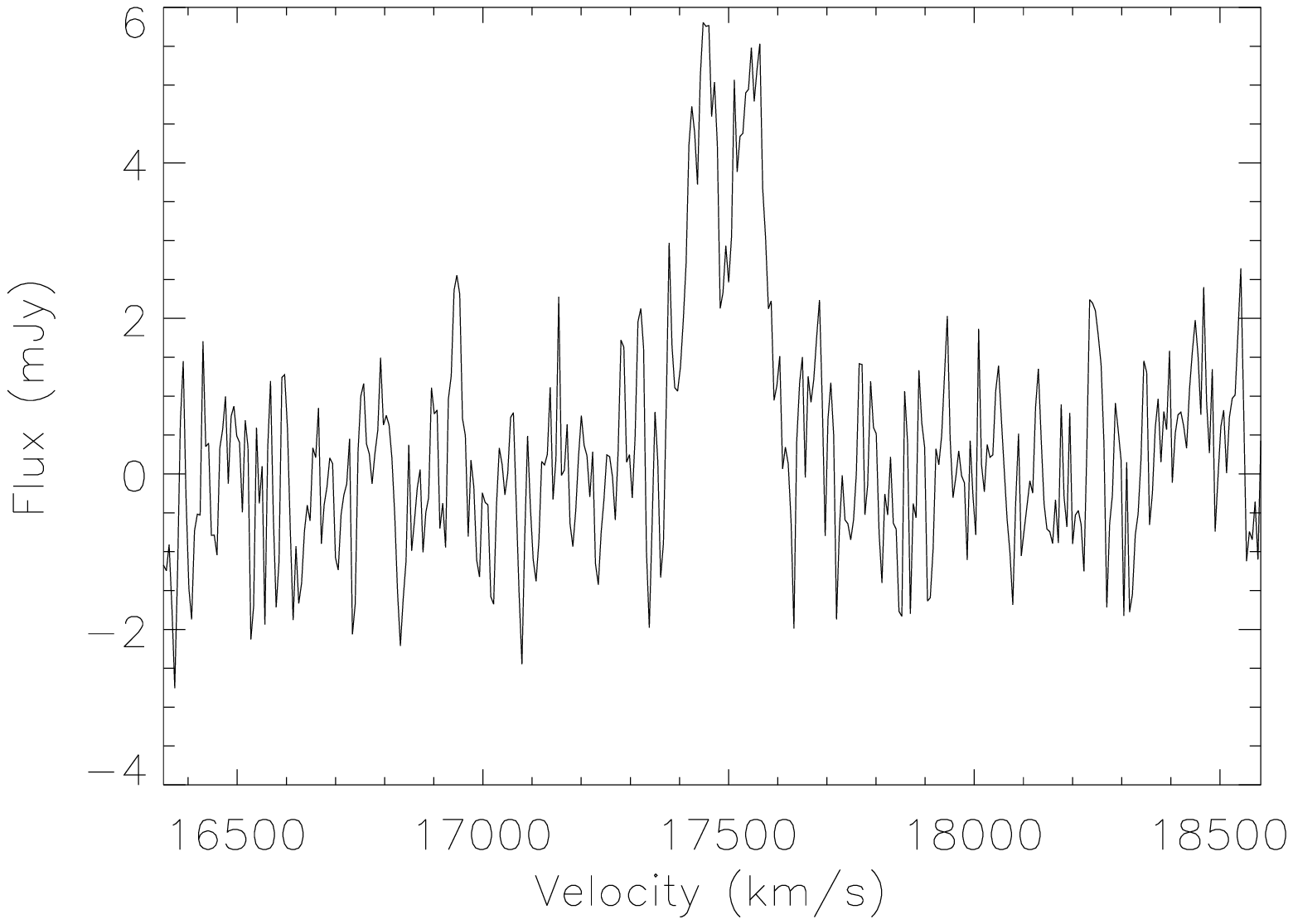}}
 \subfigure{
    \includegraphics[width=0.32\textwidth]{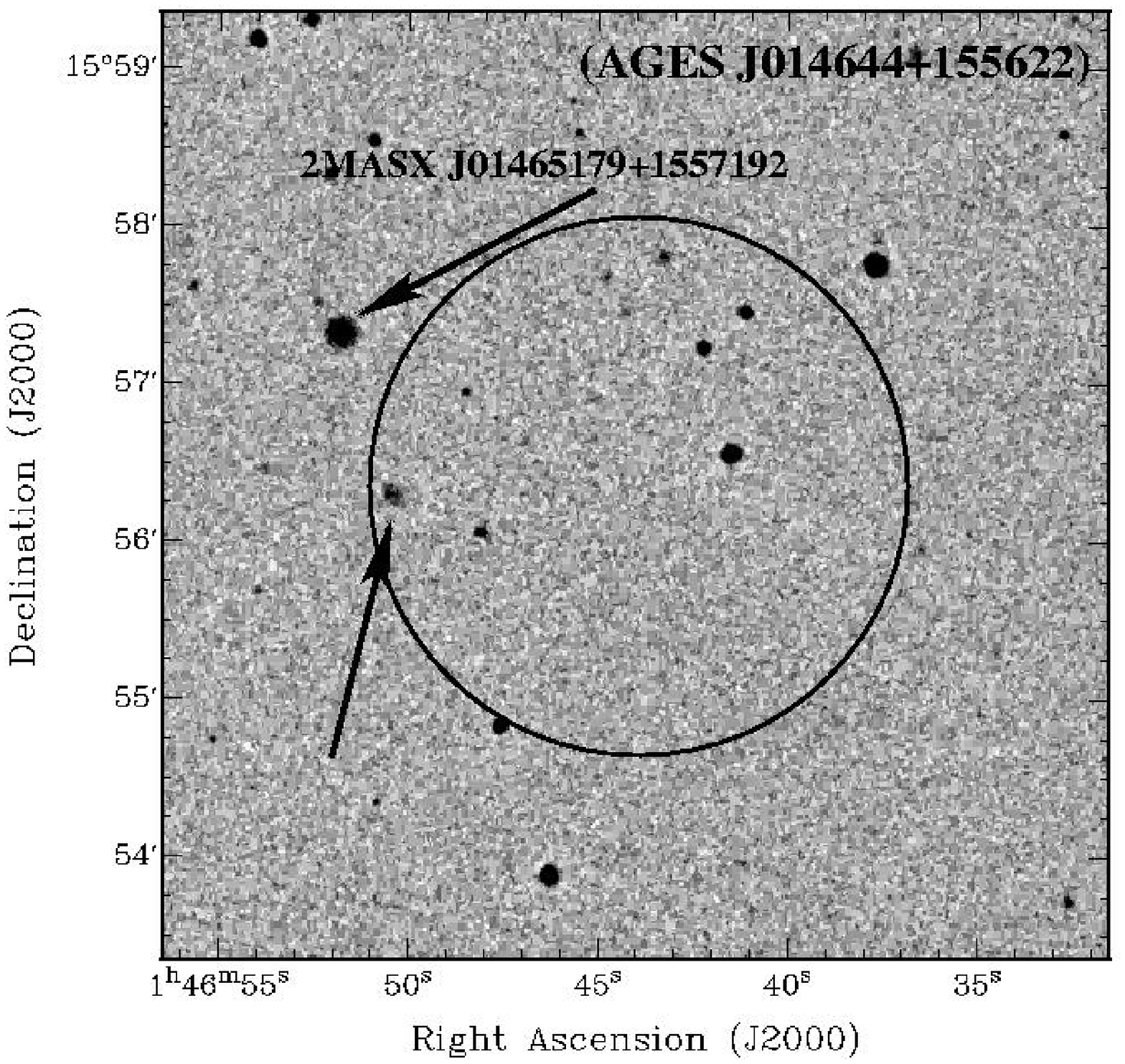}}
  \subfigure{
    \includegraphics[width=0.45\textwidth]{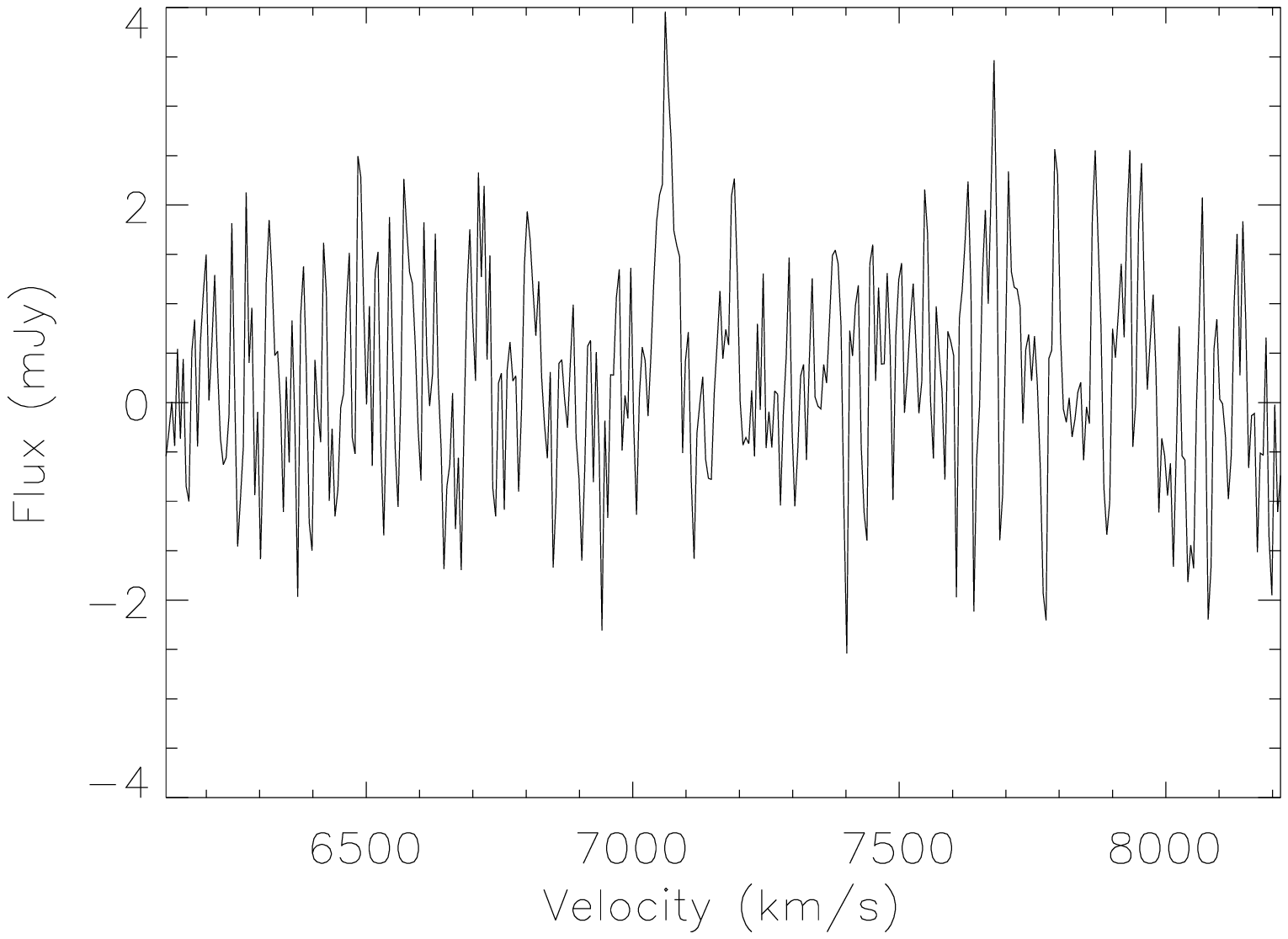}}
  \subfigure{
    \includegraphics[width=0.32\textwidth]{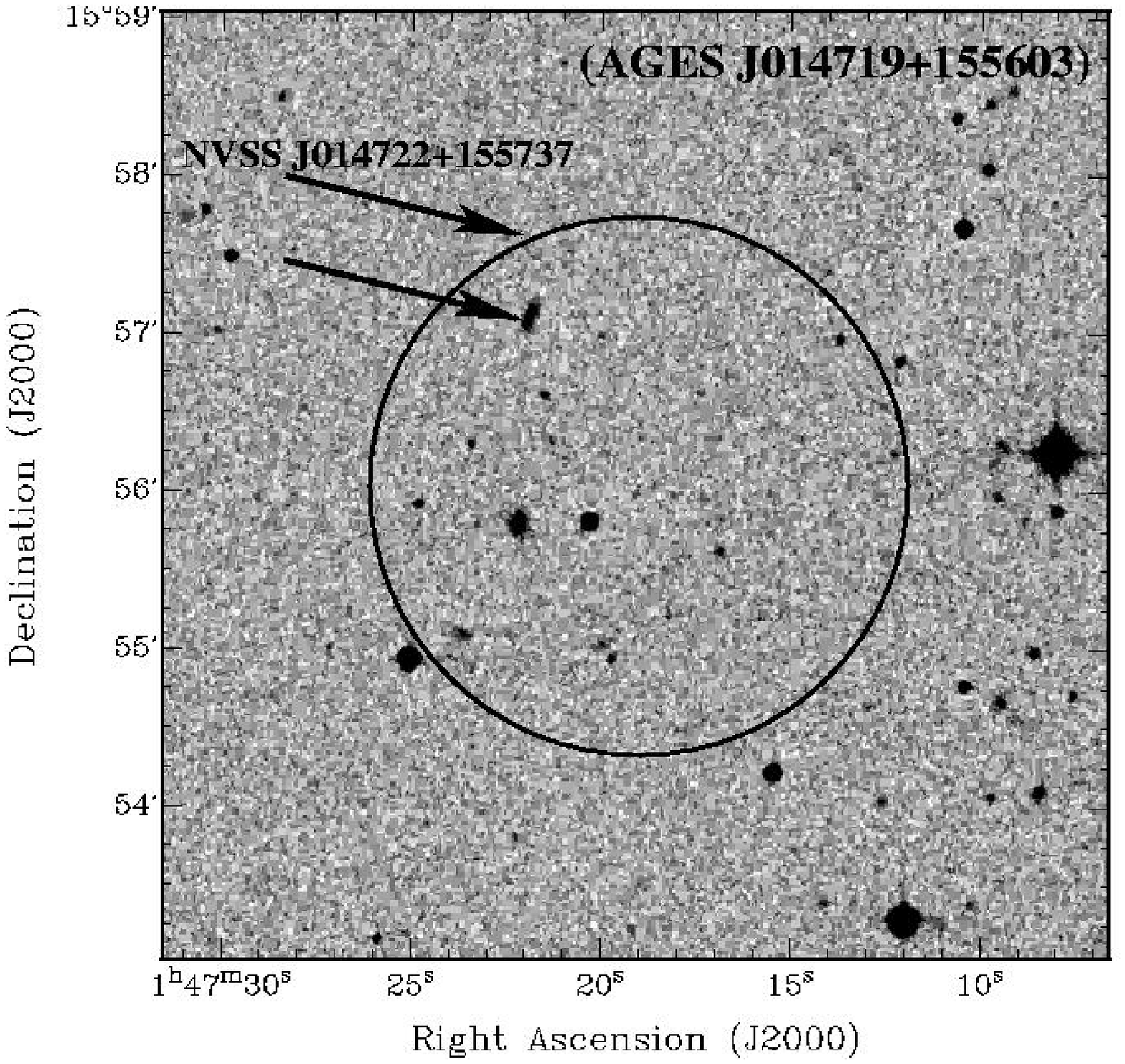}}
  \subfigure{
    \includegraphics[width=0.45\textwidth]{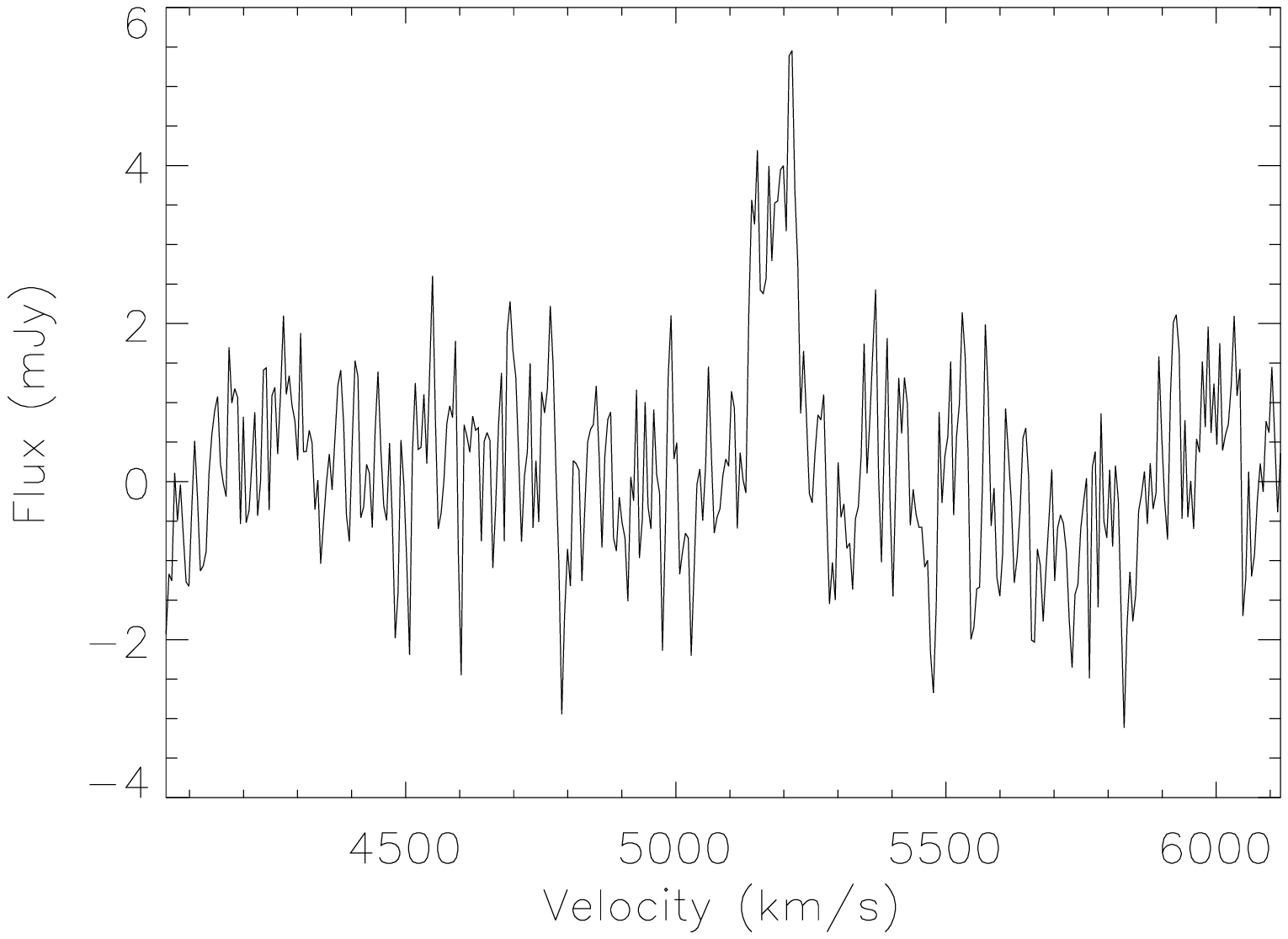}}
  \subfigure{
    \includegraphics[width=0.32\textwidth]{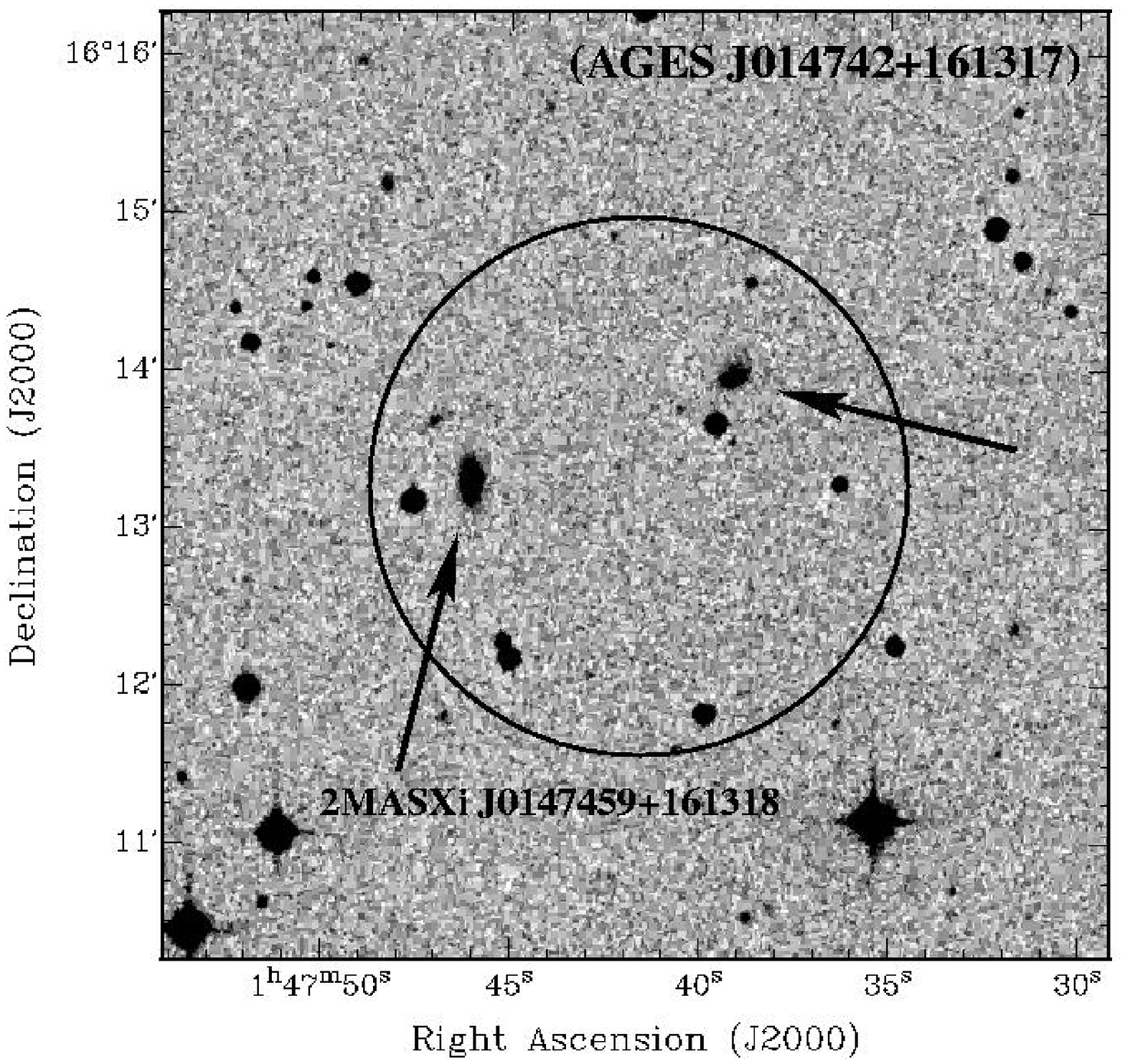}}
  \subfigure{
    \includegraphics[width=0.45\textwidth]{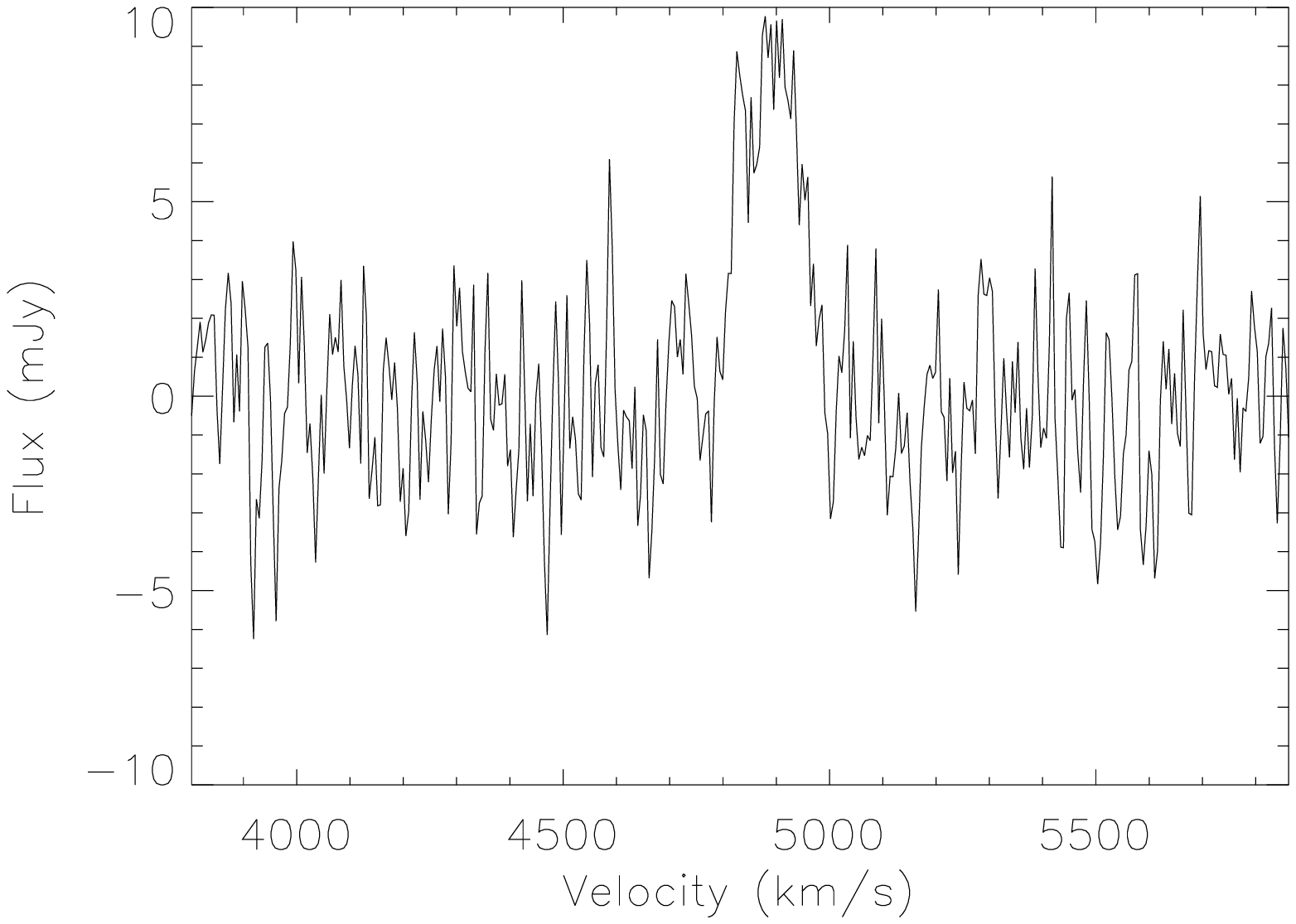}}
  
  \contcaption{Top to bottom: {\it B} band images and accompanying \hi\/
  spectra for objects J014630+154332, J014644+155622, J014719+155603 and J014742+161317\label{fig14}}
\end{figure*}
\begin{figure*}
\centering
  \subfigure{
    \includegraphics[width=0.32\textwidth]{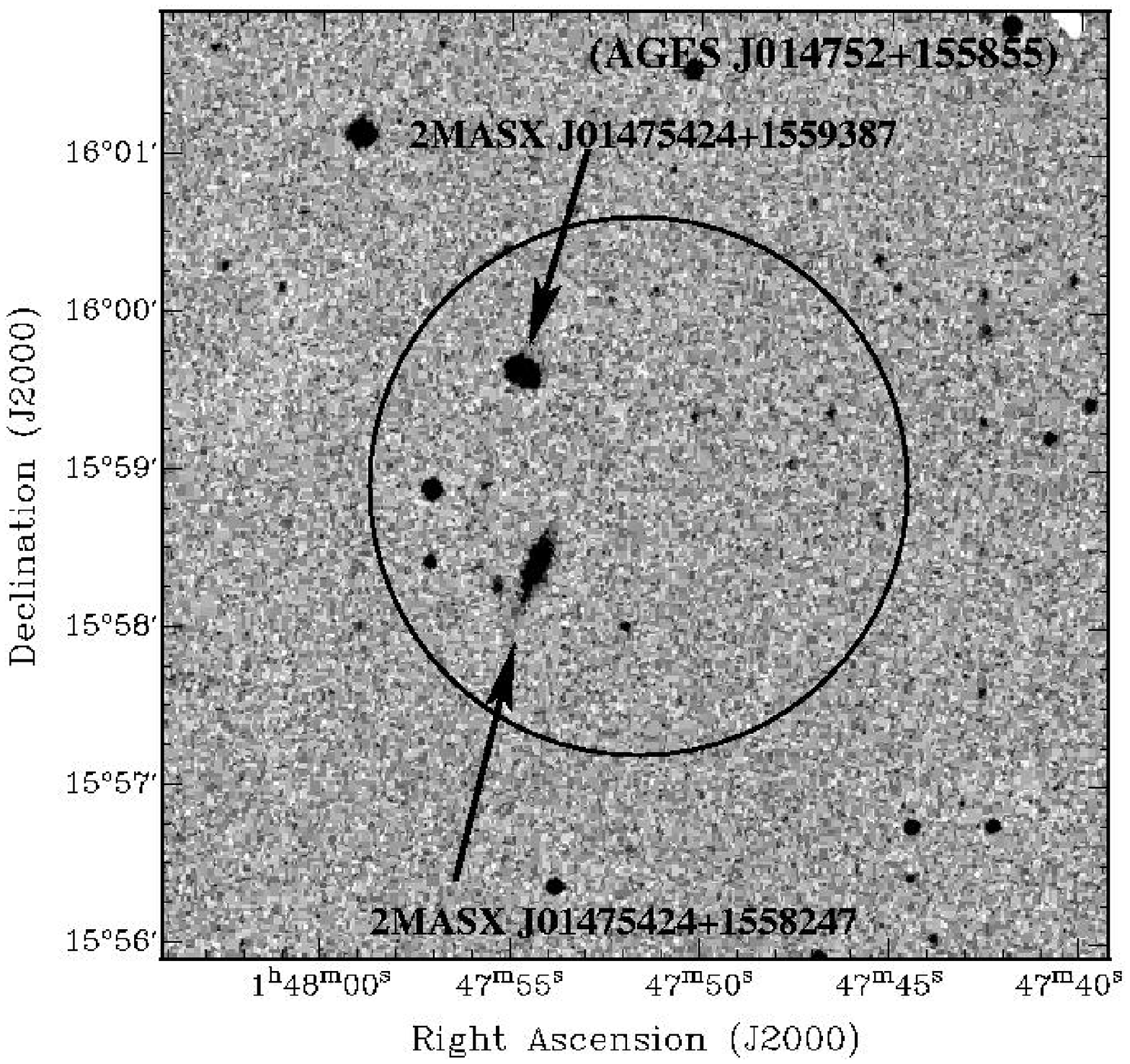}}
  \subfigure{
    \includegraphics[width=0.45\textwidth]{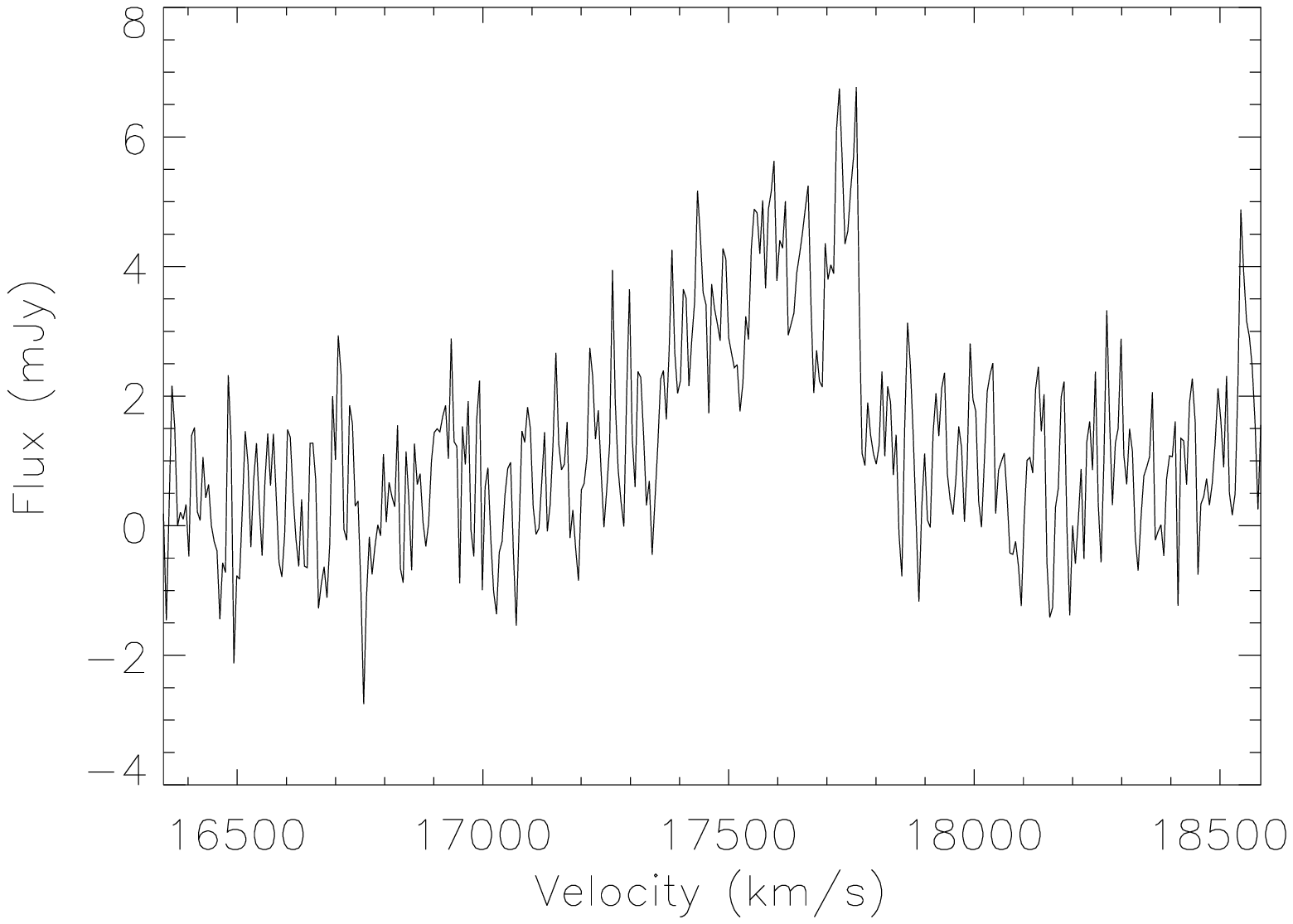}}
  \subfigure{
    \includegraphics[width=0.32\textwidth]{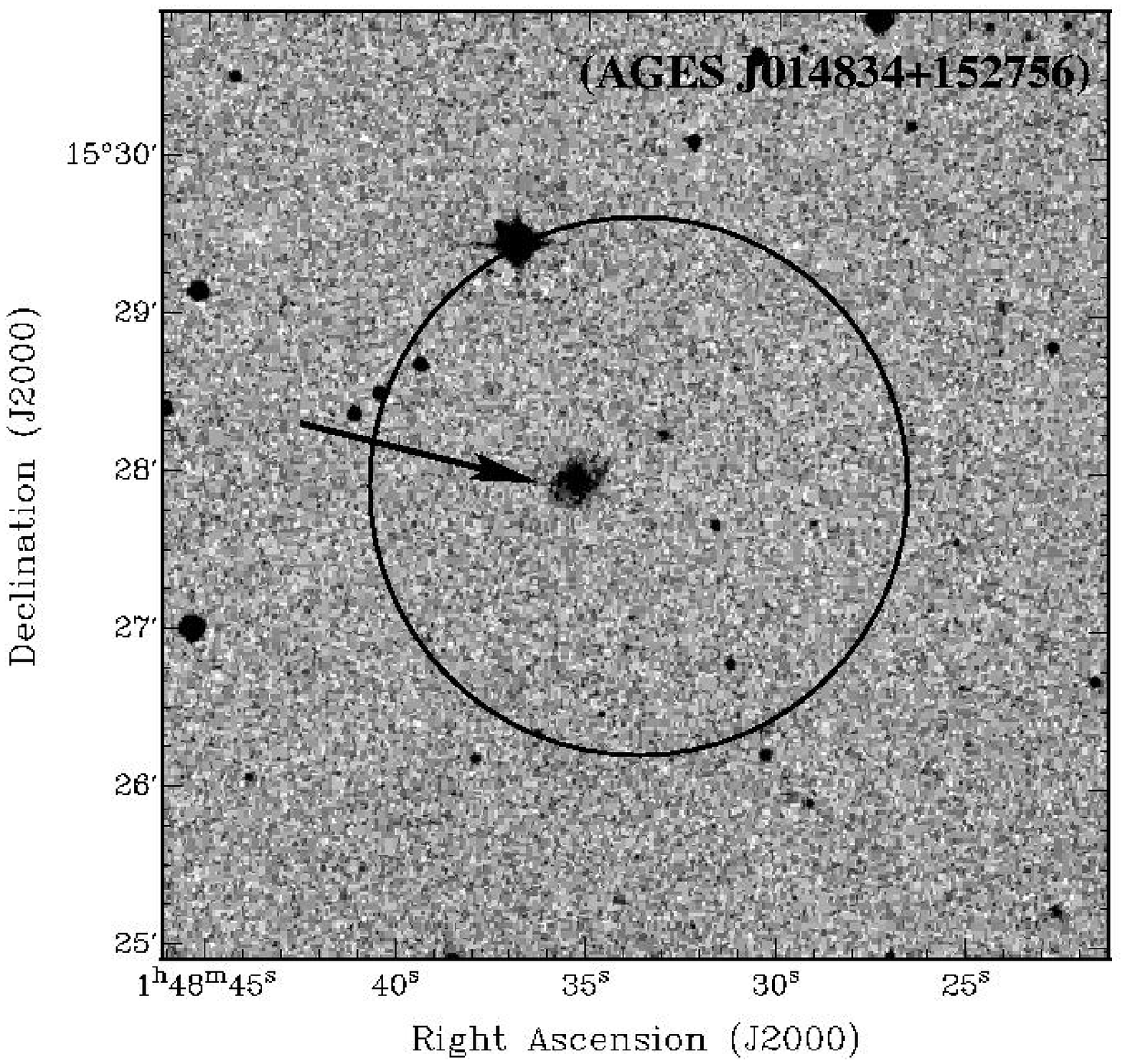}}
  \subfigure{
    \includegraphics[width=0.45\textwidth]{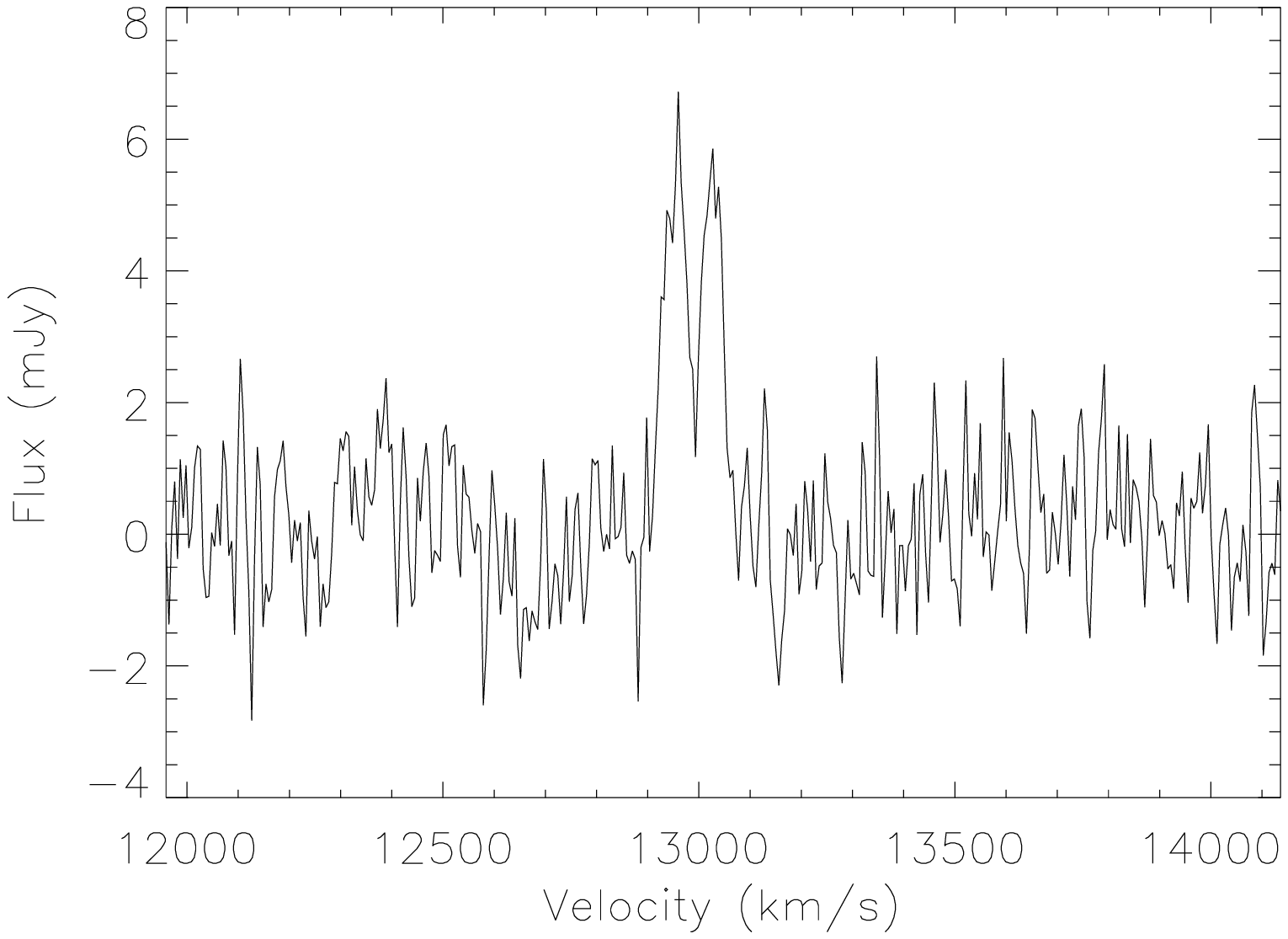}}
  
  \contcaption{Top to bottom: {\it B} band images and accompanying \hi\/
    spectra for objects J014752+014752 and J014834+152756\label{fig15}}
\end{figure*}

\begin{figure}
  \begin{center}
    \includegraphics[width=60mm,height=40mm]{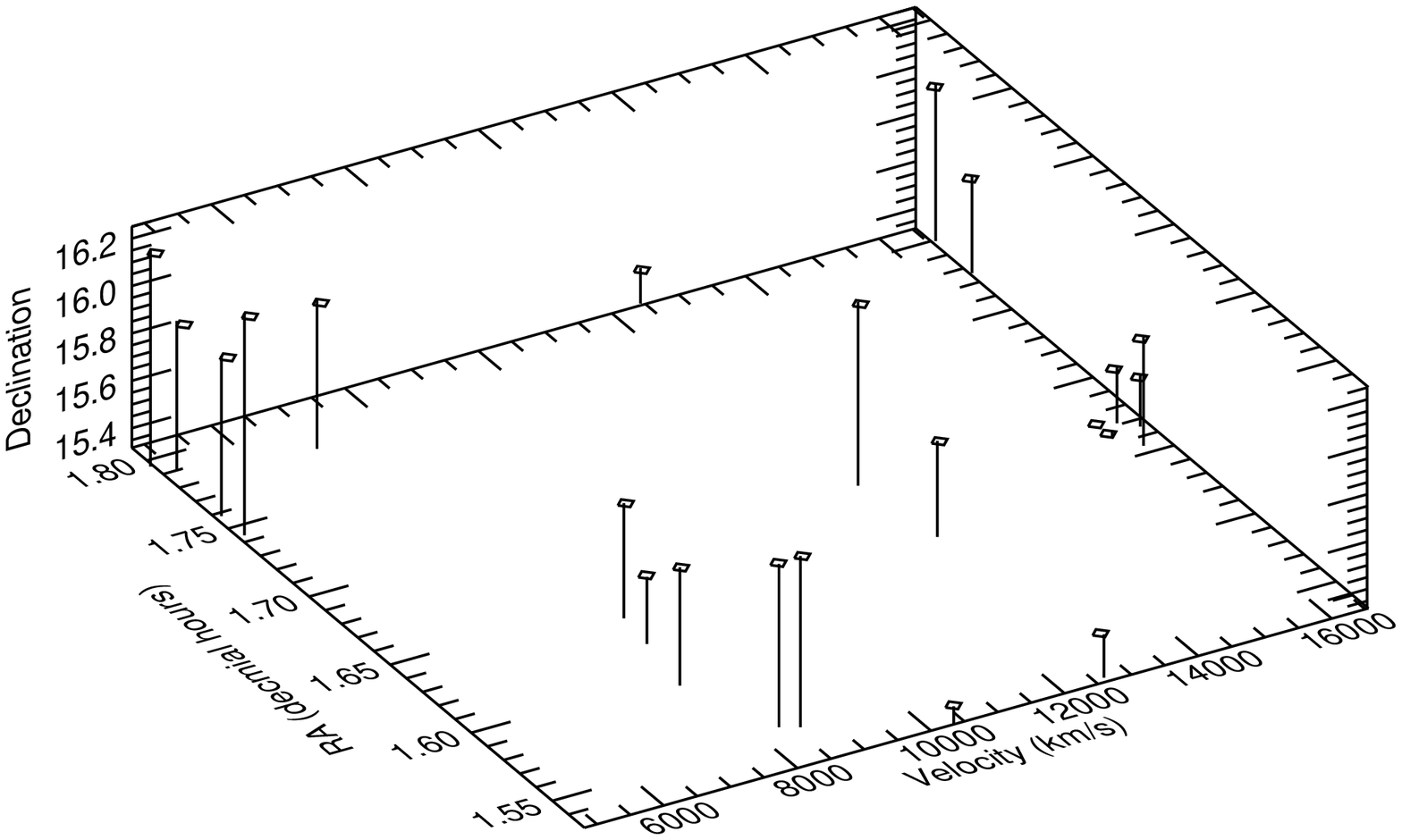}
    \caption{A 3D plot of the positions of the galaxies beyond the
    NGC 628 group with their \hi\/ redshifts.\label{fig19}}
  \end{center}
\end{figure}

 There are a further 15 objects whose signals were considered too
 marginal (S/N $\la 3\sigma$ and $\Delta v<$ 30 \kms) 
 to be included in this paper pending further follow-up observations
 from Arecibo and several other large radio telescopes. 
From the velocity information alone there appear to be four group associations at around 6000, 8000,
12500 and 16000 \kms. J013149+152353 is the one exception being the
only galaxy that has no other neighbour within 1700
\kms. Fig. \ref{fig19} is a 3D plot of the galaxies, position and
redshift. From this it is possible to distinguish 3 groups: 1 group at
$\sim$5000 \kms, one at $\sim$8000 \kms\/ and one at $\sim$17000 \kms. The
objects at $\sim$12500 \kms\/ seem too scattered in RA to be considered
a group, although they may be the massive tracers of a filament
running at right-angles to our line of sight at this velocity.

\subsection{Detection efficiency}

The precursor survey covered an area of about 4.6 sq.deg. with an rms
noise of 1.1 mJy (for 10 \kms\/ resolution) and another 0.9 sq.deg. with
an rms noise of 1.8 mJy. Using these noise values we tested the
distribution of detections beyond the NGC 628 group using a
$V/V_{max}$ test following the procedures described in greater detail
by Rosenberg \& Schneider (2002). Assuming the line-width dependence
for detections that they found along with a roll-off in completeness
near the detection limit, the expected mean value of $V/V_{max}$
should be 0.61. By adjusting the signal-to-noise level at which there
is a 50\% completeness in the detection rate, we found that at an
effective S/N = 7 (for a source of 300 \kms\/ line width), the correct
value of $V/V_{max}$ was recovered. This sensitivity level is very
similar to those derived from $V/V_{max}$ tests of past blind \hi\/ 
surveys.

Given these sensitivity estimates, we modelled the predicted number of
sources that should be detected within the precursor survey
volume. This analysis is complicated by the fact that the precursor
survey covered an area centred on a known group with a substantial
void region behind it. We populated the volume with galaxies based on
the HIMF derived from an analysis of the Parkes data
(Zwaan \etal\/ 2003) down to an \hi\/ mass of $3\times10^6 $M$_{\odot}$. The
model galaxies were given rotation speeds consistent with the range
normally observed for their \hi\/ mass, and then given a random
orientation as well as a standard deviation of 300 \kms\/ relative to
the Hubble flow. We then `observed' the volume at the measured
sensitivity and completeness levels of the survey and determined which
galaxies would be detected, repeating this process 1000 times to
estimate the likely number of galaxies we should detect.

Based on the Parkes HIMF function we would have expected to detect
49.2$\pm$6.8 galaxies in total. This is almost twice as high as the
actual number detected, 27, but only about 3$\sigma$ from the
predicted number. Since the nearby volume contains both a known group
and a void, we can look instead at the results for the more distant
portion of the volume---indeed no galaxies were detected with
velocities between that of the NGC 628 group and $cz = 4890$ \kms. If
we consider the volume beyond 4500 \kms, where we might
anticipate that the volume is relatively unbiased by local large scale
structure, the predicted number of detections is 42.3$\pm$6.3. This is
again about 3$\sigma$ more than the number detected in this volume,
22.

The predicted number of sources for the Parkes HIMF is higher
than would be predicted with most other mass functions because of the
relatively high normalisation factor ($\Phi_\star = 0.0086$
Mpc$^{-3}$), which indicates the relative number of galaxies near the
`knee' in the mass function at log(M$_{HI\star}$/M$_{\odot}$) =
9.79. Both the HIMFs of Rosenberg \& Schneider (2002) and
Zwaan \etal\/ (1997) have lower normalisation factors yielding a
predicted number of detections of 37.2$\pm$6.2 and 28.4$\pm$5.3
respectively. Given the limited area covered by the precursor study,
this result may indicate more about cosmic variance than about the
HIMF. It should also be noted that because of the relatively 
low sensitivity of the Parkes survey, its results are based on a
relatively nearby volume of space. The full AGES survey will provide a
much larger and deeper volume for testing the variance of the HIMF.

\subsection{$H$-band properties}

In order to preform a preliminary study of the stellar properties 
of our \hi\/ selected sample we use the 2MASS (Skrutskie \etal\/ 2006) 
data available for the candidate 
NIR counterparts listed in Table \ref{tbl6}.
The near infrared data are ideal for this kind of analysis since 
they are less affected by extinction and they are a good indicator 
of stellar mass (Gavazzi \etal\/ 1996; Rosenberg, Schneider \&
Posson-Brown 2005).

\begin{figure}
  \begin{center}
    \includegraphics[width=0.48\textwidth]{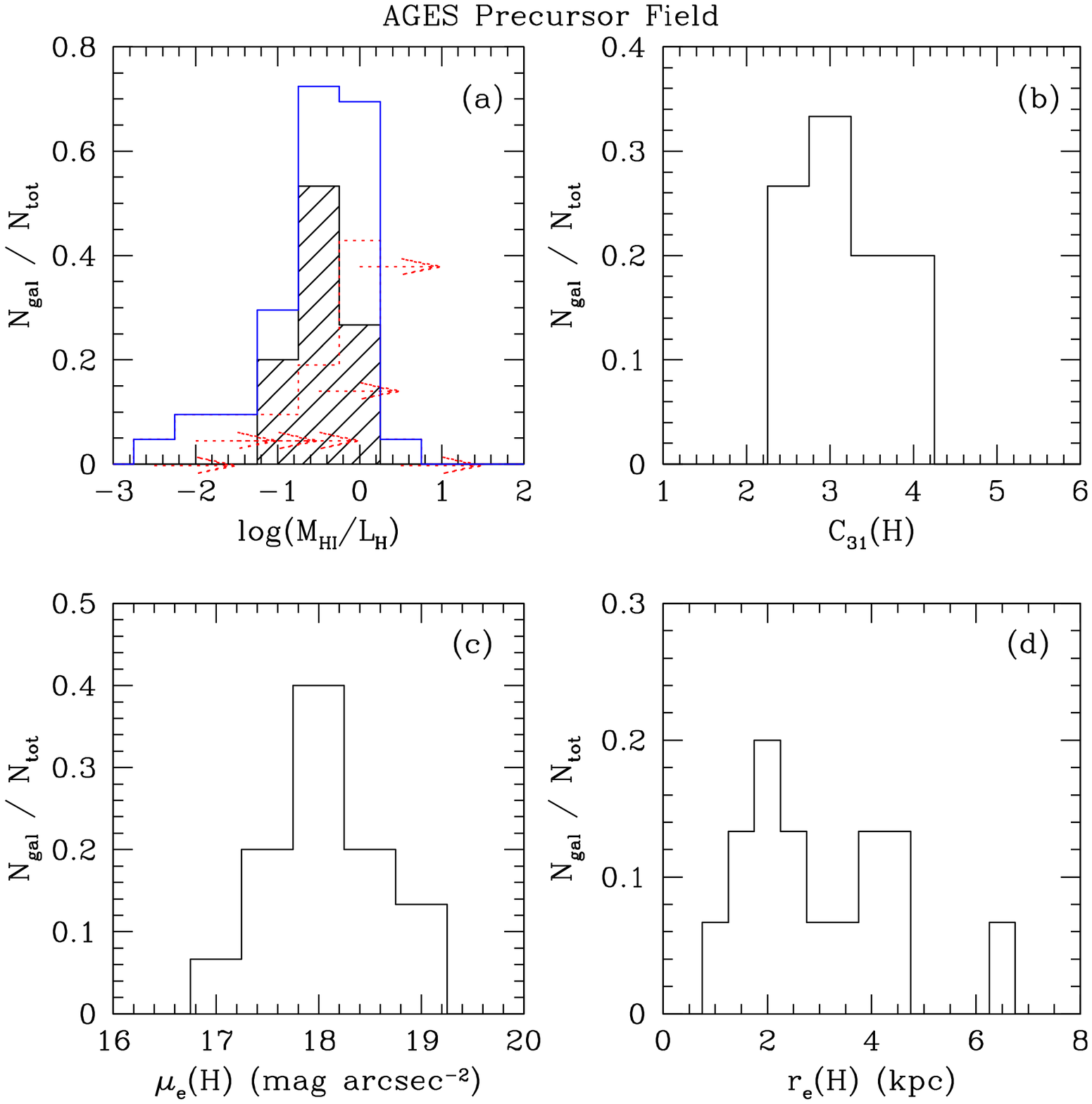}
    \caption{AGES precursor sample \hi\/--$H$-band properties. (a) M$_{HI}/$L$_H$ distribution. The shaded region represents measured values, the dotted
    lines represent estimated lower limits for those galaxies that
    were not detected in $H$-band. The white region is the sum of the
    measured values and the estimated values. (b) $H$-band concentration
    parameter distribution. Bulge
    galaxies are those galaxies with a value greater than 4. (c)
    $H$-band effective surface brightness distribution. (d) $H$-band effective radius. \label{fig20}}
  \end{center}
\end{figure}

\begin{figure}
  \begin{center}
    \includegraphics[width=0.48\textwidth]{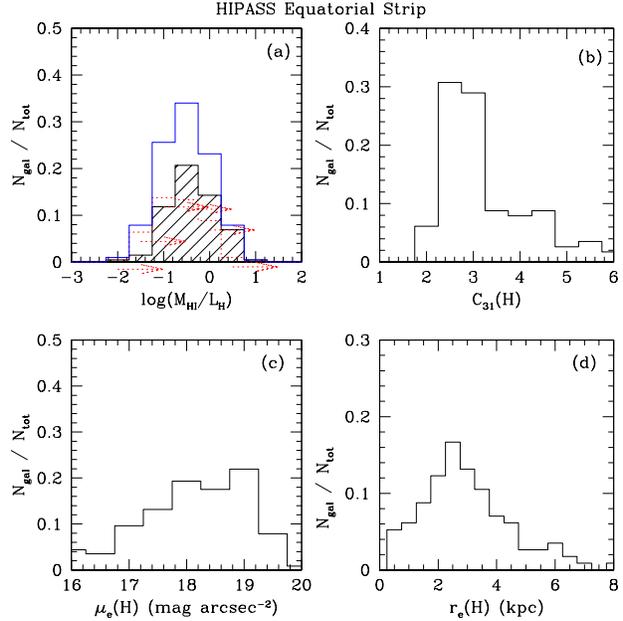}
    \caption{HIPASS Equatorial Strip \hi\/--$H$-band properties. (a) M$_{HI}/$L$_H$ distribution. The shaded region represents measured values, the dotted
    lines represent estimated lower limits for those galaxies that
    were not detected in $H$-band. The white region is the sum of the
    measured values and the estimated values. (b) $H$-band concentration
    parameter distribution. Bulge
    galaxies are those galaxies with a value greater than 4. (c)
    $H$-band effective surface brightness distribution. (d) $H$-band effective radius. \label{fig21}}
  \end{center}
\end{figure}

2MASS data (extracted from the extended source catalogue) 
were available for only 15 candidate counterparts of our 27 \hi\/ sources.
Fig. \ref{fig20} shows the distribution of the M$_{HI}/L_{H}$ ratio, of 
the $H$-band effective radius ($r_{e}$: the radius containing 50\% of the 
light), effective surface brightness ($\mu_{e}$: the mean surface brightness 
within $r_{e}$) and concentration index ($C_{31}$: the ratio of 
the radii containing 75\% and 25\% of the light.) for our sample.   
As a reference, Fig. \ref{fig21} shows the same distribution for a
sample of 114 galaxies belonging to the HIPASS equatorial strip
(Garcia-Appadoo, 2005). The two samples occupy a similar parameter
space being composed by `normal' disk-like gas-rich galaxies. Any deeper
analysis of the stellar content of the \hi\/ sources must be
postponed to future works due to the lack of sufficiently deep near
infrared and optical data. We only point out that the distribution of
the $M_{HI}/L_{H}$ ratio shown in Fig. \ref{fig20} has to be considered as a
lower limit of the real distribution of AGES sources, since it is
based on the 15 brightest $H$-band candidate counterparts in our sample.

\section{Conclusions}
\label{sec:conc}

The precursor observations of NGC 628 and its companions have shown
that the observing strategy and the data reduction pipeline are all
working successfully. The detection rate is close to what we would expect for
this volume. With 192s integration time it has been possible
to reach a column density limit of 2$\times10^{18}$cm$^{-2}$. Studying the
NGC 628 group has revealed no intergalactic gas down to this column
density limit. 

The detection of dw0137+1541 has shown that the survey is sensitive to
low mass \hi\/ objects. We have also demonstrated the ability of AGES to
detect galaxies out to a velocity of 17500 \kms. 3 groups have been
identified beyond the NGC 628 group. Of the 22 detections outside of
the NGC 628 group, 9 are new detections and 3 galaxies with
redshift data available from SDSS compare well with the \hi\/
redshifts. 15 low signal-noise detections have also been identified and we
have applied for time at several large radio telescopes for follow-up
observations to confirm these sources.

From the point of dark galaxy detection, superCOSMOS images have
revealed that there is an optical detection within each \hi\/ beam,
but these objects cannot be assumed to be associated with the \hi\/
emission on grounds of spatial coincidence alone. Optical redshift
data, and possibly higher resolution \hi\/ observations will be
required to confirm the associations. To this end we are applying for
follow-up observations on several other telescopes.

One of the project goals is to study the HIMF in different
environments. With only 27 objects this sample, is much too small to
perform any meaningful analysis of the HIMF. However, we have
shown that the quality of this data is good enough that the
galaxies can be included with those from future AGES fields to
compile a HIMF.

\section*{Acknowledgements}
The AGES group wish to thank the NAIC director, Bob Brown, for granting us time on 
the Arecibo radio telescope. The Arecibo Observatory is
  part of the National Astronomy and Ionosphere Center, which is
  operated by Cornell University under a cooperative agreement with
  the National Science Foundation. We would also like to thank all the
  staff at the observatory who ensured the observations went as
  smoothly as possible. Individual thanks go to Mikael Lerner, Phil
  Perillat and Jeff Hagen for their excellent work on the data
  collection system and the telescope operating software. We are very
  grateful to the ALFALFA and AUDS working groups for sharing their
  knowledge and experiences with us and helping us to avoid potential
  pitfalls. Thanks also to Carl Heiles for allowing us to use his
  images. J. L. Rosenberg is funded under NSF grant ast-0302049. This
  research has made use of the NASA/IPAC Extragalactic Database (NED)
  which is operated by the Jet Propulsion Laboratory, California
  Institute of Technology, under contract with the National
  Aeronautics and Space Administration.

\bibliographystyle{mnras}

\bibliography{precursor-mnras}
\nocite{1}
\nocite{2}
\nocite{3}
\nocite{4}
\nocite{5}
\nocite{6}
\nocite{7}
\nocite{8}
\nocite{9}
\nocite{10}
\nocite{11}
\nocite{12}
\nocite{14}
\nocite{15}
\nocite{16}
\nocite{17}
\nocite{18}
\nocite{19}
\nocite{20}
\nocite{21}
\nocite{22}
\nocite{23}
\nocite{24}
\nocite{25}
\nocite{26}
\nocite{27}
\nocite{28}
\nocite{29}
\nocite{30}
\nocite{31}
\nocite{32}
\nocite{33}
\nocite{34}
\nocite{35}
\nocite{36}
\nocite{37}
\nocite{38}
\nocite{39}
\nocite{40}
\nocite{41}
\nocite{42}
\nocite{43}
\nocite{44}
\nocite{45}

\end{document}